\documentclass{aa}
\usepackage{graphicx}
\usepackage{txfonts}
\usepackage{amstext}
\usepackage{hyperref}
\usepackage{float}
\usepackage{placeins}
\hypersetup{
  colorlinks   = true, %Colours links instead of ugly boxes
  urlcolor     = blue, %Colour for external hyperlinks
  linkcolor    = blue, %Colour of internal
  citecolor    = blue  %Colour of citation}
}

\begin{document}

\newcommand{\va}{ V923~Aql }
\newcommand{\ve}{V923~Aql}
\newcommand{\bu}{$\bullet$ }
\newcommand{\Mnom}{\hbox{$\mathcal{M}^{\mathrm N}_\odot$}}
\newcommand{\Rnom}{\hbox{$\mathcal{R}^{\mathrm N}_\odot$}}
\newcommand{\Lnom}{\hbox{$\mathcal{L}^{\mathrm N}_\odot$}}
\newcommand{\GMnom}{\hbox{$\mathcal{GM}^{\mathrm N}_\odot$}}
\newcommand{\spefo}{{\tt SPEFO} }
\newcommand{\spefoe}{{\tt SPEFO}}
\newcommand{\phoebe}{{\tt PHOEBE} }
\newcommand{\phoebee}{{\tt PHOEBE}}
\newcommand{\fotel}{{\tt FOTEL} }
\newcommand{\fotele}{{\tt FOTEL}}
\newcommand{\korel}{{\tt KOREL} }
\newcommand{\korele}{{\tt KOREL}}
\newcommand{\binsyn}{{\tt BINSYN} }
\newcommand{\binsyne}{{\tt BINSYN}}
\newcommand{\pyt}{{\tt PYTERPOL} }
\newcommand{\pyte}{{\tt PYTERPOL}}
\newcommand{\hecdd}{{\tt HEC22} }
\newcommand{\hecdde}{{\tt HEC22}}

\newcommand{\tria}{\hbox{$\bigtriangleup$}}
\newcommand{\ubv}{\hbox{$U\!B{}V$}}
\newcommand{\ubvr}{\hbox{$U\!B{}V\!R$}}
\newcommand{\bv}{\hbox{$B\!-\!V$}}
\newcommand{\ub}{\hbox{$U\!-\!B$}}
\newcommand{\uvby}{\hbox{$uvby$}}
\newcommand{\hp}{$H_{\rm p}$}

\newcommand{\p}{\hbox{$\pm$}}
\newcommand{\arcm}{$^\prime$}
\newcommand{\arcs}{$^{\prime\prime}$}
\newcommand{\m}{$^{\rm m}\!\!.$}
\newcommand{\D}{$^{\rm d}\!\!.$}
\newcommand{\F}{$^{\rm P}\!\!.$}
\newcommand{\kms}{km~s$^{-1}$ }
\newcommand{\ks}{km~s$^{-1}$}
\newcommand{\ms}{M$_{\odot}$}
\newcommand{\rs}{R$_{\odot}$}
\newcommand{\oc}{$O-C$}
\newcommand{\tef}{$T_{\rm eff}$}
\newcommand{\lgg}{$\log~g$}
\newcommand{\vsin}{$v\sin~i$}

\newcommand{\ha}{H$\alpha$ }
\newcommand{\hb}{H$\beta$ }
\newcommand{\hg}{H$\gamma$ }
\newcommand{\hae}{H$\alpha$}
\newcommand{\hbe}{H$\beta$}
\newcommand{\hge}{H$\gamma$}
\newcommand{\hde}{H$\delta$}
\newcommand{\he}{\ion{He}{i}~6678 }
\newcommand{\hea}{\ion{He}{i}~6678}
\newcommand{\cii}{\ion{C}{ii} }
\newcommand{\Am}{\ANG~mm$^{-1}$ }
\newcommand{\Ame}{\ANG~mm$^{-1}$}

\newcommand{\hip}{$Hipparcos$}
\newcommand{\ond}{Ond\v{r}ejov}
\newcommand{\ova}{Ostrava}
\newcommand{\valmez}{Vala\v{s}sk\'e Mezi\v{r}\'{\i}\v{c}\'{\i}}
\newcommand{\jil}{J\'{\i}lov\'e}
\newcommand{\hra}{Hradec Kr\'alov\'e}
\newcommand{\pec}{Pec pod Sn\v{e}\v{z}kou}
\newcommand{\tso}{Trhov\'e Sviny}
\newcommand{\pro}{Prost\v{e}jov}
\newcommand{\mbo}{Mlad\'{a} Boleslav}

\title{Long-term, orbital, and rapid variations \\ of the Be star V923 Aql = HD~183656
\thanks{Based on new spectral and photometric observations from the following
observatories: Canakkale, Dominion Astrophysical Observatory, ESO La Silla, Haute Provence Chiron, Hvar, Mt. Palomar, Ond\v{r}ejov, San Pedro M\'artir, Sierra Nevada, Toronto, Trieste, Tubitak National Observatory, ASAS and ASAS-SN services, BeSS spectra database, and Kamogata-Kiso-Kyoto Wide-field Survey.
Tables 1 -- 4 are available in electronic form at {\tt www.aanda.org} } }
\author{M. Wolf~\inst{1}\and
P. Harmanec\inst{1}\and
H.~Bo\v{z}i\'c~\inst{2}\and
P.~Koubsk\'y~\inst{3}\and
S.~Yang\inst{4}\and
D.~Ru\v{z}djak\inst{2}\and
M.~\v{S}lechta\inst{3}\and \\
H.~Ak\inst{5}\and
H.~Bak\i\c{s}~\inst{6}\and
V.~Bak\i\c{s}~\inst{6}\and
A.~Opli\v{s}tilov\'a~\inst{1}\and
K.~Vitovsk\'y~\inst{1}
                    }
\institute{Astronomical Institute, Faculty of Mathematics and Physics,
Charles University, V~Hole\v{s}ovi\v{c}k\'ach~2, CZ-180~00~Praha~8, \\
Czech Republic, \email{wolf@cesnet.cz, hec@sirrah.troja.mff.cuni.cz}
\and
Hvar Observatory, Faculty of Geodesy, Zagreb University,
Ka\v ci\'ceva~26, HR-10000 Zagreb, Croatia
\and
Astronomical Institute, Czech Academy of Sciences, Fri\v{c}ova~298,
CZ-251~65~Ond\v{r}ejov, Czech Republic
\and
Physics \& Astronomy Department, University of Victoria,
    PO Box 3055 STN CSC, Victoria, BC, V8W 3P6, Canada
\and
Erciyes University, Science Faculty, Astronomy and Space Sci. Dept.,
38039~Kayseri, Turkey
\and
Department of Space Sciences and Technologies, Akdeniz University, Faculty of Science, Antalya, Turkey
}
\date{Received \today}

\abstract {We present the latest results of a long-term observational project
aimed at observing, collecting from the literature, and homogenising
the light, colour, and spectral variations of the well-known emission-line Be star \ve. Our analysis
of these parameters confirms that all of the observables exhibit
cyclic changes with variable cycle length between about 1800 and 3000
days, so far documented for seven
consecutive cycles. We show that these
variations can be qualitatively understood within the framework of
the model of one-armed oscillation of
the circumstellar disk, with a wave of
increased density and prograde revolution in space. We confirm
the binary nature of the object with
a 214.716 day period and estimate the
probable system properties. We also confirm the presence of rapid light,
and likely also spectral changes. However, we cannot provide any firm conclusions regarding their nature.
A quantitative modelling study of long-term
changes is planned as a follow-up to this work.
}

\keywords{binaries: spectroscopic --
  stars: emission-line, Be --
  stars: early-type --
  stars: individual: V923 Aql  --
  stars: fundamental parameters }

\maketitle

\section{Introduction}

Be stars are early-type B-type stars whose spectra have
exhibited emission in the Balmer lines at least once over the course of their recorded history.
In particular, the \ha emission line is typically the dominant feature in the spectra
of such stars and many authors have studied the time evolution of the Balmer emission
line profiles in order to understand the Be star phenomenon better. In spite of the substantial efforts of several generations of astronomers, the true reason for the repeated appearance of the
emission lines remains unknown. For further details, we refer  to a detailed review of
the Be star properties, variability, and modelling by \citet{rivi2013}.

Renewed interest in the studies of long-term spectral and light
variations of Be stars started after \citet{lee91} proposed a model
of a viscous decretion disk (VDD). It was investigated and developed in
a number of studies, for example, by \citet{telting94,carci2006,carci2009,carci2012,haub2012}, and \citet{ghore2018}.
Quite often, such studies are based on a modelling of the time evolution
of one or a limited set of parameters, such as the secular light changes  of $\omega$~CMa in the $V$ passband \citep{ghore2018}. However, to
restrict a certain class of models, it is necessary to study the time
evolution of a number of well-observed variables. The importance of
such an approach has been demonstrated, for instance, in a recent study of
$\beta$~Lyr \citep{broz2021}.  Systematic studies of individual well-observed Be stars over an extended time interval and an investigation of the mutual behaviour of various observables is, therefore, of utmost importance.
Individual Be stars exhibit wide variety of complex variability patterns,
some of them exhibiting the most substantial light changes in the visual region, while
others such as \ve, exhibit the most significant light changes in the ultraviolet
region. Also, the mutual relation between various spectral and photometric
observables differs from one star to another. The collection of a representative
number of well-documented cases may therefore provide a good starting point
for critical tests of the  models that have been proposed thus far. This study represents
one such attempt, that is, a carefully documented analysis of
spectral and photometric observations for the well-known bright Be star
\va (HR~7415, HD~183656, HIP~95929, MWC~988; $V$=6\m0-6\m2 var.), which belongs to the type of Be stars exhibiting long-term radial-velocity (RV) and a $V/R$ ratio of the double Balmer emission lines, as well as light and colour changes.

\section{History of investigations}
\subsection{Spectral variations}
Initially, \citet{harp37} reported variable RVs and the presence of
narrow Balmer lines and strong \ion{Fe}{ii} lines. These were interpreted by \citet{bidel50} as shell lines. He also noted the presence of diffuse \ion{He}{i} lines. \citet{merrill52} studied the RV variations of Balmer and metallic shell lines from ten spectra secured between 1949 and 1952 and  concluded that they varied with a~6.5~year period.
\citet{gulli76} studied a collection of spectra of the star secured between 1949 to 1975, including the spectra used by \citet{merrill52}. He confirmed the cyclic long-term RV variations and found also parallel $V/R$ changes, with $V/R$ and RV maxima coinciding. In reference to the binary hypothesis of the Be star origin by \citet{kh75}, he remarked that it would be desirable to measure small RV changes on high-dispersion spectra. Another spectroscopic study was published by \citet{ring81}, in which the authors studied a few spectra and demonstrated that the RV of the broad emission wings of \ha differs from the shell RV on the spectra taken on JD 2444450 - 52.
\citet{alduseva80} measured the RVs and emission strengths of the double
\ha emission based on two spectra from 1979.
\citet{ring84} obtained optical and IUE spectra in July 1981 and measured and published their RVs, concluding that the high-excitation UV lines are formed in a transition region between the stellar photosphere and the disk where the shell lines are formed.
\citet{zarf13} collected all the RVs available in the astronomical literature and measured new RVs on the spectra from several observatories. They found that over the whole interval of 60 years covered by the available data, the RVs varied cyclically, with an average cycle length of about 5.8~years and a full amplitude of about 80~\ks, noting, however, the variable length of the individual cycles. After prewhitening the RVs for these cyclic
changes with the help of spline functions, they found that the RV residuals vary periodically, with a 214\fd7 period and a semi-amplitude of 6.4~\ks ; they interpreted this as the orbital motion in a binary with a circular orbit.
\citet{deniz94} studied optical and near-infrared spectra of \ve. They confirmed variable Balmer \ha line profiles with shell components and
calculated the parameters of the envelope. \citet{arias2004} analysed UV and visual spectra. They used \ion{Fe}{ii} lines to derive temperatures and locations of the line-forming regions. Their results indicate that the dimensions of the circumstellar envelope vary with the phase of
the orbital period of 214.75 days. In addition, they also estimated a~cycle
length of 6.8 years for the cyclic $V/R$ cyclic changes of \hae, which was
also found in the RVs and photometry.

\subsection{Light and colour changes}
The light variability of \ve, with a~characteristic timescale of 0\fd85 and
a~secularly variable amplitude, was discovered by \citet{lynds60}.
He provided the ephemeris of the light minima, that is:
\begin{equation}
    T_{\rm min.light}={\rm JD}~2436458.66+0\fd8518\cdot E \,.
\end{equation}
\citet{percy88} reported small-amplitude light variations on several
time scales and noted a light decrease on JD~44810.
\citet{mennic94} studied systematic ESO \uvby \ photometry of \va spanning
about 3000~days. They found cyclic variations with a cycle length of about
seven~years and amplitudes of 0\m07 in $y$, and 0\m25 in $u$. They also
found rapid variations on a time scale of days, reaching up to 0\m1, but
without any obvious periodicity. However, in the residuals from the
long cycle, they detected a possible periodicity of 261~d. \citet{hvar5}
studied systematic \ubv \ observations secured at Hvar and spanning
about 4000~d. They basically confirmed the results found by
\citet{mennic94}. A similar behaviour was also reported by \citet{percy97}.
\citet{balona95} reported a period of 0\fd882, but without giving any further details.
 The Hipparcos \hp \ photometry \citep{esa97} was analysed
by \citet{hubert98}, who found a period of 0\fd652, and by \citet{percy02},
who was unable to detect any short period. \citet{gutie2007} analysed \uvby \ photometry
from Observatorio de Sierra Nevada (OSN) secured in 2002 and 2005. For 2002
data, they found a period of 0\fd4566, which is, as they remarked, a one-day alias of the period found by \citet{lynds60}. Removing a systematic trend
from the 2005 observations, they found a short period of 0\fd2755.

\section{Available observational data and their reduction}
Throughout this paper, we use the abbreviated
form for heliocentric Julian dates, RJD = HJD-2400000.0,
to avoid overlooking the 0.5~d shift in case of MJD. Whenever the heliocentric Julian dates were not provided in the original papers, we calculated them ourselves.

\begin{figure}[!]
%\centering
\resizebox{1.0\hsize}{!}{\includegraphics{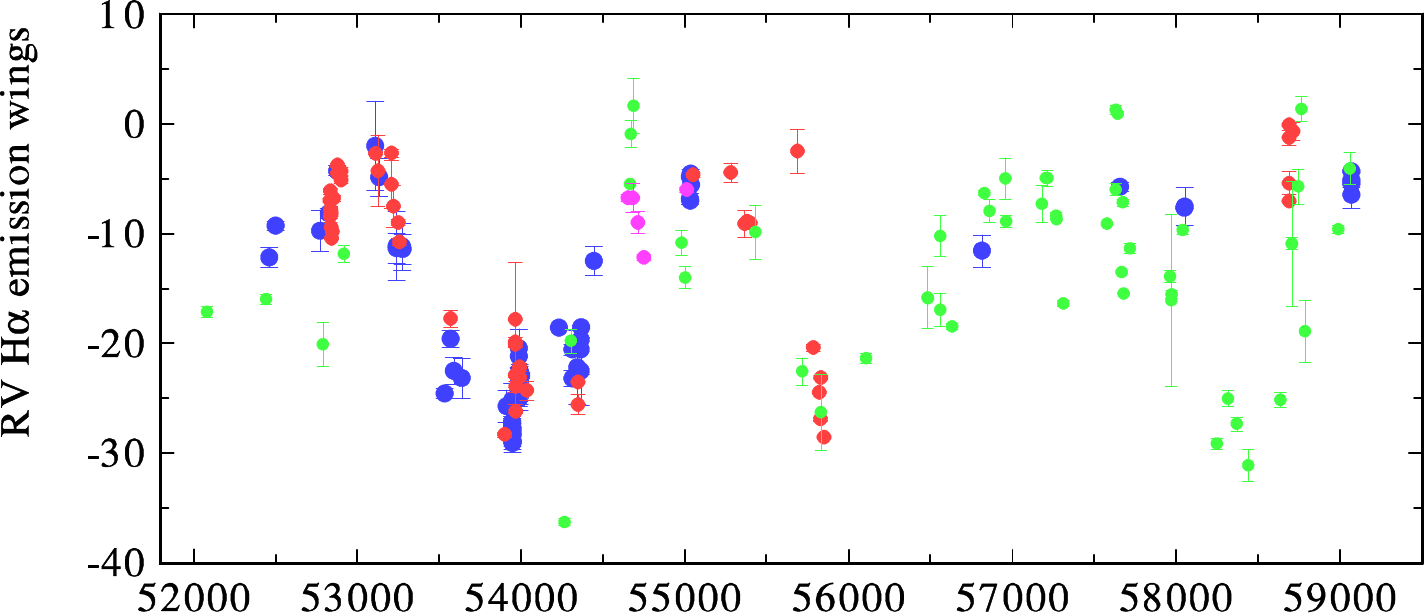}}
\resizebox{1.0\hsize}{!}{\includegraphics{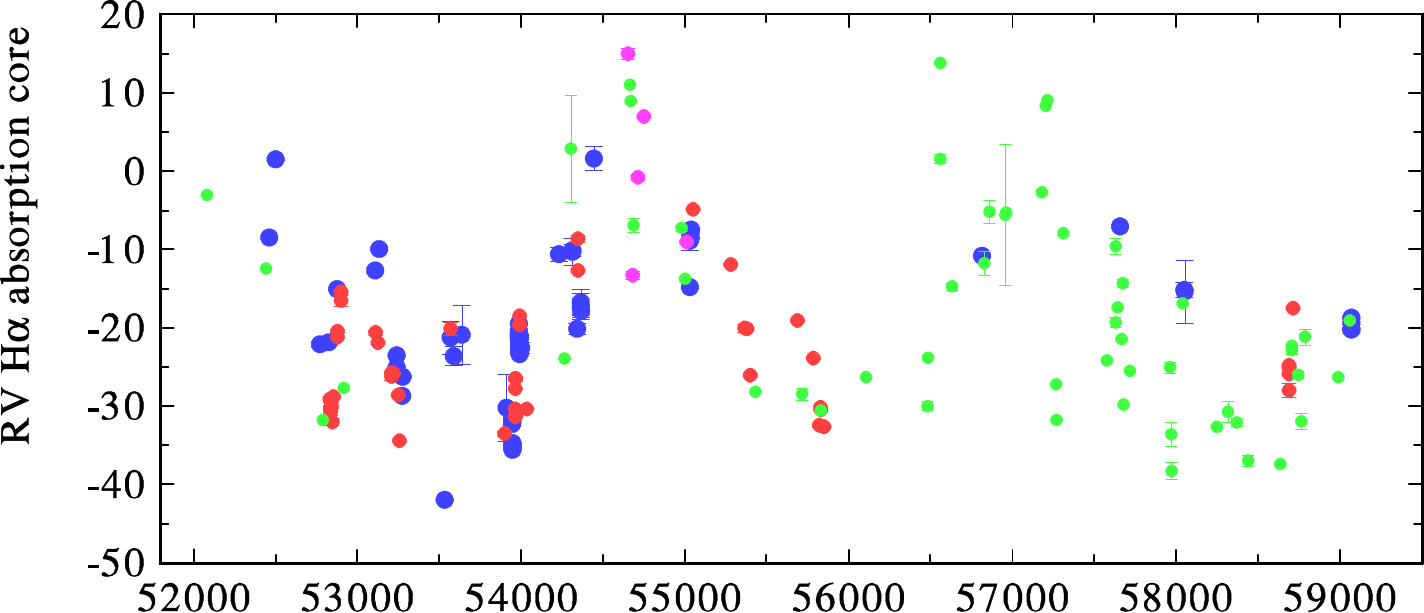}}
\resizebox{1.0\hsize}{!}{\includegraphics{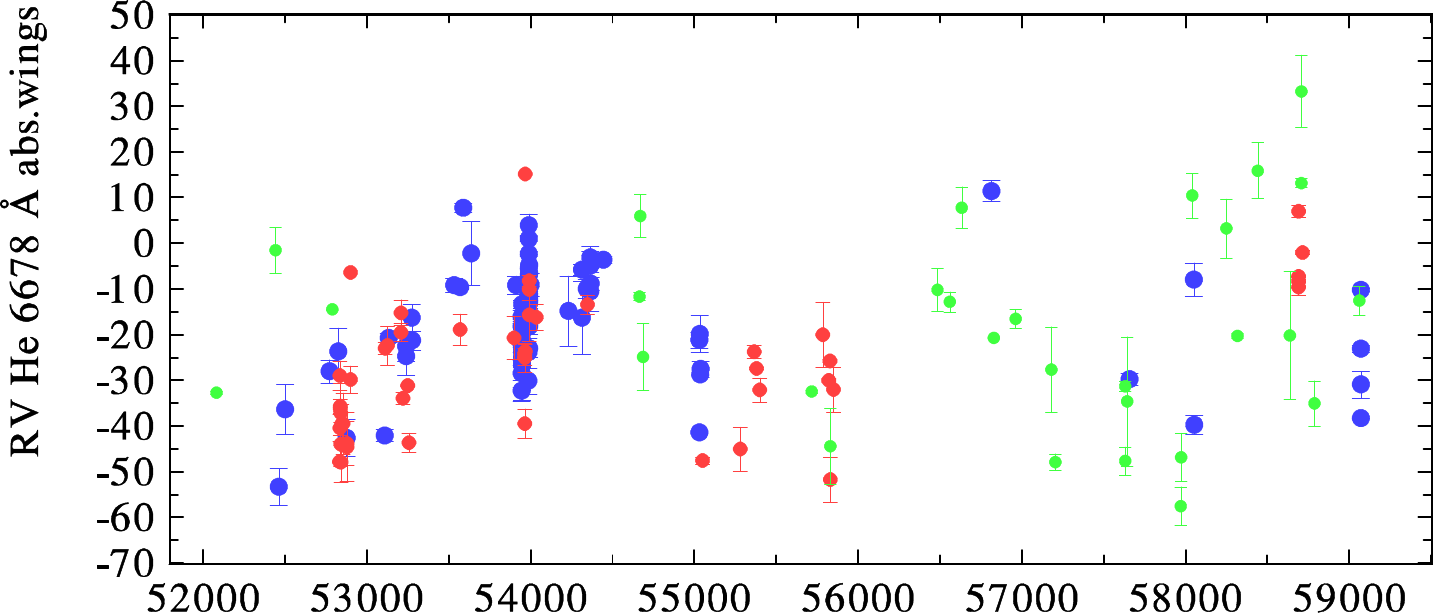}}
\resizebox{1.0\hsize}{!}{\includegraphics{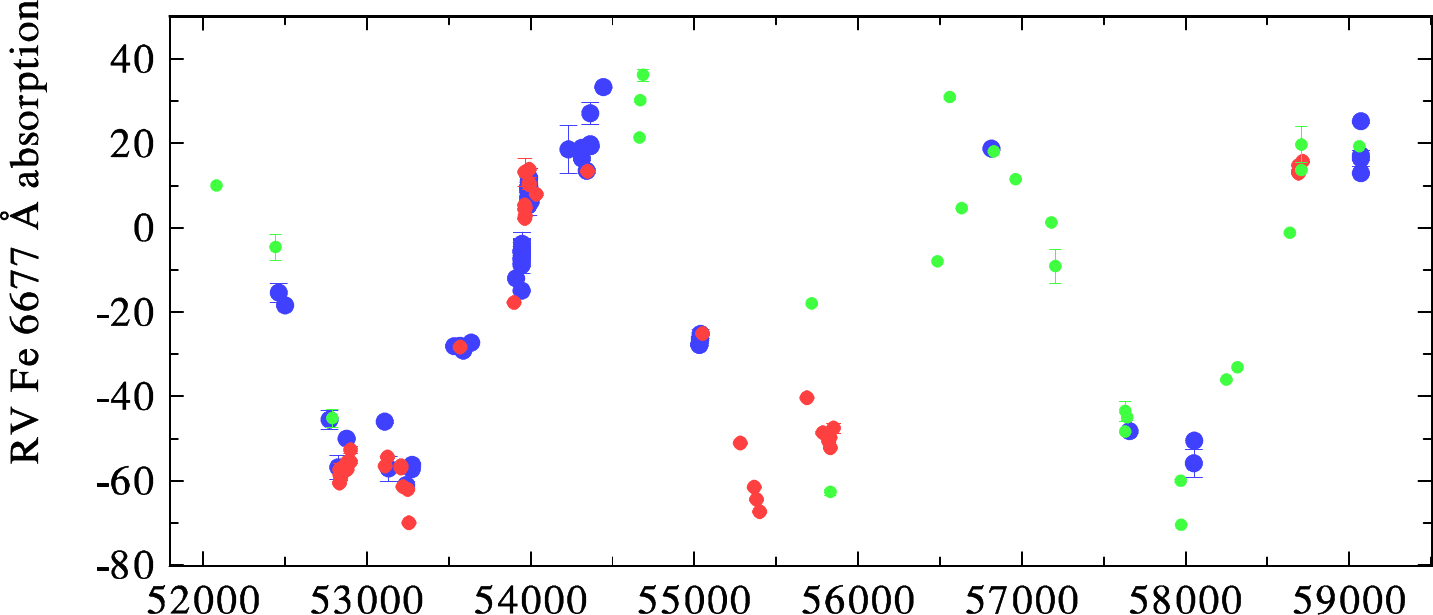}}
\resizebox{1.0\hsize}{!}{\includegraphics{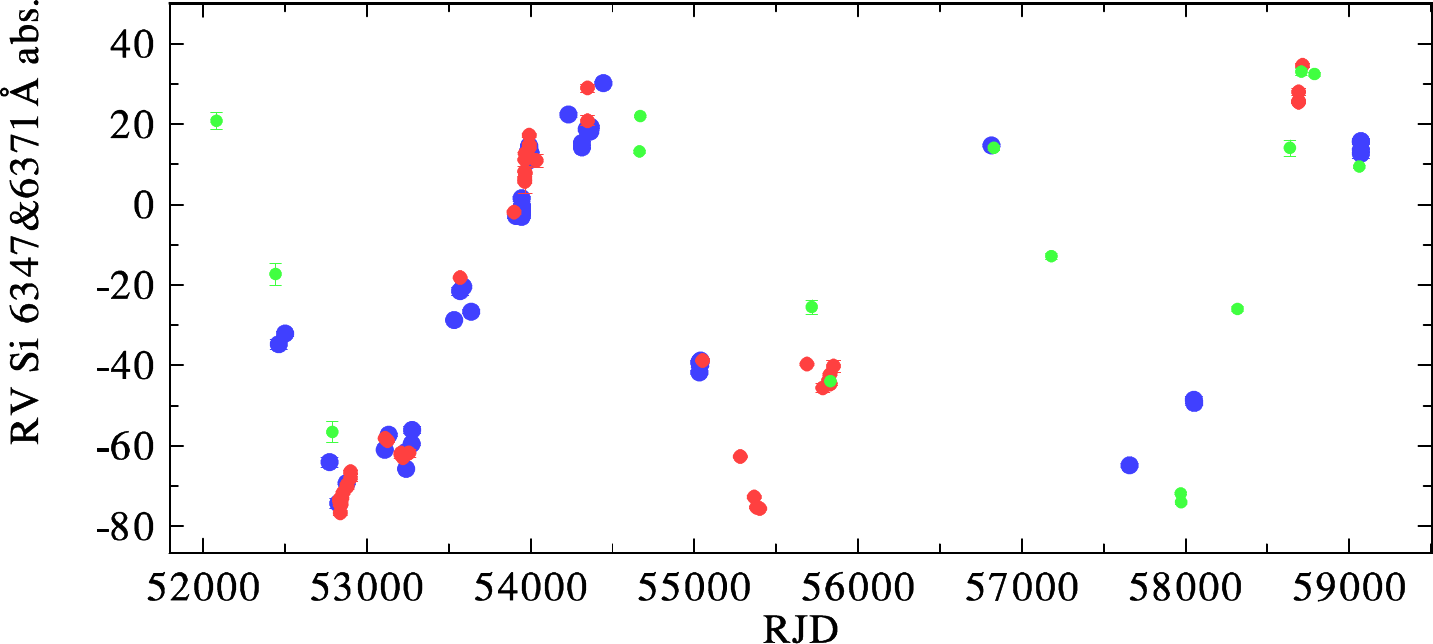}}
\caption{Time plots of RVs (in \ks)
of individual measured features based on our new measurements.
From top to bottom: \ha emission wing, \ha absorption core, He~6678~\AA \ absorption wings,
\ion{Fe}{ii} absorption core, and the mean \ion{Si}{ii}~6347\&6371~\AA.
The RVs taken from the higher-resolution Elodie, Feros, and DAO are shown as
blue circles, those from OND in red, those from the BeSS amateur spectra in green, and those from our re-measurements of the Trieste \ha spectra are shown as magenta circles.}
\label{rvnew}
\end{figure}

\begin{table}[h]
\caption{All RVs from the literature, edited and corrected for misprints.
Data sources are identified by spectrographs numbers (Spg.No.) identical to those in Table D.2 .}
\begin{center}
\label{radvel1}
\begin{tabular}{cccccc}
\hline\hline\noalign{\smallskip}
RJD & RV & Spg. No. \\
    &(km s$^{-1}$) \\
\hline\noalign{\smallskip}
25159.6571 &  -20.90   &  1\\
25159.6571 &  -20.90   &  1\\
25829.7890 &  -53.30   &  1\\
25836.8138 &  -48.20   &  1\\
25839.7766 &  -58.50   &  1\\
... & ... & ... \\
\hline
\end{tabular}
\tablefoot{ This table is available in its entirety in machine-readable form.}
\end{center}
\end{table}

\begin{table*}[h]
\caption{All RVs with their rms errors (in \ks) measured on
electronic spectra at our disposal. Data sources are identified by letters identical to those from table D.1. Missing values are denoted by 9999.99.}
\begin{center}
\label{radvel2}
\begin{tabular}{crrrrccccccc}
\hline\hline\noalign{\smallskip}
 & \multicolumn{4}{c}{H$\alpha$}  &  \multicolumn{2}{c}{ He II 6678} & \multicolumn{2}{c}{Fe II 6677} & \multicolumn{2}{c}{Si II 6347\&6371} & File \\
  RJD   &  \multicolumn{2}{c}{wings of emission} & \multicolumn{2}{c}{absorption core} &  \multicolumn{2}{c}{absorption wings} & \multicolumn{2}{c}{absorption core } &  \multicolumn{2}{c}{absorption cores} \\
   &  RV     &  rms   &    RV   &    rms   &   RV  &    rms    & RV    &    rms   &  RV    &   rms \\
\hline\noalign{\smallskip}
52461.8429  & -12.15  &   0.93  &  -8.49  &   0.17  & -53.30  &   3.98  & -15.38  &   2.19  & -34.70  &   1.31 & A \\
52500.4383  &  -9.25  &   0.44  &   1.51  &   0.10  & -36.33  &   5.41  & -18.45  &   0.26  & -32.10  &   0.52 & B \\
52771.9048  &  -9.74  &   1.84  & -22.12  &   0.39  & -28.00  &   2.51  & -45.57  &   2.31  & -64.02  &   1.29 & C \\
52824.8554  &  -8.15  &   0.47  & -21.89  &   0.34  & -23.65  &   5.10  & -56.85  &   2.85  & -74.36  &   1.27 & C \\
52875.8624  &  -4.25  &   0.45  & -15.10  &   0.10  & -42.61  &   3.97  & -50.04  &   0.15  & -69.31  &   0.26 & C \\
    ...  & ...  & ...   & ... & ...  & ...   & ...  & ... & ... & ...  & ...& ... \\
\hline\noalign{\smallskip}
\end{tabular}
\tablefoot{ This table is available in its entirety in machine-readable form.}
\end{center}
\end{table*}

\begin{table*}[h]
\caption{Spectrophotometry of \ha line from various published papers.
Whenever needed, we measured intensities on the enlarged
plots of published Figures with \ha profiles. Spectra published by
Catanzaro (2013) were downloaded and re-measured by us.}
\begin{center}
\label{spectrofot1}
\begin{tabular}{cllcccccc}
\hline\hline\noalign{\smallskip}
 RJD  &  V  &   R  &  (V+R)/2  & V/R  &   Ic     &   EW     &   rms    &  Source \\
\hline\noalign{\smallskip}
43007.7000 & 1.371 & 1.574 & 1.4725 & 0.8710 & 0.674 & -1.27 & 0.04 & Fontaine et al. (1982) \\
43008.7000 & 1.371 & 1.583 & 1.4770 & 0.8661 & 0.736 & -4.02 & 0.07 & \\
43010.7000 & 1.328 & 1.534 & 1.4310 & 0.8657 & 0.734 & -2.21 & 0.04 & \\
44052.9000 & 1.80 &  1.47  & 1.6350 & 1.2245 & 0.629 &   --  & --   & Alduseva \& Kolotilov (1980)\\
44054.9000 & 1.75 &  1.25  & 1.5000 & 1.4000 & 99.99 &   --  & --   &   \\
 ...& ...& ...& ...& ...& ...& ...& ...& ... \\
\hline\noalign{\smallskip}
\end{tabular}
\tablefoot{ This table is available in its entirety in machine-readable
form.}
\end{center}
\end{table*}

\begin{table*}[h]
\caption{Spectrophotometric quantities measured on electronic
spectra at our disposal for all five considered spectral lines.
No meaningful FWHM values can be obtained for the double
emission line \ha and the blend of He I 6678 / Fe II 6677.
Sources of spectra are identified by letters in the column labelled 'File',
which correspond to those in table D.1.  Missing measurements
are denoted as 9.9999. }
\begin{center}
\label{spectrofot2}
A. \ha emis. wings \\
\medskip
\begin{tabular}{cccccccc}
\hline\hline\noalign{\smallskip}
 RJD  &   EW(A)  &  V  &  R  &   Ic  &  V/R & (V+R)/2   &   File  \\
\hline\noalign{\smallskip}
52461.8429 & -4.8879 & 2.1463 & 2.0886 & 0.2187 & 1.0276 & 2.1175 & A \\
52500.4383 & -4.9124 & 2.1478 & 2.1079 & 0.2901 & 1.0189 & 2.1279 & B \\
52771.9048 & -4.5095 & 1.6976 & 2.3696 & 0.2180 & 0.7164 & 2.0336 & C \\
52824.8554 & -4.7266 & 1.6615 & 2.6543 & 0.3226 & 0.6260 & 2.1579 & C \\
52875.8624 & -4.8426 & 1.5730 & 2.6995 & 0.2720 & 0.5827 & 2.1363 & C \\
  ...      & ...     & ...    & ...    & ...    & ...    & ...    & ...\\
\hline\noalign{\smallskip}
\end{tabular}
\tablefoot{This table is available in its entirety in machine-readable form.}
\end{center}
\end{table*}

\subsection{Spectroscopy}
New spectra used in this study consist of 46 \ond\ (OND) CCD spectra
secured in the red spectral region between 2003 and 2011 and five similar
spectra taken with another CCD detector in the summer of 2019,
78 CCD spectra from the Dominion Astrophysical Observatory (DAO),
one Haute Provence Observatory (OHP) Elodie echelle spectrum, one
ESO Feros echelle spectrum, and 54 amateur spectra extracted from
the BeSS database.
The BeSS database~\footnote{\tt http://basebe.obspm.fr} \citep{neiner11}
contains spectra of all known Be stars obtained by a~number of amateur and professional observers and is an ideal tool for long-term studies of their spectral variation.

We also extracted five published normalised \ha spectra from Trieste Observatory
acquired with the 0.91 m telescope by \citet{catan2013} and reduced by the authors.
An overview of all new electronic spectra that we used is given in Table~\ref{new}.
Besides, we critically collected published RVs from many
sources. The journal of all RV data sets is in Table~\ref{jourv}.

 We also collected records of published \ha profiles and used enlarged plots of them to measure the peak intensities of the double emission
 $I_V$ and $I_R$, and central intensity of the shell absorption line $I_{\rm c}$, all in the units of continuum level. The same quantities plus the equivalent width of the \ha line
were measured in all electronic spectra used. We also collected records
of the \ha equivalent width published by previous investigators. The journal of these
\ha spectrophotometric measurements is in Table~\ref{jouewic}.
Inconsistencies and misprints, which we noted in some published papers,
are listed in Appendix~\ref{apa}.

Initial reductions of OND, and DAO spectra was carried out in IRAF
(by M.\v{S}, and S.Y., respectively), while the initial reductions of the spectra from the BeSS database were carried out by their authors in a standardised way. The normalisation of all spectra and their RV measurements was carried out independently twice, by M.W. and P.H., with
the new Java program {\tt reSPEFO}. \footnote{The program was
written by A.~Harmanec and is available with the User manual at
\url{https://astro.troja.mff.cuni.cz/projects/respefo/}}
We recall that the RV measurements in {\tt reSPEFO} are based
on the comparison of the direct and flipped line profile on the
computer screen. This technique is also known as the tracing paper method. It was devised by K.~Kordylewski in Cracow in 1924 to enable the determination of the times of the minima of eclipsing binaries. It was first described by \citet{szaf48}. Its application to the RV measurements
in {\tt reSPEFO} is carried out in such a way that the user specifies
the laboratory wavelength of each measured spectral line, $\lambda_0,$
and the wavelength range around it, $d\lambda$. The spectrum for the
interval ($\lambda_0 - d\lambda$,\,$\lambda_0+d\lambda$) is defined
by discrete data points $\lambda_1$, $\lambda_2$, ... $\lambda_N$.
It is then re-normalised into representation with discrete steps linear in RV, using a three-fold finer step,
\begin{equation}
    \tria RV = {1\over{3}}c{(\lambda_2-\lambda_1)\over{\lambda_1}}
,\end{equation}
\noindent than what would correspond to the original resolution in the wavelength scale. The re-normalisation of the whole spectrum is carried
out with the recursive formula:
\begin{equation}
s_k=s_{k-1}(1+{\tria RV\over{c}}),
\end{equation}
\noindent where $c$ is the speed of light in vacuum and $s_k$ are new
pixel numbers in the RV scale ($k=1...M$). The flipped spectrum
in the RV scale, $f_j$, is calculated for the line neighbourhood as:
\begin{equation}
    f_j= 2.s_c-s_k
,\end{equation}
\noindent and displayed on the screen in a different colour. The user
can move the flipped spectrum on the computer screen interactively to achieve the best agreement
between the original and flipped profile  for the part of the line to be measured. The
radial velocity RV$_{\rm meas}$ is derived from the shift $\tria s$
in the $s$ scale via
\begin{equation}
    RV_{\rm meas}=\tria RV{\tria s\over{2}}.
\end{equation}
The advantage of this type of RV measurements over some automatic
procedures is that the user clearly sees the quality of the profile and
can avoid flaws or blends by telluric lines in the red parts of spectra.
An automatic process of RV measurement for the \ha emission profile
was compared to the manual one described here for the Be star
$\gamma$~Cas by \citet{zarf29} with the result that the manual
settings provided a lower scatter around the mean RV curve than the
automatic one.

For the absorption lines, we made the settings
on the line cores, while for the \ha emission, we measured the RV
on the broad line wings. An illustration of this can be seen in Fig.~C.2 of
\citet{hec2020}. To obtain some estimate of the uncertainties
involved, all such RV measurements were carried out independently
by M.W. and P.H. and the average values of these were used.

\begin{figure}
\centering
\resizebox{1.0\hsize}{!}{\includegraphics{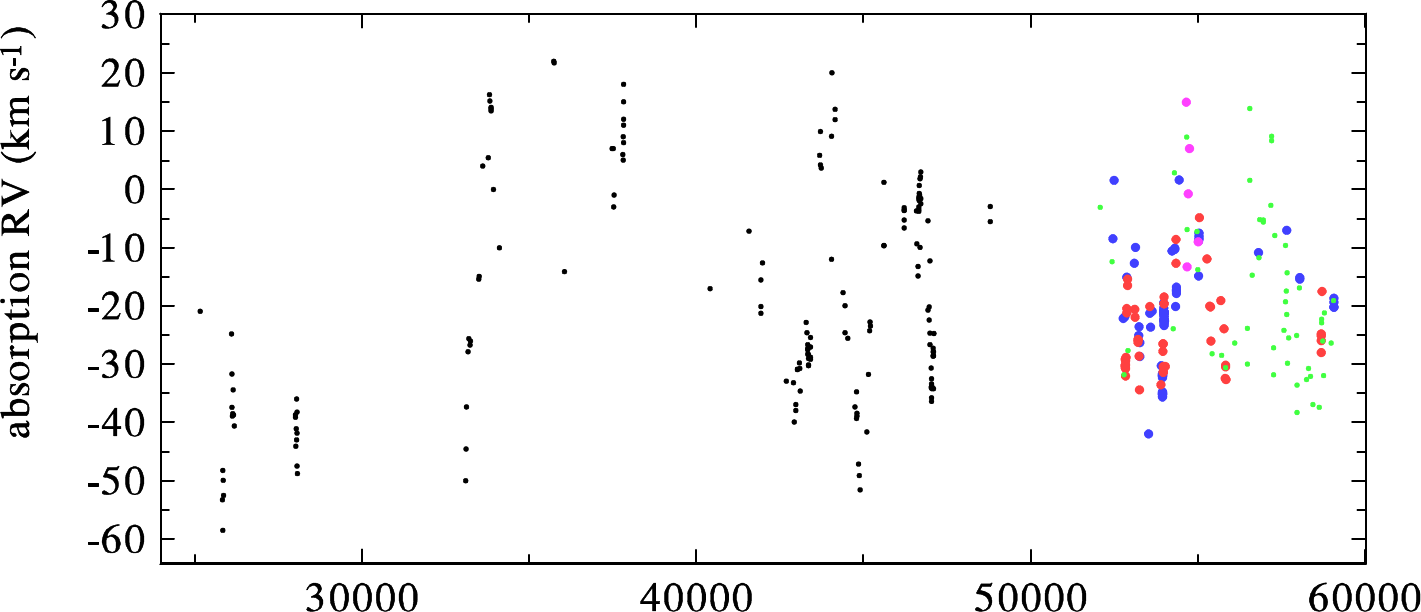}}
\resizebox{1.0\hsize}{!}{\includegraphics{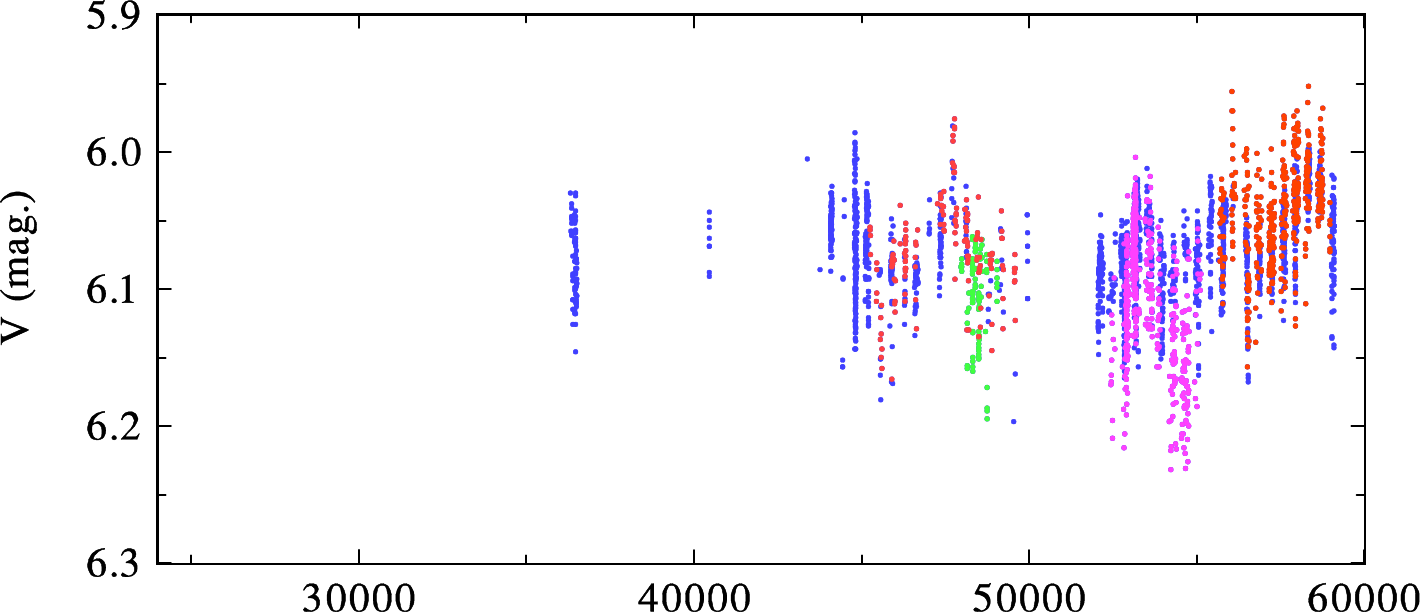}}
\resizebox{1.0\hsize}{!}{\includegraphics{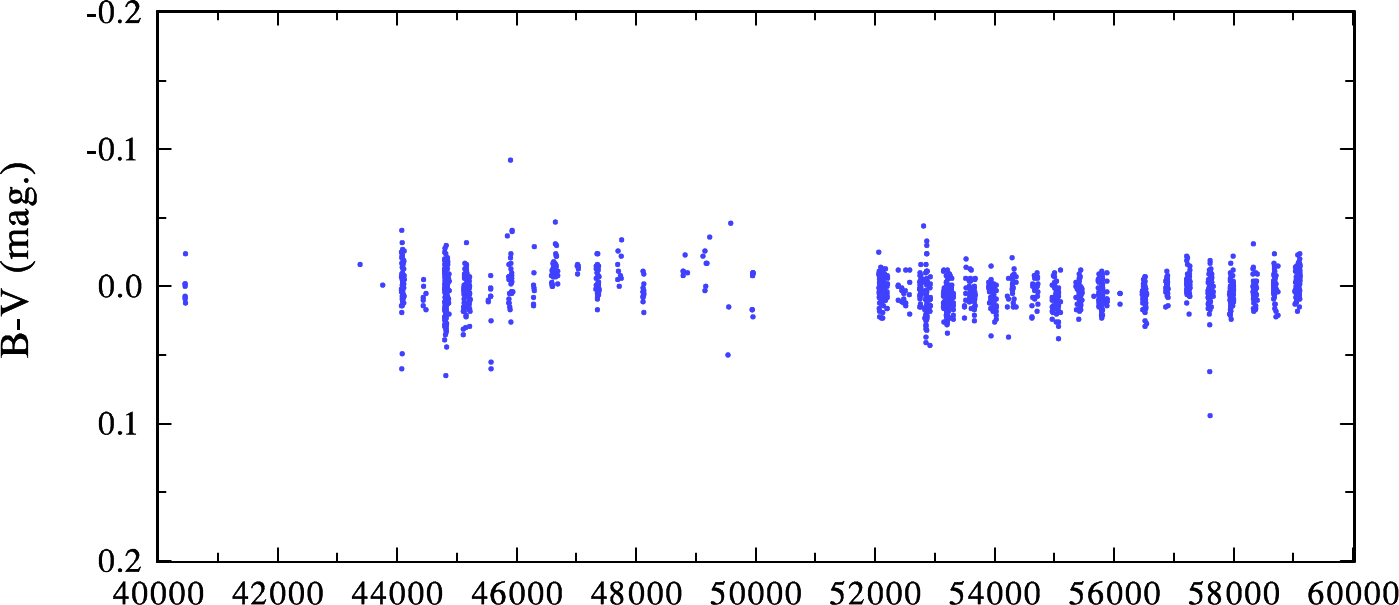}}
\resizebox{1.0\hsize}{!}{\includegraphics{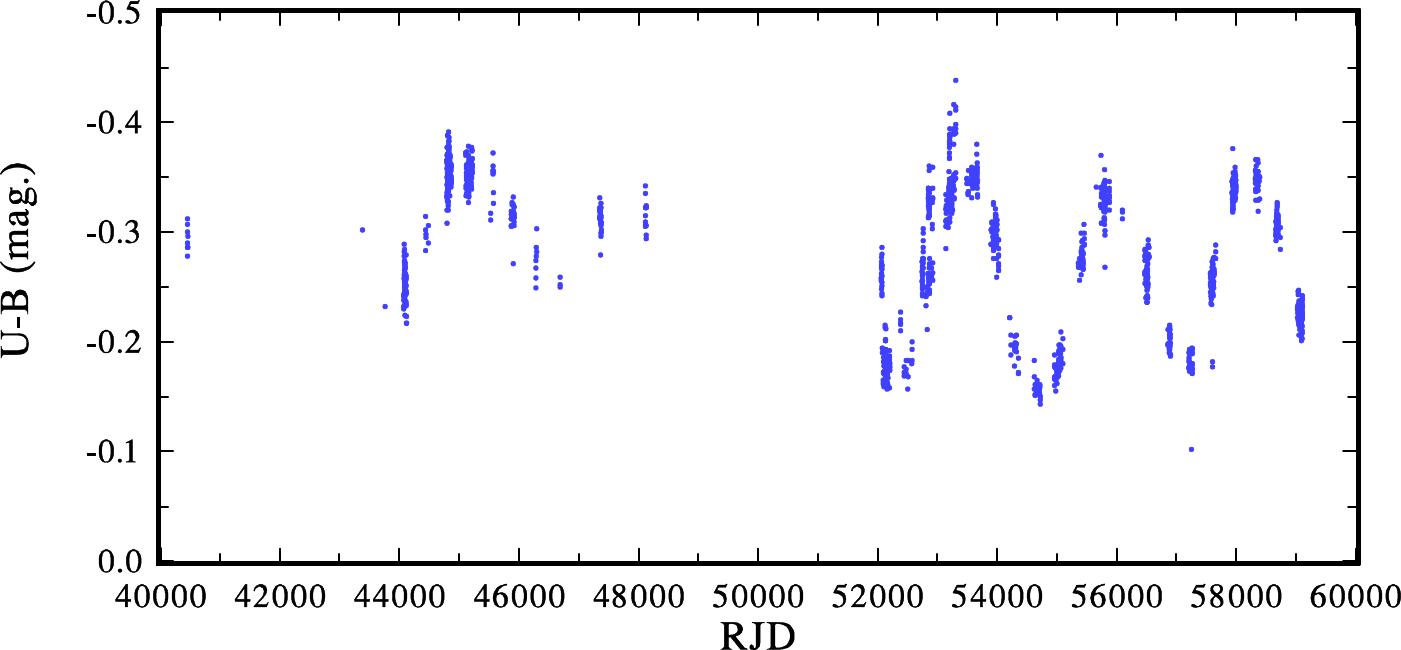}}
\caption{Time evolution of absorption RV (mean RV from the literature, and the \ha core RV for all the new spectra at our disposal), the $V$ magnitude,
and \bv, and \ub \ colour indices over the whole time interval covered by
available observations. In the RV plot,
black dots show the RVs from the literature, blue circles are from our high signal-to-noise (S/N)
spectra, red ones from OND spectra, and green ones from the BeSS spectra.
In panels with photometry, blue dots denote calibrated \ubv \ values, red the $y$ magnitude of the Str\"omgren system, green the Hipparcos \hp \ magnitudes transformed to Johnson $V$, magenta the ASAS3 $V$ magnitude, and brown the KWS $V$ magnitude.
We note the shorter time interval covered by the observed colour changes.}\label{time}
\end{figure}

\begin{figure*}[t]
\centering
\includegraphics[angle=0,scale=0.63]{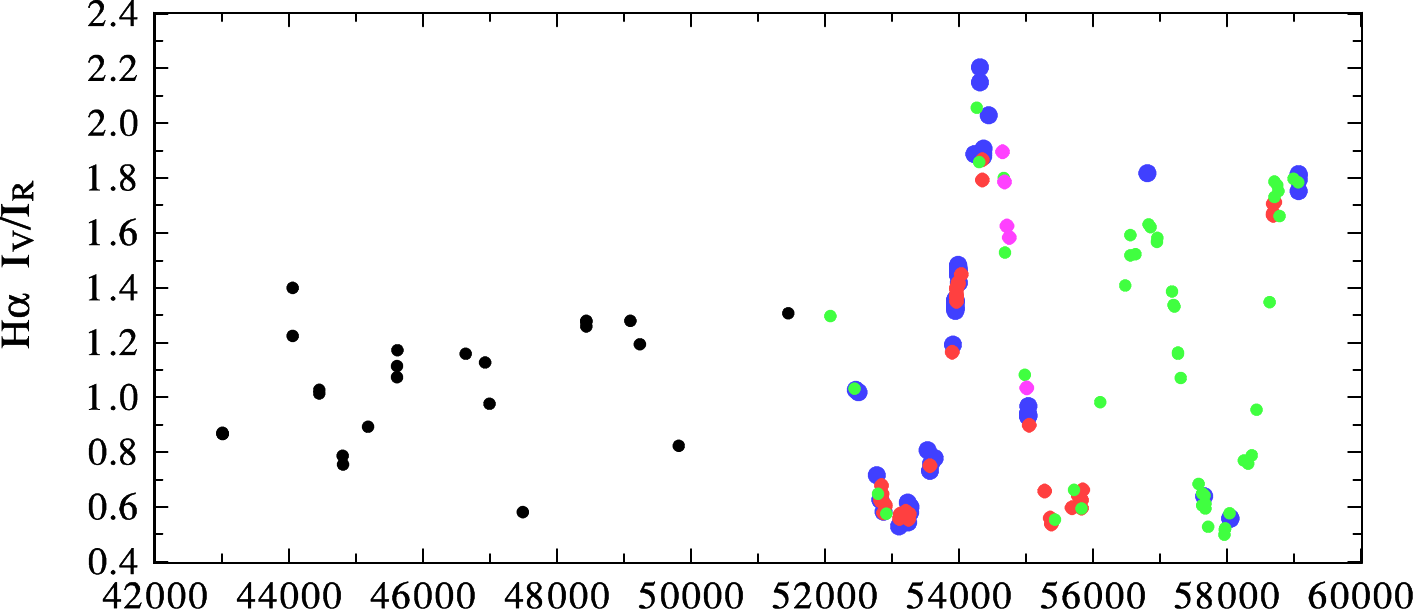}
\includegraphics[angle=0,scale=0.63]{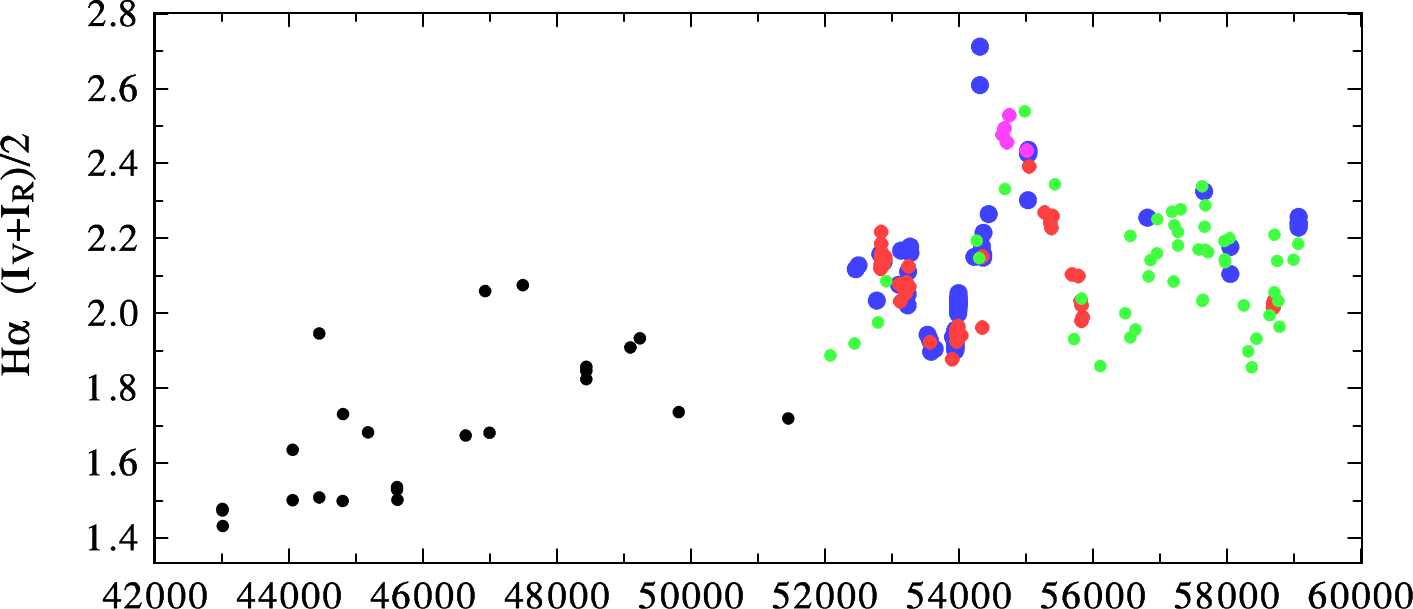}
\includegraphics[angle=0,scale=0.63]{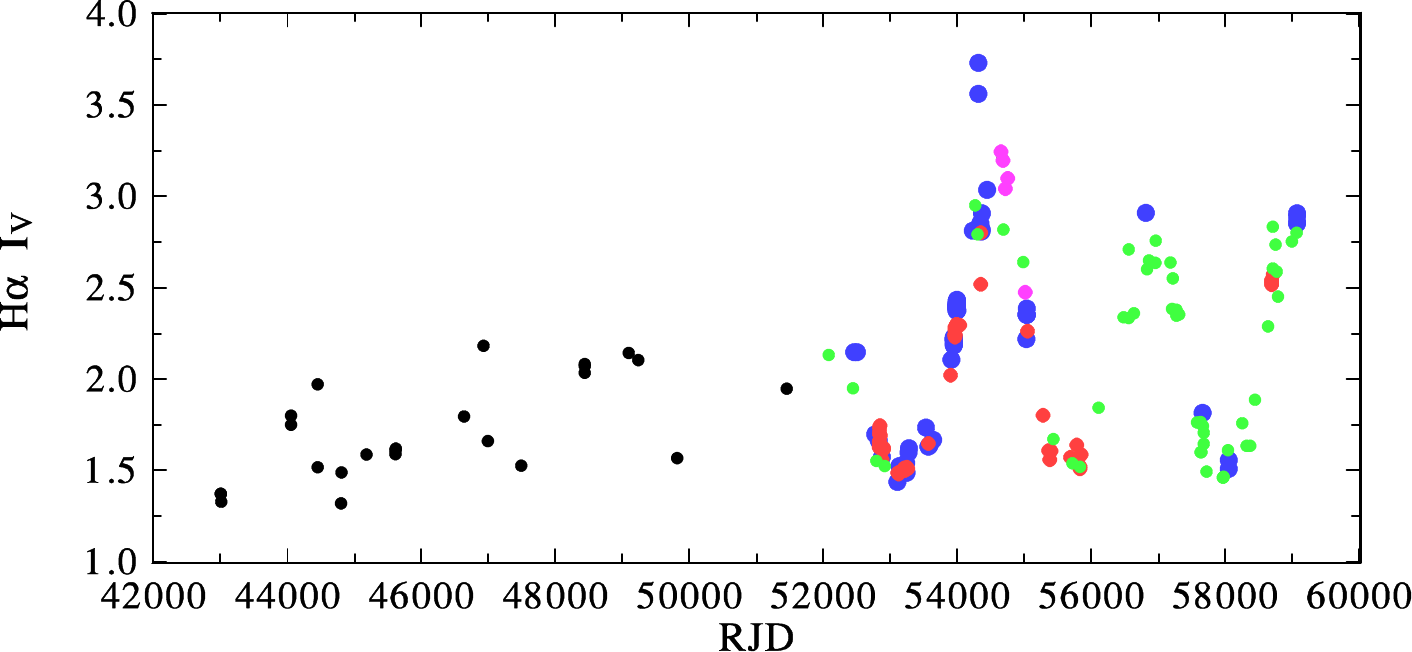}
\includegraphics[angle=0,scale=0.63]{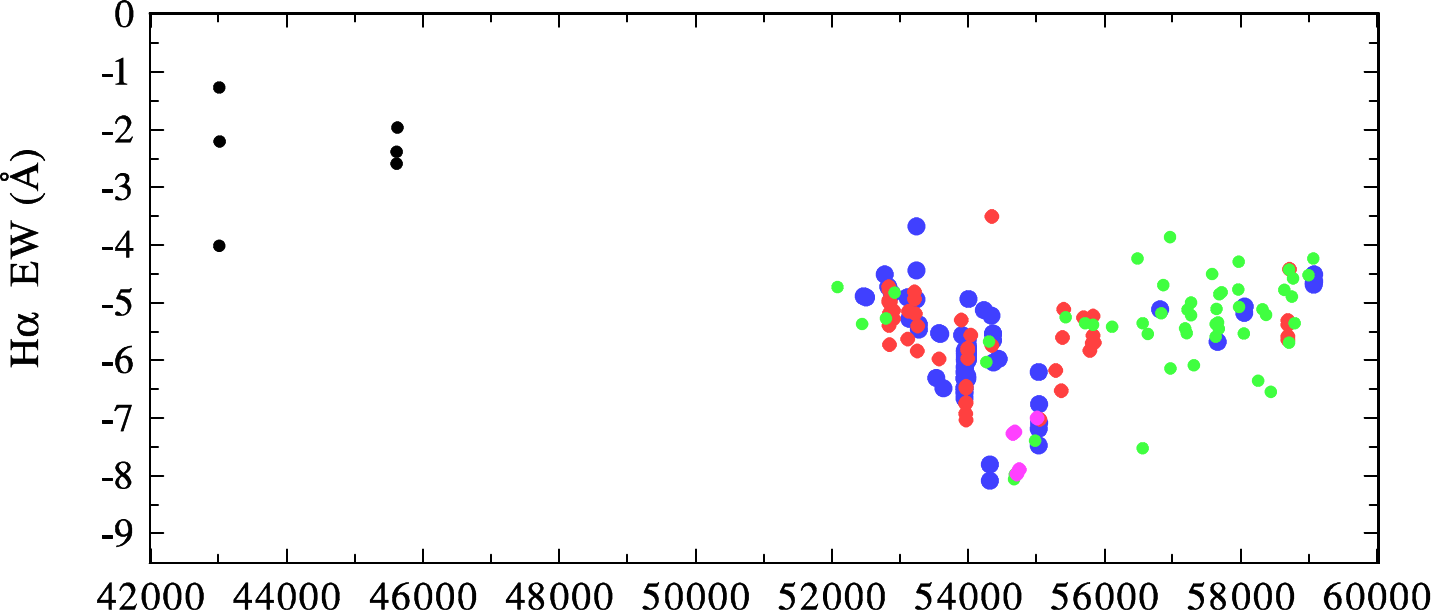}
\includegraphics[angle=0,scale=0.63]{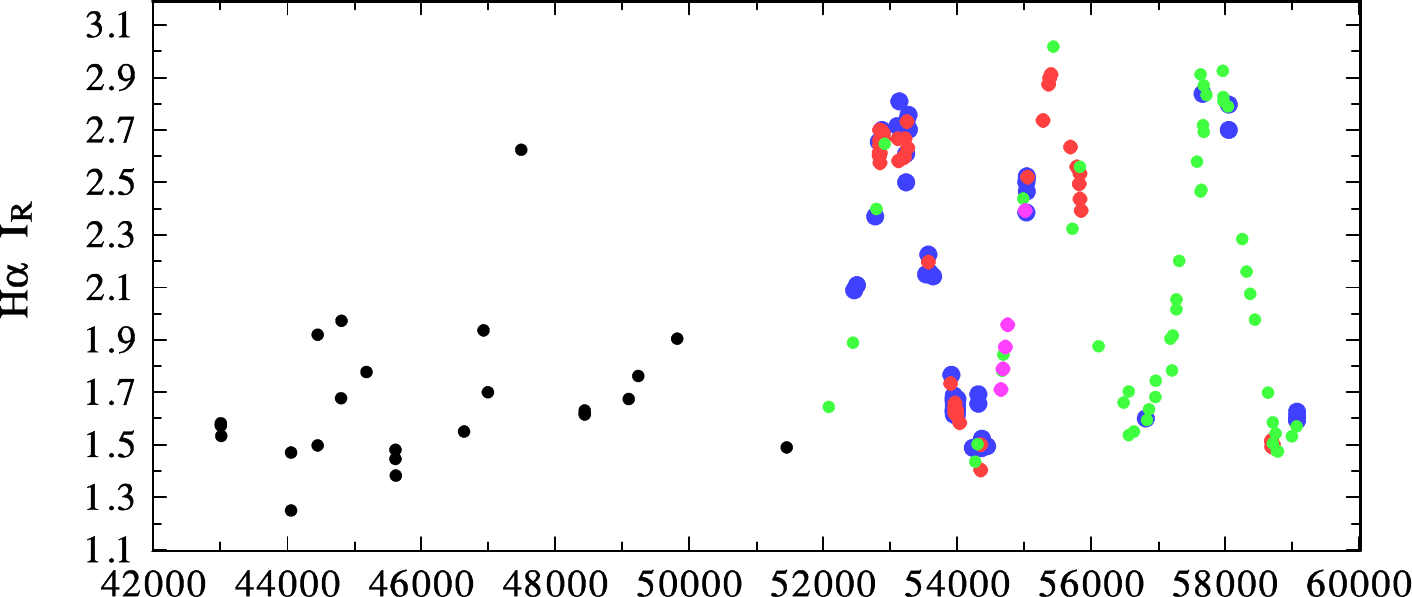}
\includegraphics[angle=0,scale=0.63]{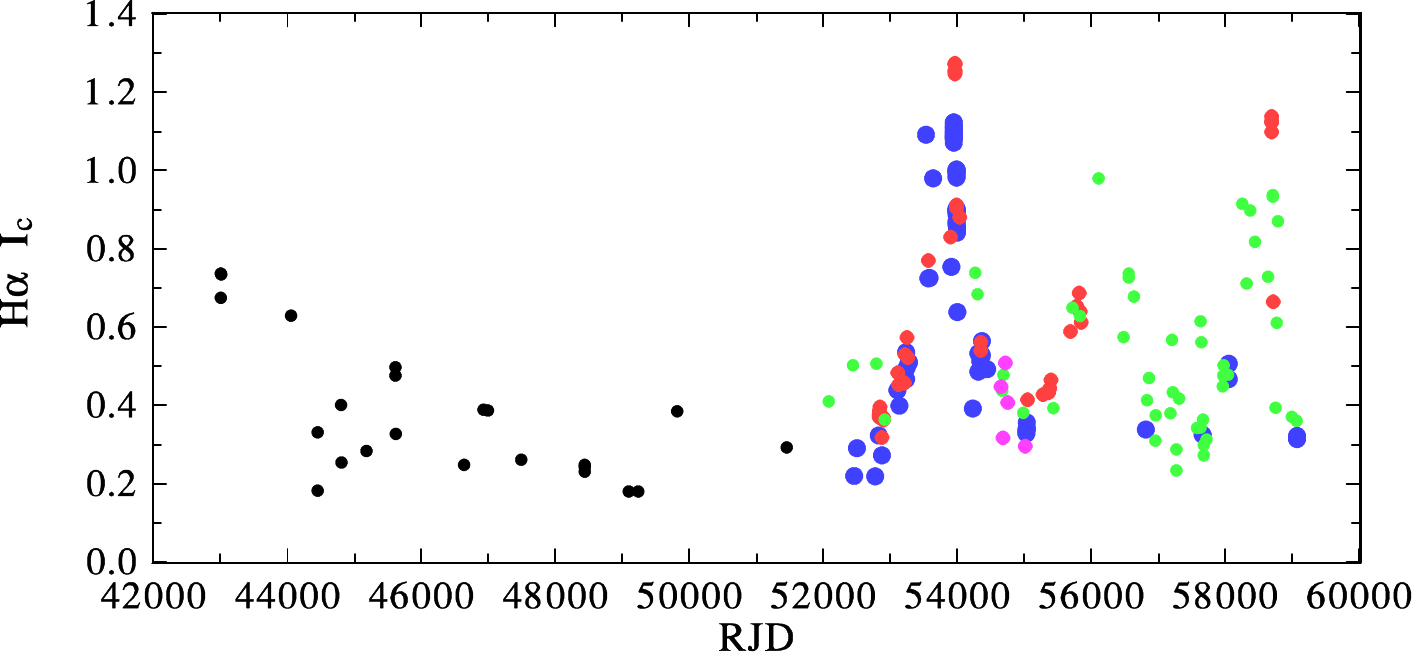}
\caption{Long-term spectral variations of the \ha line profile. From top to
bottom: Left panels show the ratio of the violet and red emission peak strength, 
normalised flux of the $V$ peak, and normalised flux of the $R$ peak.
Right panels show the mean of the normalised fluxes in the $V$ and $R$ peak,
equivalent width of the whole line, and central intensity of the core absorption.
Data from the higher-resolution Elodie, Feros, and DAO are shown by blue
circles, those from OND by the red circles, those from the BeSS amateur
spectra by the green circles, those from the Trieste spectra
re-reduced by us by magenta circles, and those from the literature by
the black circles.}\label{eqwh3}
\end{figure*}

All individual RVs collected from the literature and corrected for
misprints are provided in Table~\ref{radvel1}; our RV measurements of the \hae,
\ion{He}{i}~6678~\AA, \ion{Si}{ii}~6347 and 6371~\AA\ doublet,
and \ion{Fe}{ii}~6677~\AA\ lines, along with their rms errors, 
are provided in Table~\ref{radvel2}.
Spectrophotometric quantities measured on the \ha profiles collected from
the literature and corrected for misprints are in Table~\ref{spectrofot1} and
our measurements of spectrophotometric quantities in the new electronic
spectra are in Table~\ref{spectrofot2}.
These four tables are published in their entirety only in the electronic form.

\subsection{Photometry}

The star was rather systematically observed at Hvar over a number of observing seasons,
first in the \ubv, and later in the \ubvr \ photometric systems. In addition, we tried to
collect and homogenise all numerous photometric observations with known times of observation
published by a number of investigations.
The journal of these observations is in Table~\ref{jouphot} and details of the individual data sets
can be found in Appendix~\ref{apb}.
All the individual photometric observations will be published in a separate study after
we finish the final homogenisation and standardisation of our photometric archives.

\section{Long-term changes}
As we attempted to collect and homogenise most of the available spectral and
photometric observations, we start our analysis with the description of
long-term changes in various observed quantities and their mutual relation.
This will provide a good starting point for any future quantitative modelling
and the elimination of some possible models.

  We first inspected the RV measurements on the electronic spectra at our
disposal, plotted versus time in Fig.~\ref{rvnew}. One can see that
all measured RVs undergo slow cyclic changes. Their amplitude differs
from one line to another, being smallest for the \he absorption wings
and largest for the \ion{Si}{ii} doublet. We also take note that there are phase shifts
on the part of the RVs for individual measured features. To combine our RVs with the
earlier ones, we chose the RV of the \ha absorption core, which could
be measured for all our electronic spectra.

  The time evolution of RV, $V$ magnitude and \bv \ and \ub \ colour indices
over the whole recorded history of observations is shown in Figure~\ref{time}.
In Figure~\ref{fotom}, we show additional photometric observations in several
passbands different from the Johnson \ubv \ photometry. They cover shorter time intervals.

\begin{figure*}[t]
\centering
\includegraphics[angle=0,scale=0.63]{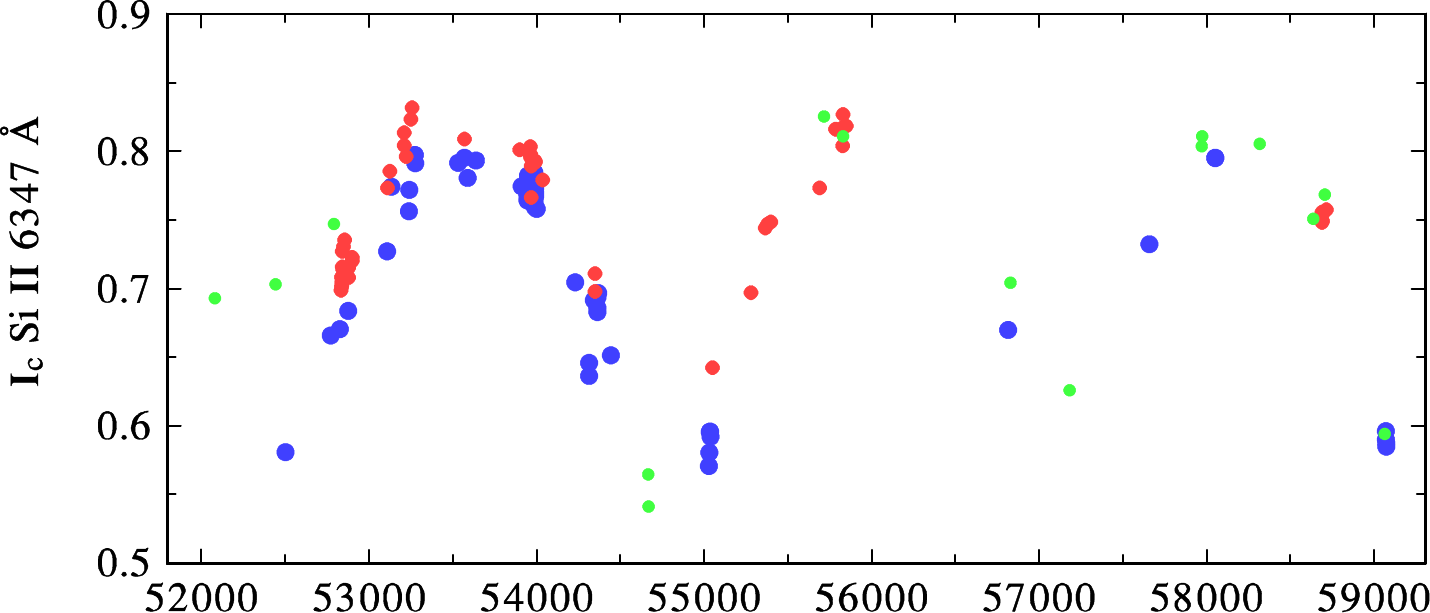}
\includegraphics[angle=0,scale=0.63]{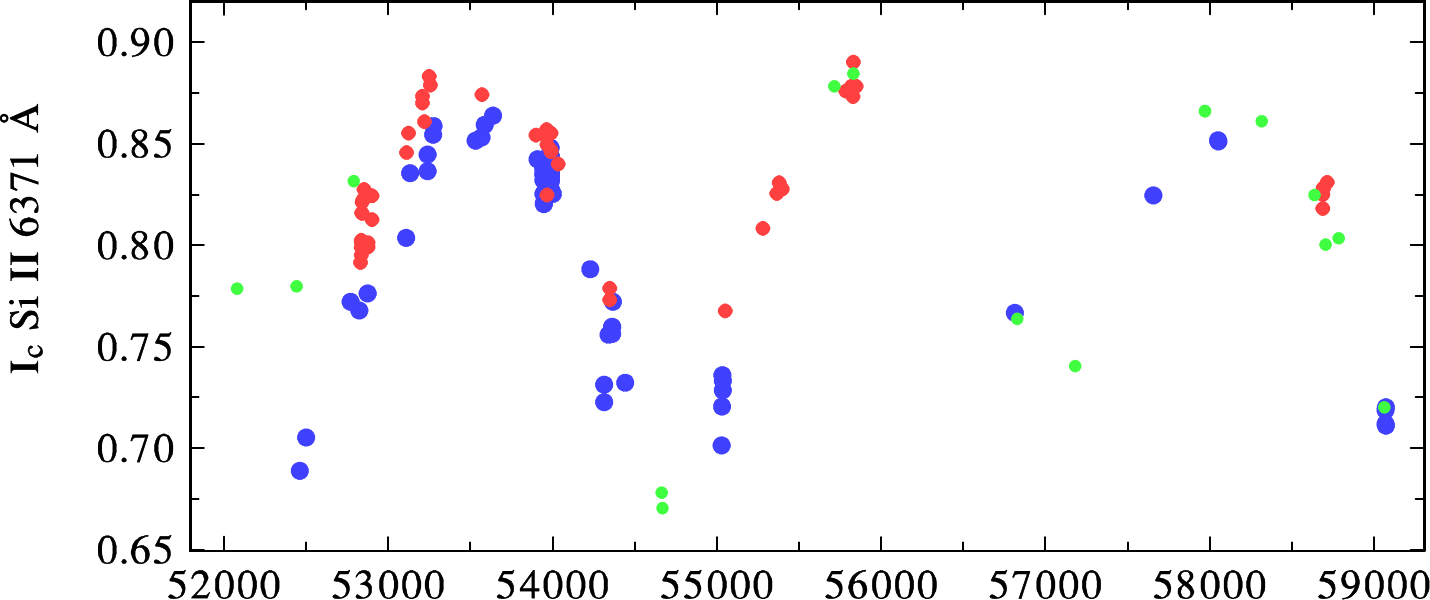}
\includegraphics[angle=0,scale=0.63]{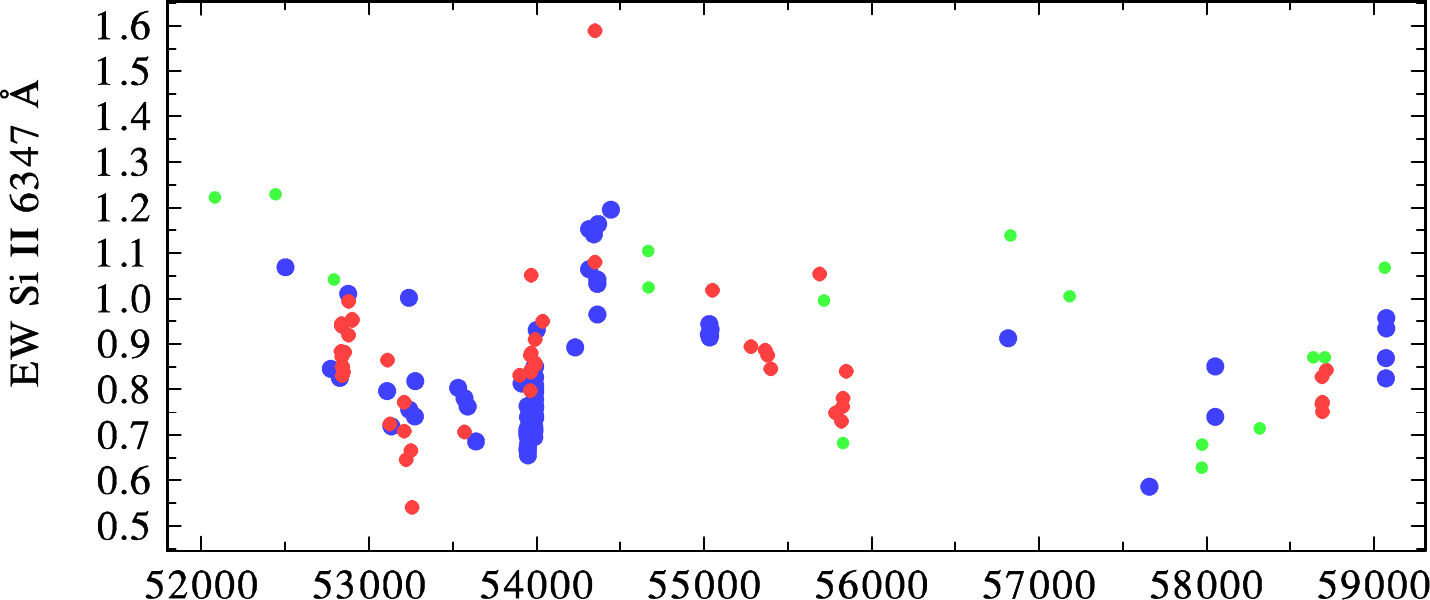}
\includegraphics[angle=0,scale=0.63]{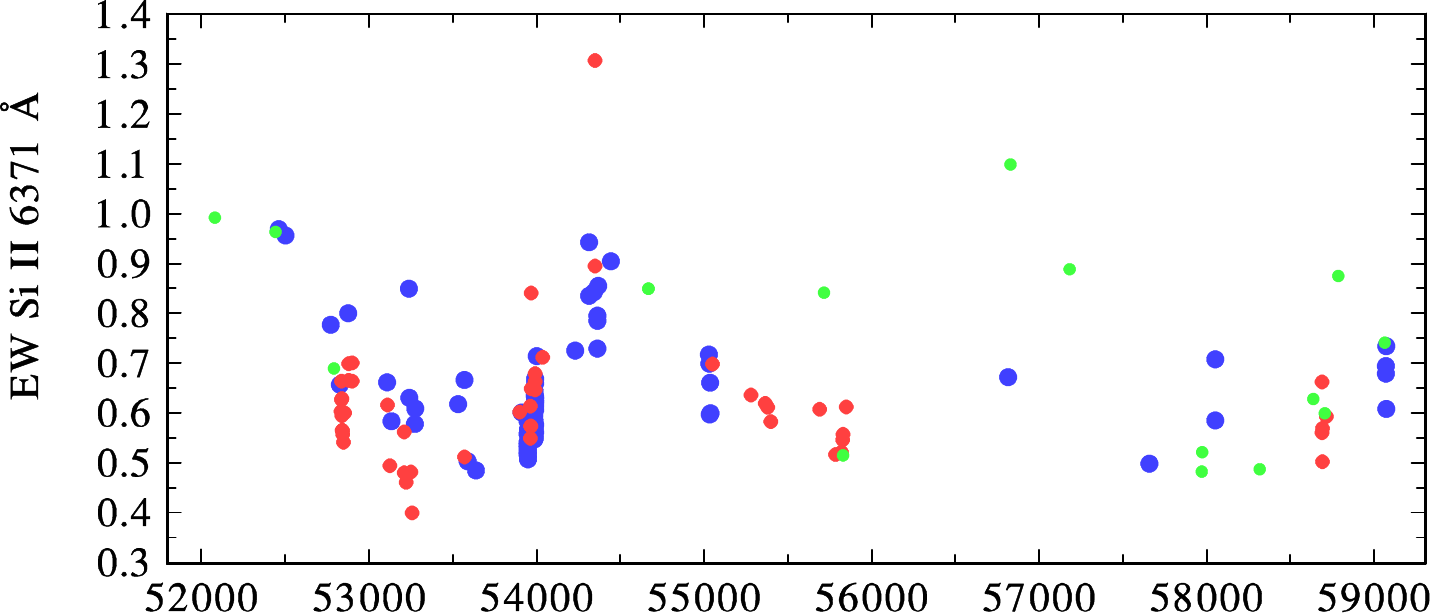}
\includegraphics[angle=0,scale=0.63]{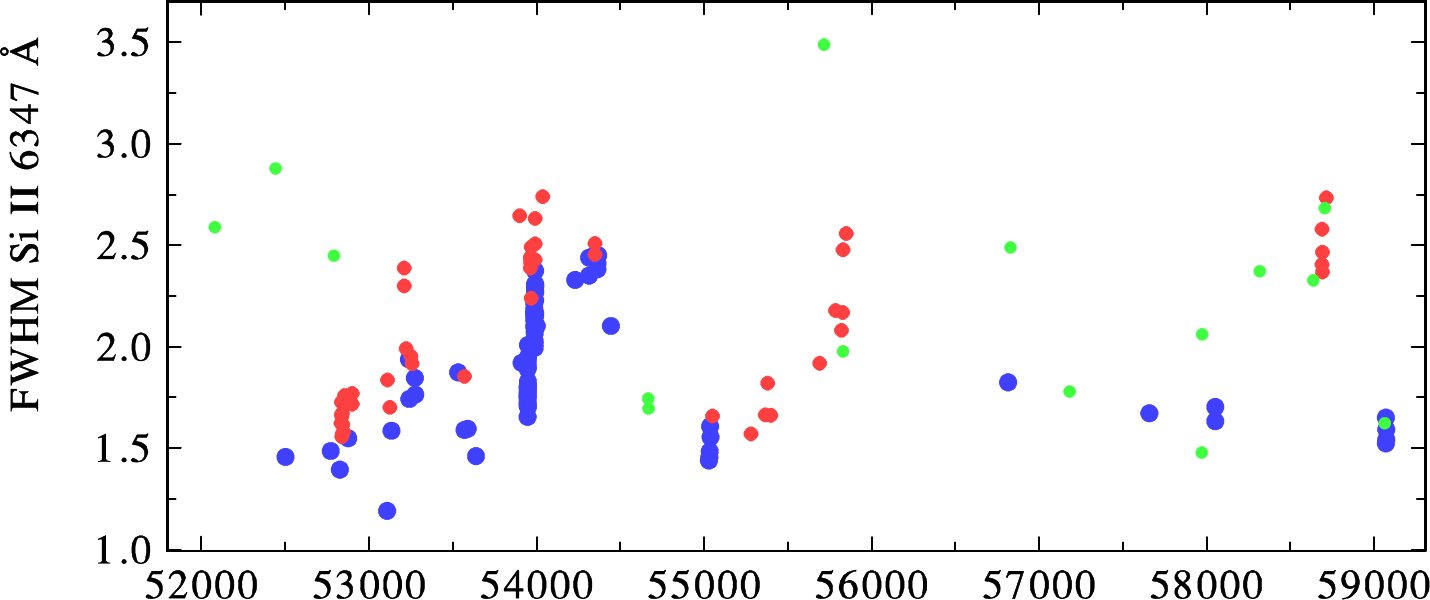}
\includegraphics[angle=0,scale=0.63]{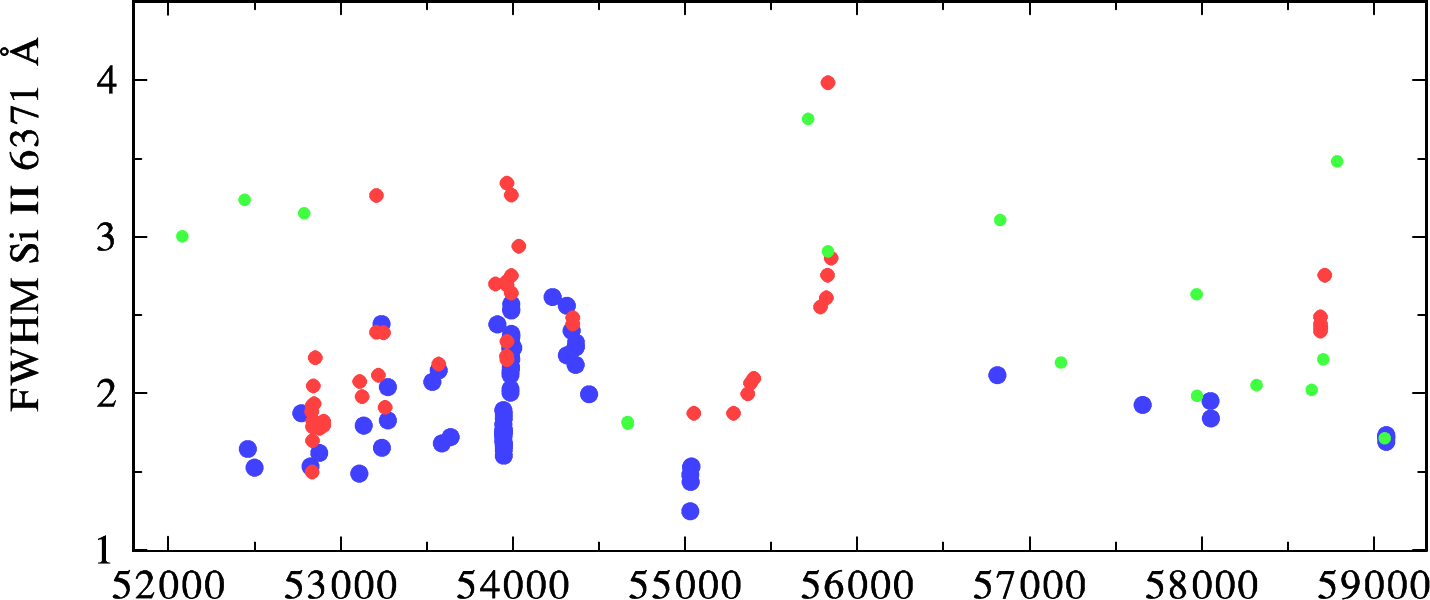}
\includegraphics[angle=0,scale=0.63]{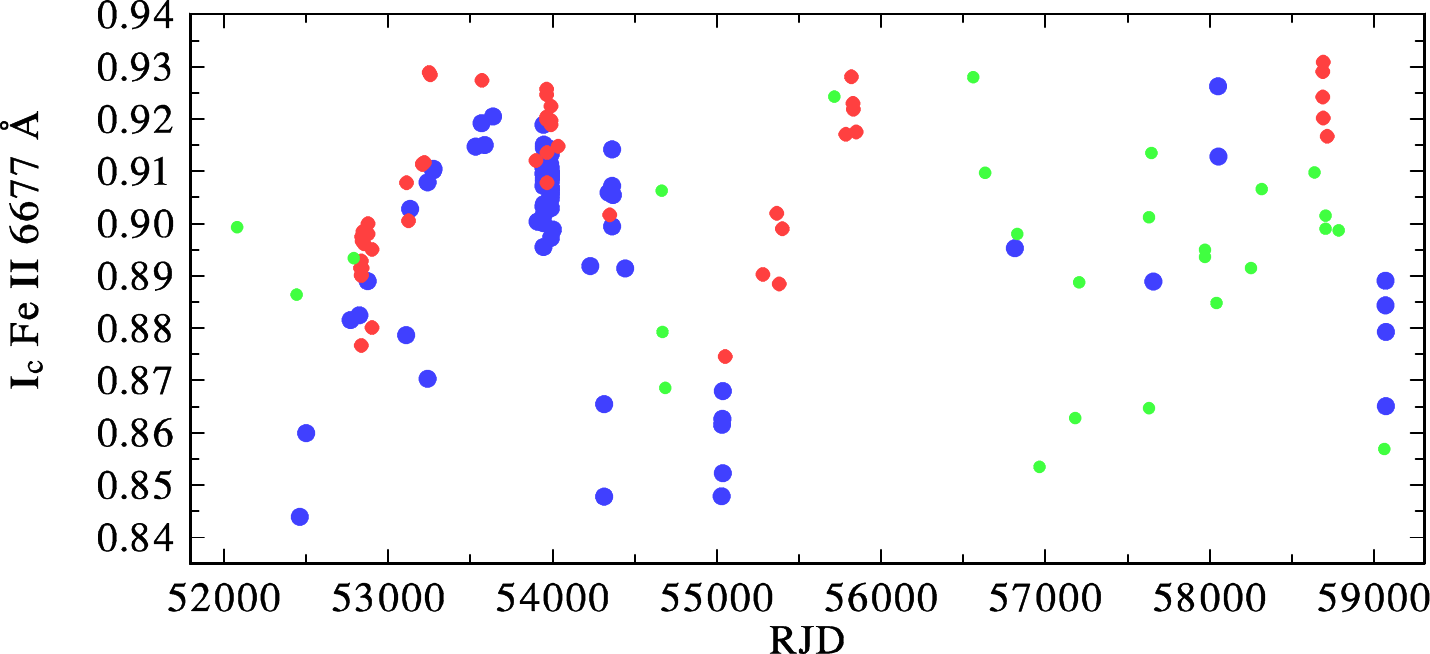}
\includegraphics[angle=0,scale=0.63]{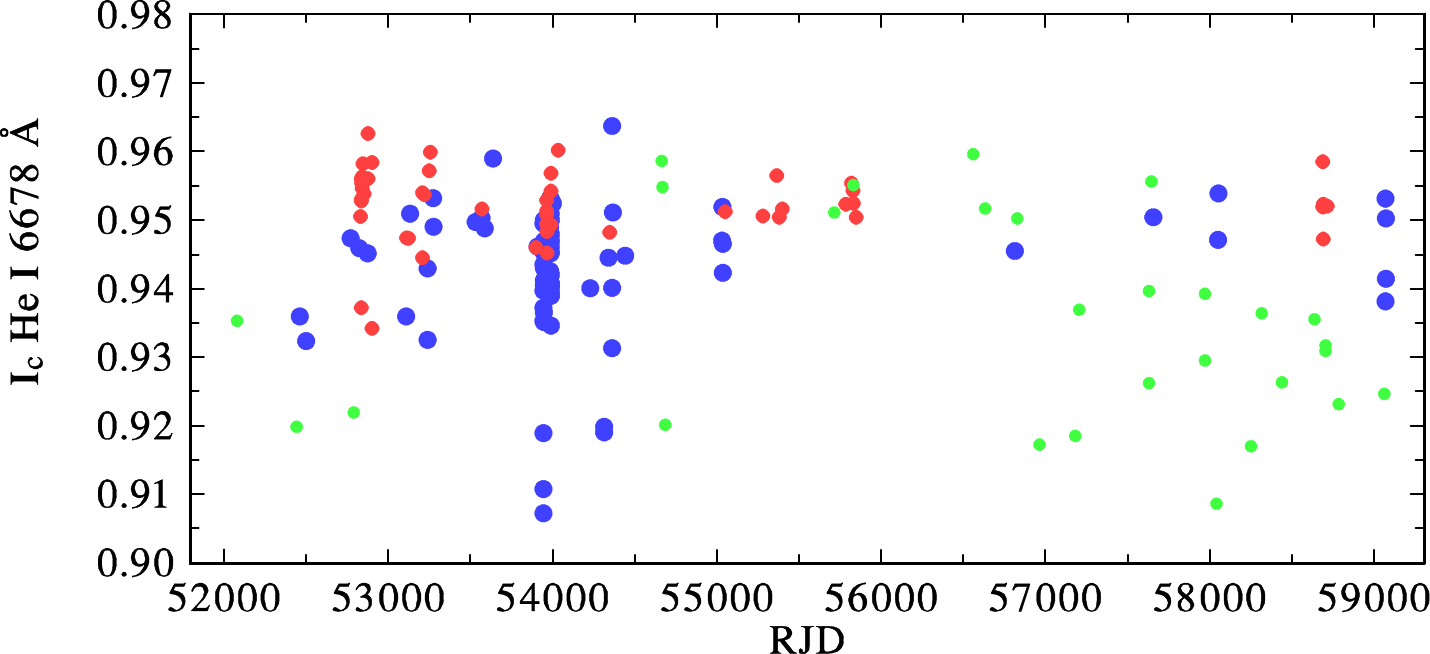}
\caption{Time evolution of the spectral changes of
the \ion{Si}{ii}~6347 and 6371~\AA, \ion{Fe}{ii}~6677~\AA, and
\ion{He}{i}~6678~\AA. Blue dots denote
data from the higher-resolution Elodie, Feros, and DAO spectra,
circles, the red ones those from OND spectra, and the green ones those
from the BeSS amateur spectra.}\label{eqw}
\end{figure*}

  Figures~\ref{eqwh3} and \ref{eqw} show the parallel time evolution of
some spectrophotometric quantities measured in the \hae, \ion{Si}{ii},
\ion{Fe}{ii}~6677~\AA, and \ion{He}{i}~6678~\AA\  line profiles.
For the two last lines, which mutually blend with each other, it was only possible to measure their central intensities.
Finally, Figure~\ref{ub-bv} shows the variations in the \ub\ versus \bv\ diagram.
Variations of another Be star V744~Her = 88~Her observed systematically
at Hvar are also shown for comparison.
We note the following:
\begin{enumerate}
\item The absorption-line RVs, $V/R$ ratio of the strength of
the violet and red peaks of the double \ha emission, star brightness, and
\ub \ colour all undergo cyclic variations during the whole documented history, whereas the \bv \ index remains secularly quite stable and close to zero
(compare Figs.~\ref{rvnew}, \ref{time}, and \ref{eqwh3}).
These variations are so regular that they look almost like a periodic
phenomenon with a 'period' of some 2400~days. We note, however, that the
quantitative fits of individual cycles carried out in Sect.~\ref{newefem} show gradually varying length of consecutive
cycles. While large parallel cyclic RV and $V/R$ changes, which come
and go, are known for many Be stars, the regular behavior observed
for \va is rather rare. \citet{mcl63} reported similar regularity for
HD~20336 = BK~Cam. Cyclic $V/R$ and RV variations were observed between
1903-1937 and never since then. In the time interval between 1916 - 1931, they
followed a regular 4.5~yr cycle, whereas outside that interval the cycles
were longer. For 105~Tau = V1155~Tau, \citet{mcl66} observed large parallel
RV and $V/R$ variations, with about a 10~yr `period' over the time interval
of 1930 - 1965.
\item The strength of the violet and red peaks of the \ha emission
also varies cyclically and in the opposite manner (see Fig.~\ref{eqwh3}.)
A notable fact is that the
amplitude of these changes is larger for the $V$ peak than for the $R$
one in one cycle, while the opposite is true for the consecutive cycles.
This indicates that the geometry of the envelope is dynamically
evolving over time.
\item There is no systematic secular trend in either RV or brightness,
perhaps with the exception of a~slight recent brightening in the $V$ band,
as Fig.~\ref{time} shows.
However, the \ha emission strength has been increasing over about 18000~days
covered by electronic spectra (cf. Fig.~\ref{eqwh3}).
A part of the increase has already been noted by \citet{ruzdjak2008}.
It is best seen in the mean peak emission, measured as
($I_{\rm V} + I_{\rm R})/2$. The same effect, but less clearly, is also seen in the EW of \hae. It is our experience that the errors in all,
the relative flux, wavelength, choice of line range due to continuum
placement for spectra with different S/N ratios, and the scatter due
to variable strength of numerous telluric lines in the \ha region,
contribute to the increased scatter of measured values of EW, making it
less useful as the indicator of changes.

\begin{figure}
\centering
\resizebox{\hsize}{!}{\includegraphics{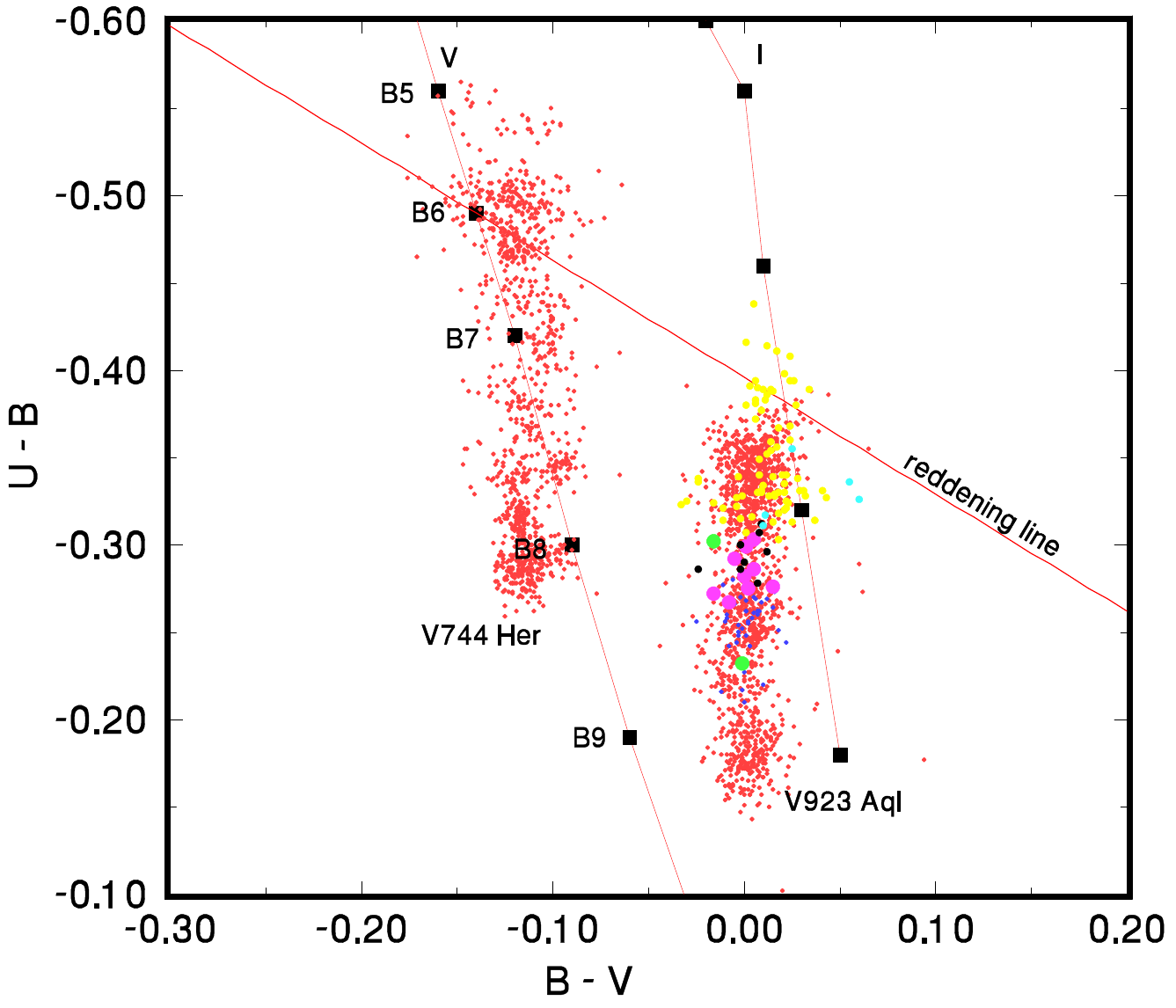}}
\caption{ \ub\ vs. \bv\ colour diagram showing the secular changes of \ve. The object is reddened and moves along the main sequence. This is indicative of an~inverse correlation between the brightness and emission-line strength. The data from
different observatories are distinguished by different colours as
follows: Hvar = red,
SPM = blue,
Tug = magenta,
Canakkale = yellow,
Haute Provence = black,
Geneve = green, and
Ond\v{r}ejov = cyan. For comparison, the Hvar photometric observations of another systematically studied Be star V744~Her = 88~Her are also shown. They are less reddened but show a remarkably similar pattern of variations.}\label{ub-bv}
\end{figure}

\begin{figure}[t!]
\centering
\resizebox{\hsize}{!}{\includegraphics{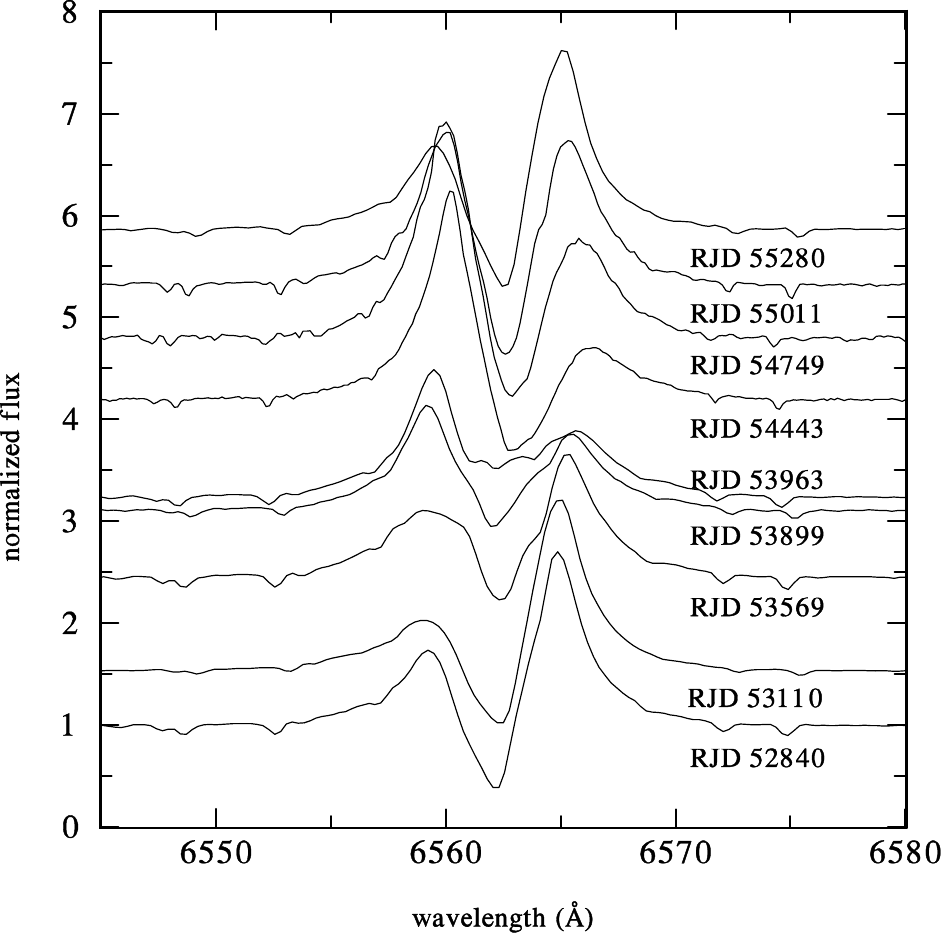}}
\resizebox{\hsize}{!}{\includegraphics{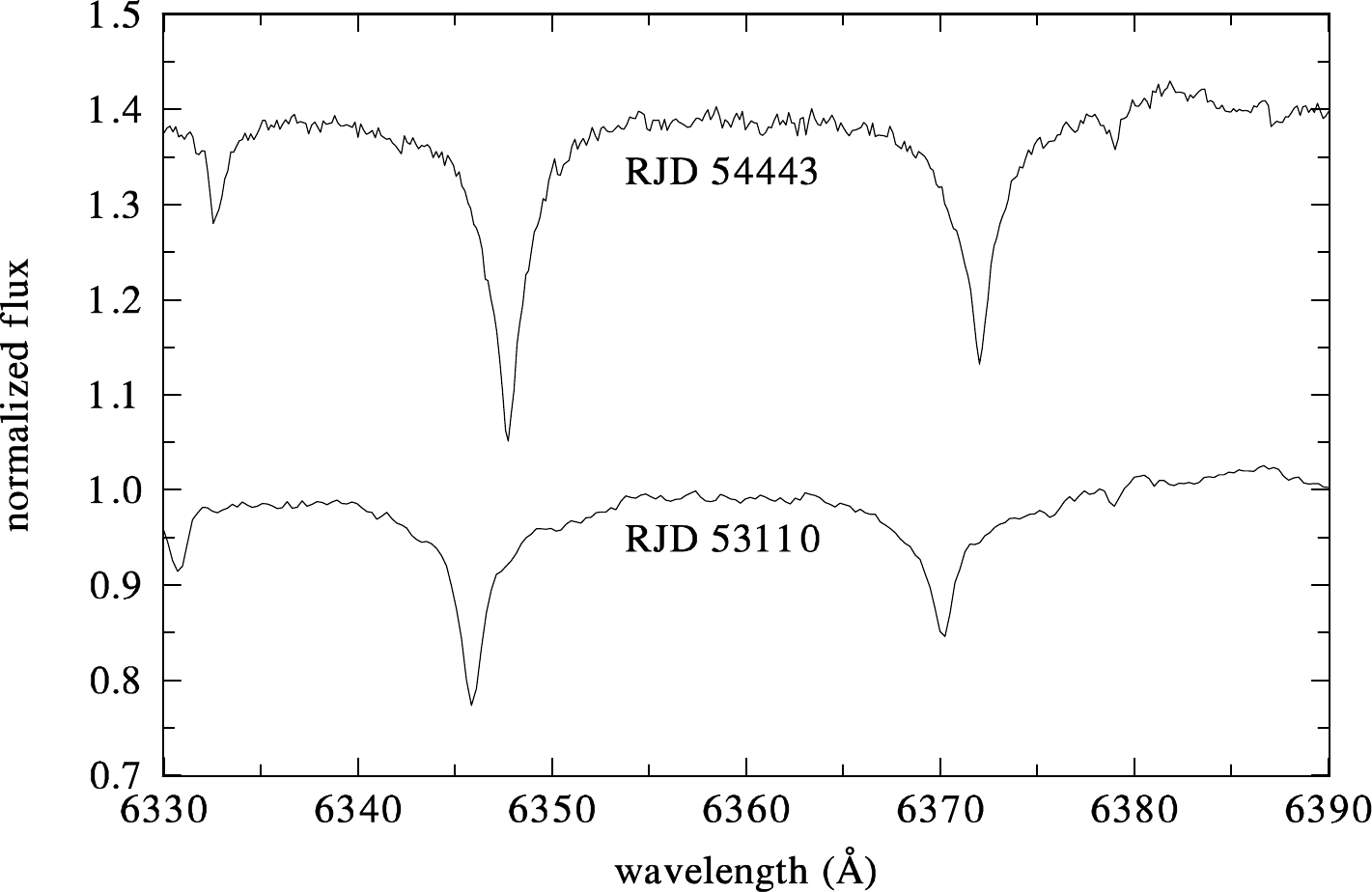}}
\resizebox{\hsize}{!}{\includegraphics{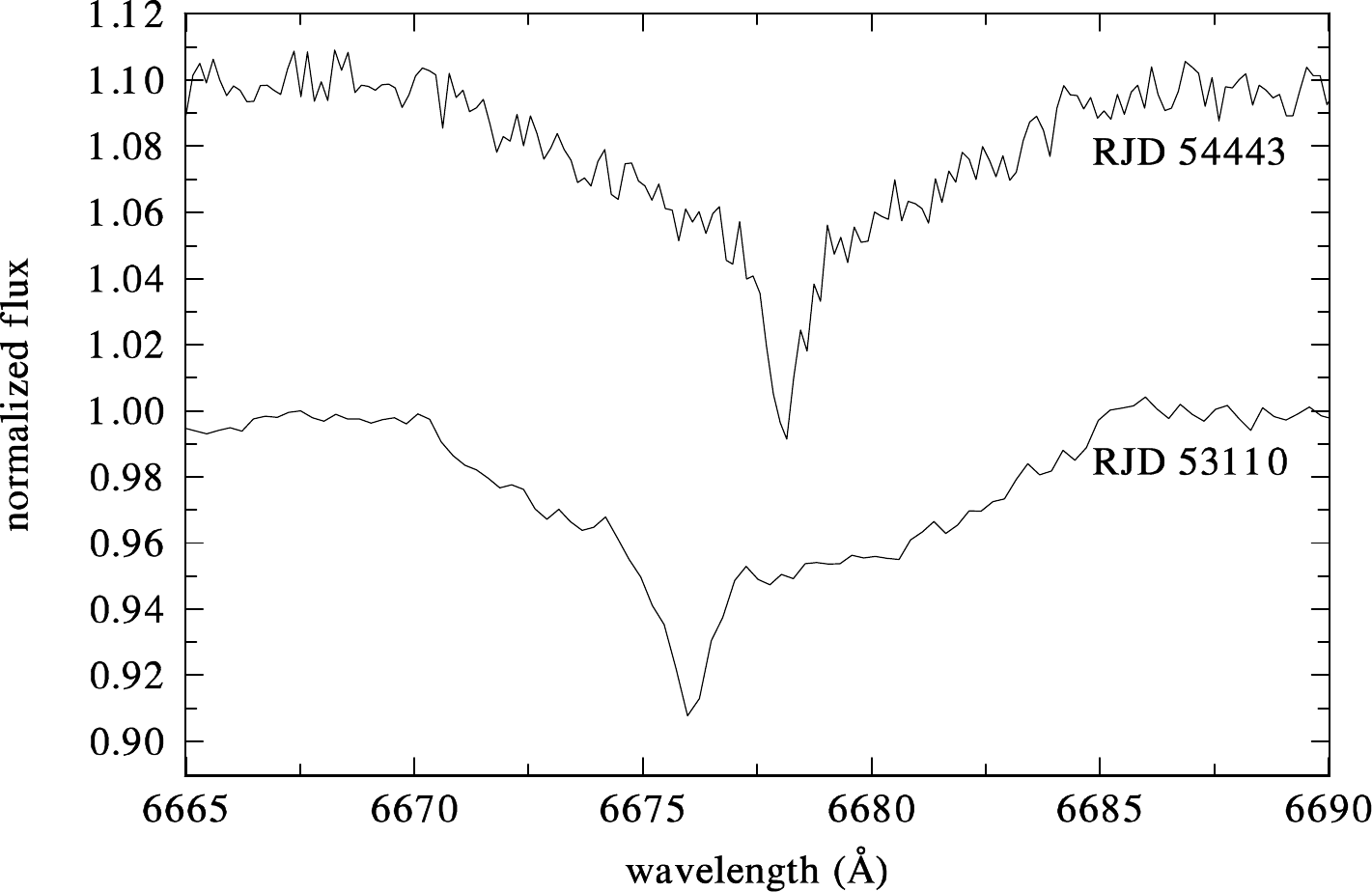}}
\caption{Selected line profiles.
Top: Sequence of \ha profiles over one whole 2400~d cycle. Shifts of continua in ordinate are scaled in such a way to correspond to differences of individual spectra in RJD.
Middle and bottom panels: Comparison of two DAO \ion{Si}{ii},
and \ion{He}{i} line profiles from the two consecutive extrema of
the long-term $V/R$ changes.}\label{lineprof}
\end{figure}

\item Variations in different photometric passbands show similar general
trends but indicate that the amplitude of more rapid changes can depend
on wavelength; see Figs.~\ref{fotom} and \ref{time52}.
\item The spectral and photometric variations are clearly correlated, but in
a complicated way, as documented by Figs.~\ref{rvnew}, \ref{eqwh3},
\ref{eqw}, and \ref{time52}. Cyclic RV variations of individual spectral features vary with mutual phase shifts and different amplitudes (growing from He to Si lines). The object becomes the brightest and
bluest in \ub \ during the He and metallic-line RV minima and vice versa.
The \ha $V/R$ ratio varies in phase with the metallic RVs. The central intensity $I_{\rm c}$  and the equivalent width $EW$ of both \ion{Si}{ii} lines also vary cyclically, with the $\sim2400$~d cycle. The lines become weakest when the \ha $V/R$ ratio changes from $V<R$ to $V>R$ and when the \ub\ index is getting redder. The full width at half maximum (FWHM) of the \ion{Si}{ii} lines is largest, when their RV is decreasing but these variations show a great deal of scatter. The central intensity of \ion{Fe}{ii}~6677~\AA\ line seems to follow
the variations of \ion{Si}{ii} lines, while the $I_{\rm c}$ of
the \ion{He}{i}~6678~\AA\ line does not exhibit any clear secular changes.
\item Variations in the \ub\ versus \bv\ diagram (Fig.~\ref{ub-bv})
are typical for the inverse correlation between the emission strength and stellar brightness as defined by \citet{hec83} and later modelled by \citet{sigut2013}. Such correlation is
observed for stars seen roughly equator-on.
\end{enumerate}

In Fig.~\ref{lineprof}, we show selected \ha profiles over one whole 2400~d cycle and two DAO line
profiles of metallic lines from the two extrema of the long-term $V/R$ changes  We note that the \ion{Si}{ii} line profiles for the spectrum with $V>R$ are notably broader than those
of the opposite extremum.

\section{Orbital and rapid changes}
The scatter in the time plots for various quantities is certainly larger
than the expected measuring errors, which is indicative of changes on shorter
time scales. However, the simultaneous presence of rapid, orbital, and long-term changes makes the analysis rather complicated.

\subsection{Rapid changes}
Before addressing the orbital variations, we attempted to find a characteristic pattern of rapid variations occurring on a time scale of days. After a number of trials, we concluded that in spite of the large amount of photometric and spectral observations we are unable to identify a clear and unique pattern of such changes. The time distribution of the available data is insufficient with respect to this goal and only a systematic, uninterrupted series of future space observations may help to shed light on the nature of these changes.

\begin{figure}
\centering
\resizebox{\hsize}{!}{\includegraphics{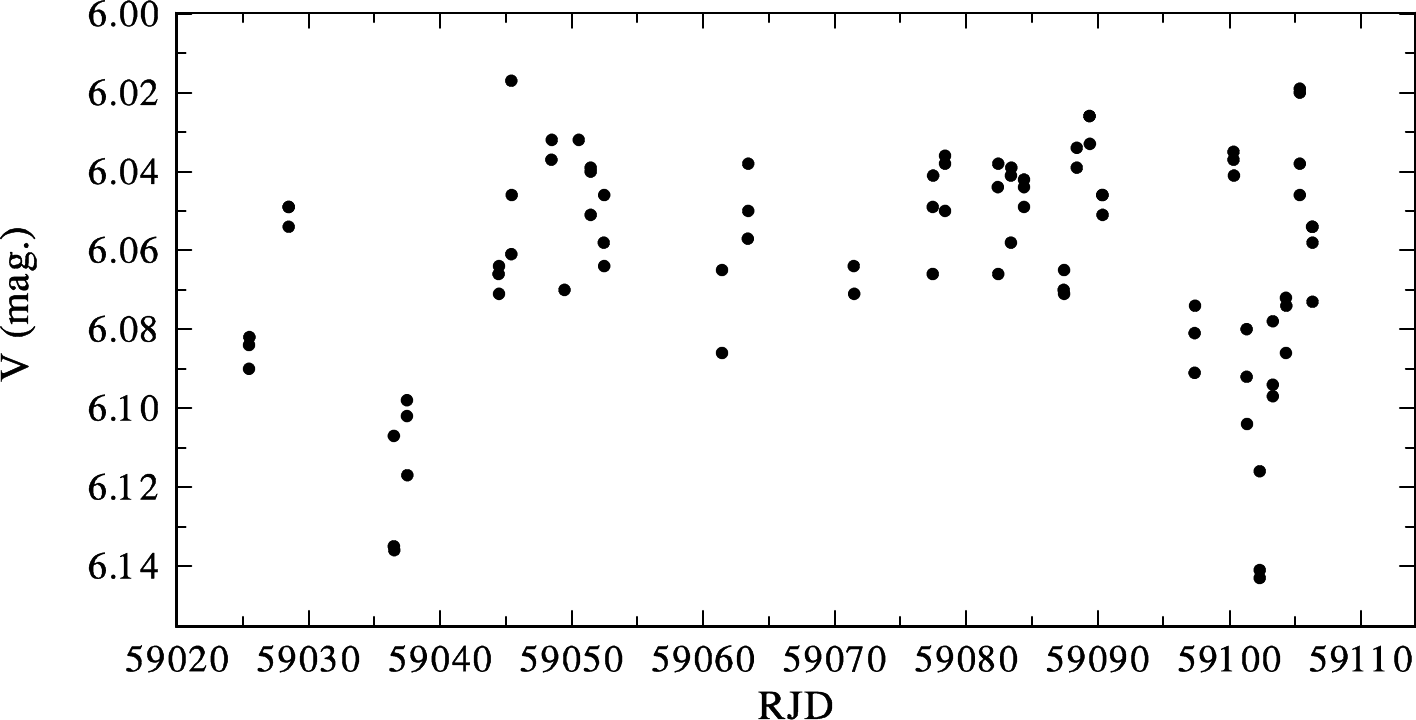}}
\resizebox{\hsize}{!}{\includegraphics{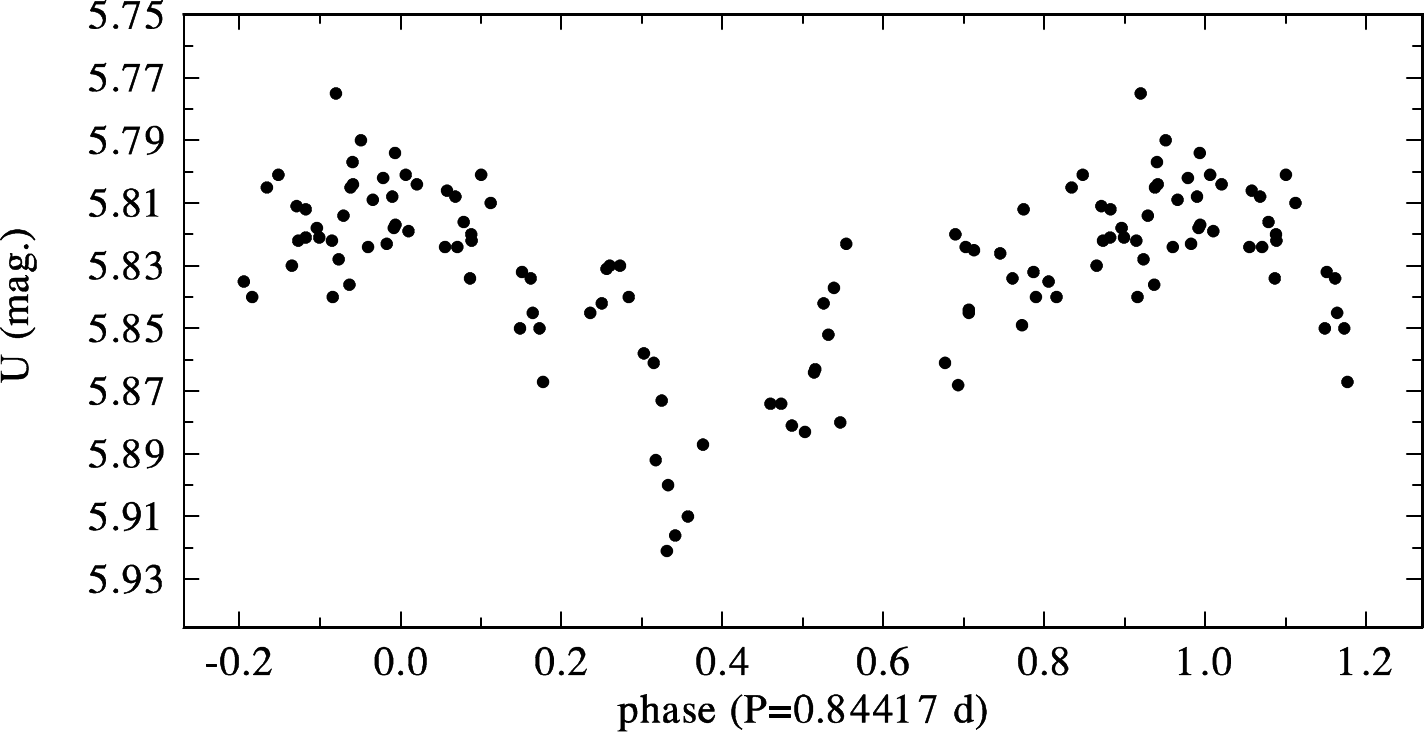}}
\caption{Top: A time plot of the Hvar $V$ photometry from the 2020 season.
Rapid variations are quite obvious.
Bottom: A phase plot of the Hvar $U$ photometry for the 0\fd84417 period,
corresponding to the highest peak in the amplitude periodogram.}\label{hvar20}
\end{figure}

In Figure~\ref{power}, we show, for the purposes of illustration, the amplitude power spectra
for three $V$ magnitude data segments without any obvious secular trends. Their power spectra clearly differ from each other.
In the upper panel of Figure~\ref{hvar20}, individual Hvar 2020 $V$
observations are plotted versus time, while the bottom plot shows a phase
curve for the $U$ magnitude (which has the largest amplitude) for the
most prominent peak of the periodogram. We note that the period of 0\fd84417 is close to the originally suggested period of 0\fd8518 \citep{lynds60}. However, this periodicity is not seen in the first Hvar season.
We also inspected several about 0\fd2 long time series of DAO spectra.
There might be some indication of rapid changes in the \ion{Si}{ii}
line profiles, but once again, we were unable to document any clear
pattern.

\subsection{A new ephemeris of the orbital motion}\label{newefem}
\begin{figure}
\centering
\resizebox{\hsize}{!}{\includegraphics{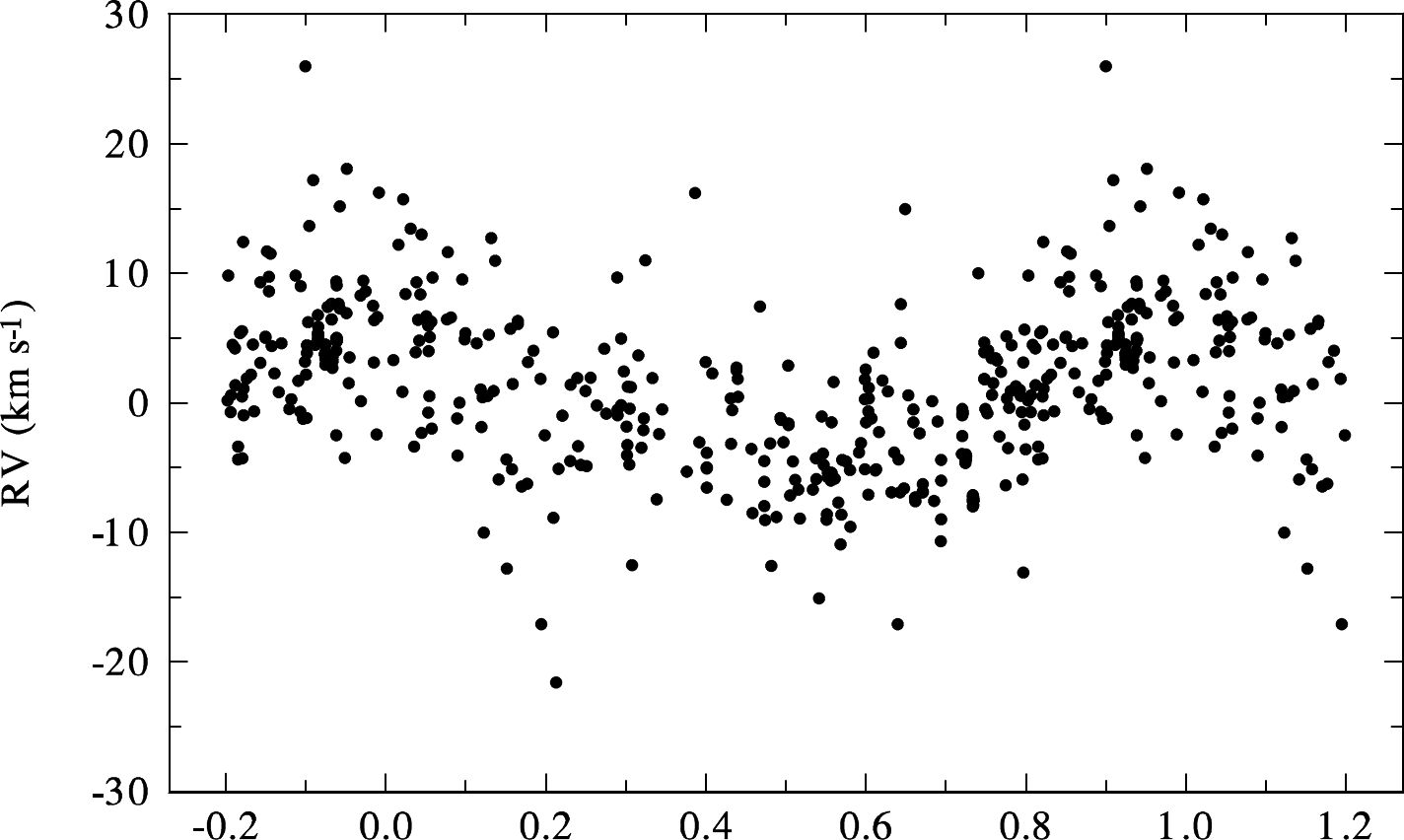}}
\resizebox{\hsize}{!}{\includegraphics{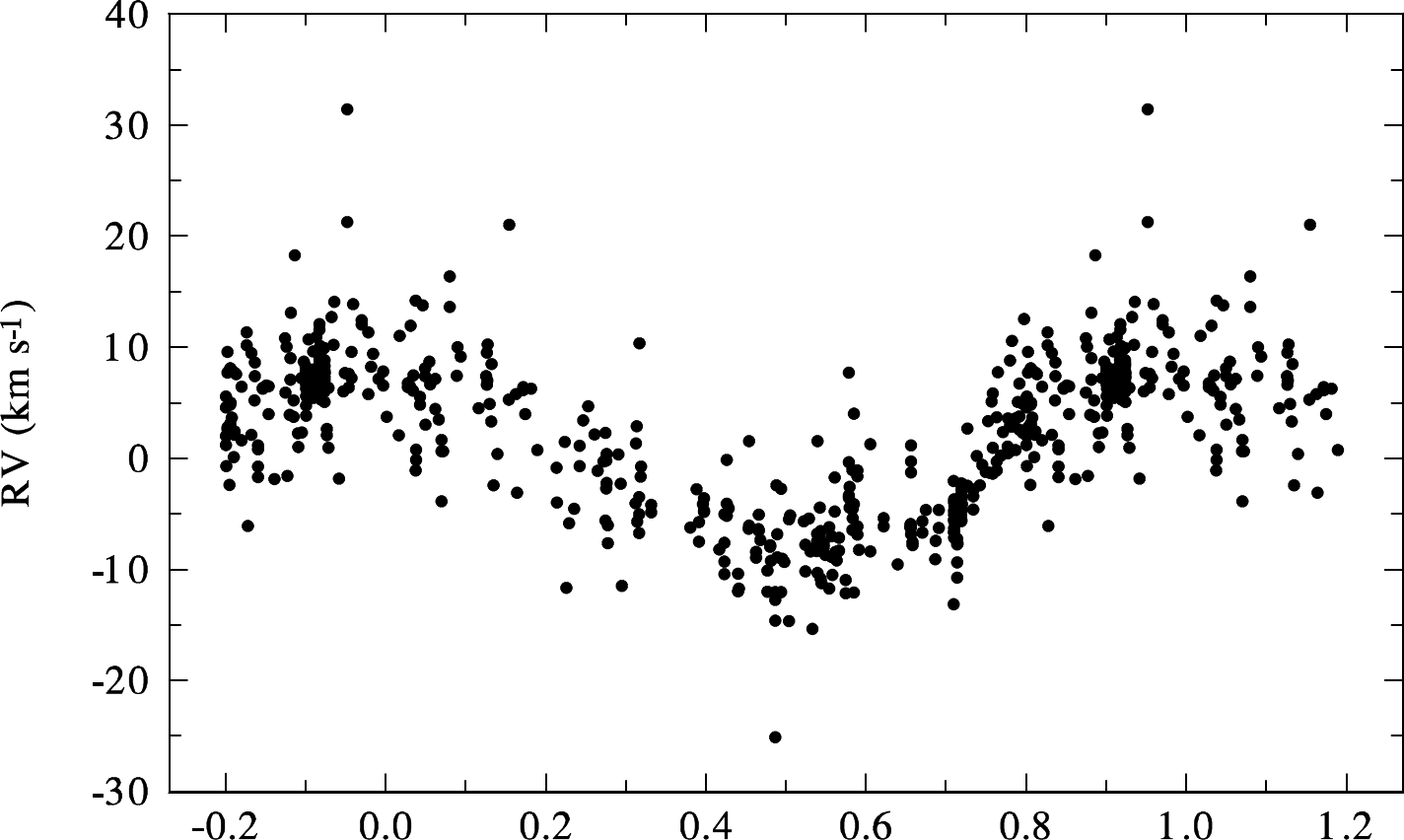}}
\resizebox{\hsize}{!}{\includegraphics{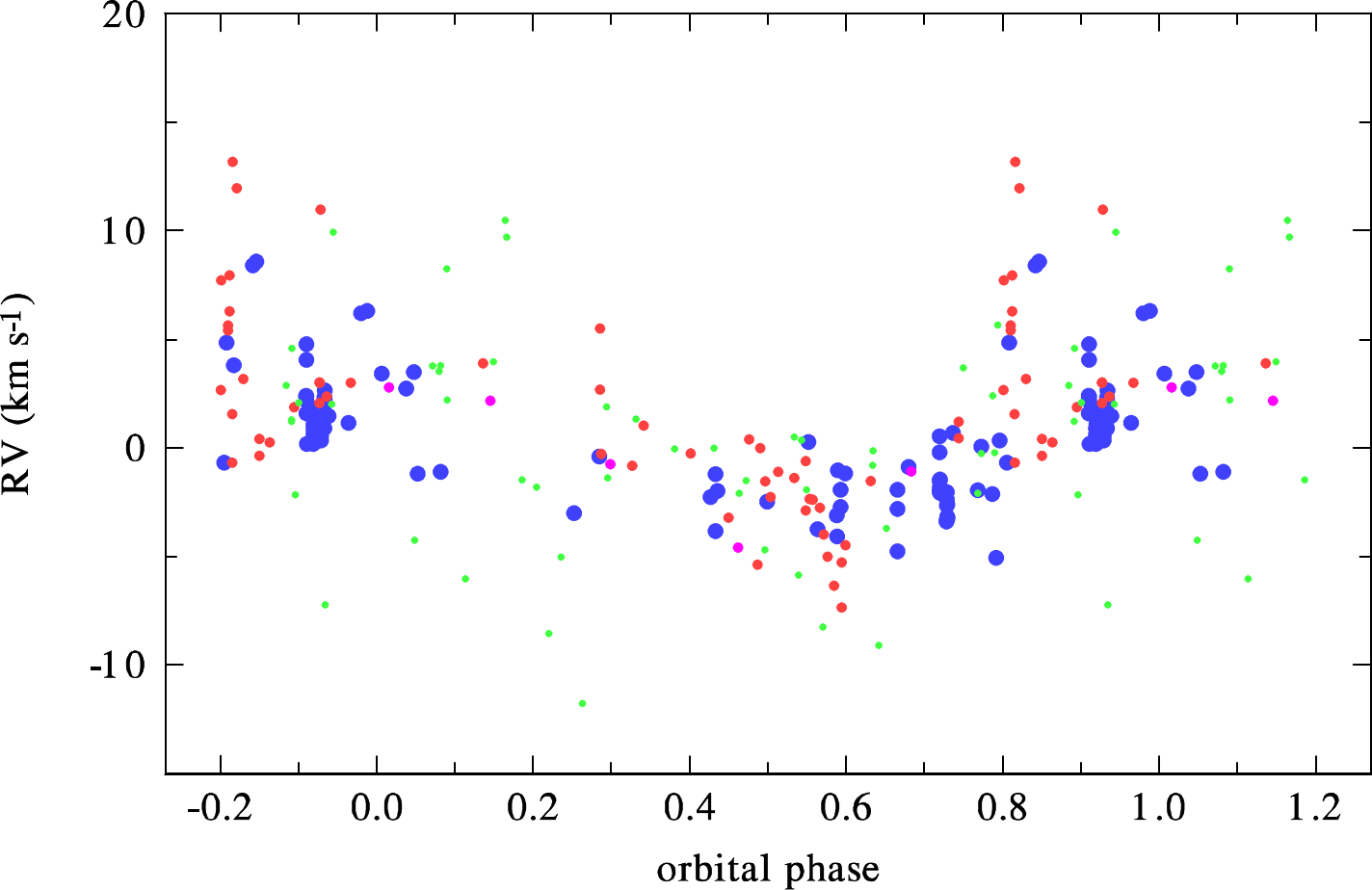}}
\caption{Orbital RV curves after removal of the long-term cyclic changes.
Top: Residuals from a smoothing by \citet{vondrak} method, using the
smoothing parameter $\varepsilon=1.0\cdot10^{-15}$.
Middle: Residuals from the local 'triple-star' solutions (see the text
for details).
Bottom: Residuals for the RVs of the \ha emission wings also based on the
'triple-star' local solutions. Data from Feros, Elodie and DAO spectra
are shown by blue circles, OND by red, Trieste by magenta, and BeSS
by green circles.}
\label{orbplot}
\end{figure}

We first extended the analysis that had previously been carried out by \citet{zarf13}.
We used all RVs listed in Table~\ref{jourv}, adopting the \ha absorption
core RVs for all new spectra that we had already reduced. We removed their long-term
changes (displayed in the top panel of Fig.~\ref{time}) with the help of
spline functions using the program {\tt HEC13}, based on the \citet{vondrak}
method.\footnote{Program {\tt HEC13} with brief instructions on how to use it is available at \url{https://astro.troja.mff.cuni.cz/ftp/hec/HEC13}.}
For a comparison with \citet{zarf13}, we used the same smoothing parameter
$\varepsilon=1.0\cdot10^{-15}$ as they did.
The RV residuals were used as the input into program \fotele, with which new
circular orbital elements were derived. The results are in Table~\ref{orbsol} and the phase plot is in the upper panel of Fig.~\ref{orbplot}.

Next, we tried another approach. Using the subsets of RVs from reasonably
well covered individual long cycles and the program \fotele, we formally calculated 'triple-star' solutions, treating the $\sim2000$~d cycles as
'an outer orbit'. This allowed us to obtain the residuals from the long
cycles and combined them into one complete data set. As seen in the middle
panel of Fig.~\ref{orbplot}, this resulted in a better-defined RV curve
and a solution with lower mean rms error (cf. Table~\ref{orbsol}).
As Table~\ref{cycles} shows, the length of individual cycles varied between about 1800 and 3000 days,
as was already noted by \citet{zarf13}.

\setcounter{table}{4}
\begin{table}
\caption{Several \fotel orbital solutions for the RV residuals prewhitened
for the long-term cyclic RV changes. The `rms' denotes the rms error of
one measurement.}
\label{orbsol}
\begin{center}
\begin{tabular}{lcccccl}
\hline\hline\noalign{\smallskip}
Element                    &Sol. 1      & Sol. 2     & Sol. 3 \\
\noalign{\smallskip}\hline\noalign{\smallskip}
$P$ (d)                    &214.667(68) &214.716(33) &214.716 fixed\\
$T_{\rm max.RV}$ (RJD)     &43057.0(3.2)&43057.6(1.6)&55079.7(4.5)\\
$K_1$ (\ks)                &4.96(30)    & 8.36(17)   &3.02(36) \\
rms (\ks)                  &5.30          &4.16      &3.32 \\
\noalign{\smallskip}\hline\noalign{\smallskip}
\end{tabular}
\end{center}
\tablefoot{Solution 1 is for the RV residuals from the spline fits with
{\tt HEC13}; solution~2 is for RV residuals from the locally derived
'triple-star' solutions; and solution~3 is the same for the \ha emission-wings
RVs.}
\end{table}

\begin{table}
\caption{Calculated duration of individual RV cycles as derived
from the 'triple-star' solutions.}
\label{cycles}
\begin{center}
\begin{tabular}{ccccccl}
\hline\hline\noalign{\smallskip}
Cycle length&Time interval& No. of RVs& Note \\
(d)         & RJD         \\
\noalign{\smallskip}\hline\noalign{\smallskip}
2012\p18 &25160--28077&24\\
1902\p49 &33103--36053&21\\
1809\p42 &43692--44745&36\\
2983\p75 &44795--48800&65\\
2562\p171&52462--53992&86&1\\
2400\p69 &52462--53992&89&2\\
2076\p49 &54002--55849&33&1\\
2063\p31 &54002--55849&33&2\\
2232\p43 &56485--59072&31&1\\
2170\p39 &57181--59072&20&2\\
\noalign{\smallskip}\hline\noalign{\smallskip}
\end{tabular}
\end{center}
\tablefoot{  1... based on \ion{Fe}{ii} RVs;
             2... based on \ion{Si}{ii} RVs.}
\end{table}

It is reassuring to note that the orbital elements obtained by the two different
methods agree mutually within the limits of their errors. Thus, for the remainder of
this study, we adopt the following linear ephemeris, based
on the more accurate solution~2 of Table~\ref{orbsol}:

\begin{equation}
T_{\rm max.RV}={\rm RJD}\,\, 43057.6+214\fd716\cdot E\,.\label{efem}
\end{equation}

\subsection{Final orbital elements and the properties of the binary}

\citet{hec2003} showed that also seemingly photospheric absorption lines
can be slightly filled by a weak emission coming from the rotating circumstellar
envelope. Given the Roche-model geometry, it seems logical that more
emission will come from the `nose' of the Roche lobe, which is
facing the secondary. This leads to a situation where, upon observing
the Be component receding from us, it exhibits a maximum RV but since the
violet wing of the line is more strongly filled by the emission,
the measured RV of the apparently absorption line is actually more
positive than what would correspond to the orbital RV at that orbital
phase. An opposite effect is observed when the Be star is approaching to the observer. As a consequence, the amplitude of the RV curve of such an absorption
line is higher than the true amplitude of the orbital motion. The effect
is modelled and discussed in more detail in Appendix~\ref{apc}. In contrast
to it, the RV of the steep wings of the \ha emission, which is formed
in the inner, more or less spherically symmetric parts of the Be disk,
describes the true orbital velocity much better
\citep[see e.g.][]{zarf21,zarf26}. This conclusion was nicely confirmed
by \citet{peters2013} who obtained the true RV curves of both components
of V832~Cyg = 59~Cyg and found that the RV curve of the Be primary agrees
with that derived from the \ha emission-wings RVs by \citet{zarf21}.
The effect was found for quite a few binary Be stars.

   We removed the cyclic RV changes of the \ha emission-wing RVs once
again via local fits of a 'triple-star' solutions and then derived
the final orbital solution~3 of Table~\ref{orbsol}, keeping the orbital
period fixed at the value from ephemeris~(\ref{efem}). The semi-amplitude
of this solution is indeed smaller than those for solutions~1 and 2, based
on absorption-line RVs.

\citet{zorec2016} published a critical study of the distribution of
rotational velocities of a group of Be stars including \ve. They obtained
observed radiative properties of the stars and their apparent \vsin\
values. They carefully analysed the effects of gravity darkening,
macroturbulence, and other effects and then attempted to estimate corrected values that would correspond to non-rotating stars having the same mass as
the studied objects. Their results for \va are summarised in
Table~\ref{radpar}. Using the corrected values, they also estimated
the mass of \va $m=6.2\p0.3$~\ms\ and inclination of the rotational axis
$i=88^\circ\p22^\circ$.

\begin{table}
\caption{Observed and corrected radiative properties of \va derived
by \citet{zorec2016}.}
\label{radpar}
\begin{center}
\begin{tabular}{lcccccl}
\hline\hline\noalign{\smallskip}
parameter & observed & corrected  \\
\noalign{\smallskip}\hline\noalign{\smallskip}
\tef (K) & 14260\p360 & 16580\p400 \\
lgg\ (cgs) & 3.08\p0.21 & 3.64\p0.21 \\
$\log L$ ($L_\odot$) & 2.954\p0.035 & 3.181\p0.043 \\
\vsin\ (\ks) & 275\p17 & 295\p25\\
\noalign{\smallskip}\hline\noalign{\smallskip}
\end{tabular}
\end{center}
\end{table}

\begin{table}
\caption{Binary properties estimated for several orbital inclinations $i$
from the mass function of solution~3 of Table~\ref{orbsol}
$f(m)=0.0006108$~\ms\ and the Be-star mass $m_1=6.2$~\ms.}
\label{masses}
\begin{center}
\begin{tabular}{ccccccl}
\hline\hline\noalign{\smallskip}
$i$   & $m_2$&$m_2/m_1$&$K_2$&a\\
($^\circ)$ & (\ms) &   & (\ks) & (\rs)\\
\noalign{\smallskip}\hline\noalign{\smallskip}
 90.0 &  0.2954 &  0.0476 & 63.32 & 281.55\\
 80.0 &  0.3001 &  0.0484 & 62.33 & 281.61\\
 70.0 &  0.3150 &  0.0508 & 59.38 & 281.83\\
 60.0 &  0.3427 &  0.0553 & 54.57 & 282.23\\
 50.0 &  0.3893 &  0.0628 & 48.05 & 282.90\\
 40.0 &  0.4676 &  0.0754 & 40.00 & 284.01\\
 30.0 &  0.6096 &  0.0983 & 30.68 & 286.01\\
\noalign{\smallskip}\hline\noalign{\smallskip}
\end{tabular}
\end{center}
\end{table}

Using the corrected values above and interpolating them in a grid of
non-rotating models of \citet{schaller92}, we find that the radius
of the star should be about 5.4~\rs, which is in a broad agreement
with the assumption that the photometric 0\fd84 period could be the
rotational period of the star. However, the uncertainties of all parameters
involved are too large to reach a firm conclusion.

Adopting the mass of 6.2~\ms\ for the Be primary and solution~3
of Table~\ref{orbsol}, we estimated the probable properties of
the binary for several possible orbital inclinations. They are
summarised in Table~\ref{masses}. It is clear that the semi-major axis
is quite insensitive to the orbital inclination within the considered
range of inclinations.

The Hipparcos \citep{esa97} and Gaia
\citep{gaia2016, gaia2018, arenou2018} satellites measured the parallaxes of \va with following results:
   $p_{\rm Hip}=0\farcs00339\pm0\farcs00083$ (distance $d$ from
237 to 391~pc), later revised by \citet{leeuw2007a,leeuw2007b} to
  $p_{\rm Hip} = 0\farcs00403\pm0.00031$ ($d = 230-269$~pc), and
  $p_{\rm GaiaDR2} = 0\farcs00367\pm0\farcs00061$
 ($d=234-327$~pc), respectively.
Adopting the Gaia DR2 value of the parallax, we estimate the angular
projection of the semi-major axis of the \va binary to be $\vartheta=0\farcs0048$. \citet{horch2020} observed \va in 2015 as a part
of their differential speckle survey and were unable to detect any
wide companion at angular separations larger than 0\farcs2.

\section{Discussion}
\citet{okazaki91} was the first to suggest that the cyclic long-term variations of Be stars could be a~consequence of a global one-armed oscillation in the Be star disk. Such an oscillation manifest itself as a density enhancement in an (essentially Keplerian)
rotating disk, which gradually revolves in space,
with a typical cycle length of several years.
\citet{oktar2009} investigated the properties
of such oscillations for the binary Be stars.  \citet{telting94}
discussed the expected parameter correlations for that model for
the objects seen roughly edge-on and predicted that for a~prograde
revolution of the density enhancement, the shell cores of the absorption
lines should get deepest during the transition from $V>R$ to $V<R$.
\citet{mennic97} attempted to interpret the observed
light and spectral variations of several Be stars
within the framework of that model. For \ve, they
had continuous \uvby\ ESO photometry covering more
than one long cycle and showing large cyclic
variations in the $u$ colour (see Fig.~\ref{fotom}
here). Using a limited set of \ha $V/R$ variations found by these authors from the literature, they tentatively suggested that the light minimum coincides with the cyclic transition from $V>R$ to $V<R$,
which indicates a~prograde revolution of the one-arm oscillation.

\begin{figure}
\centering
\includegraphics[scale=0.7]{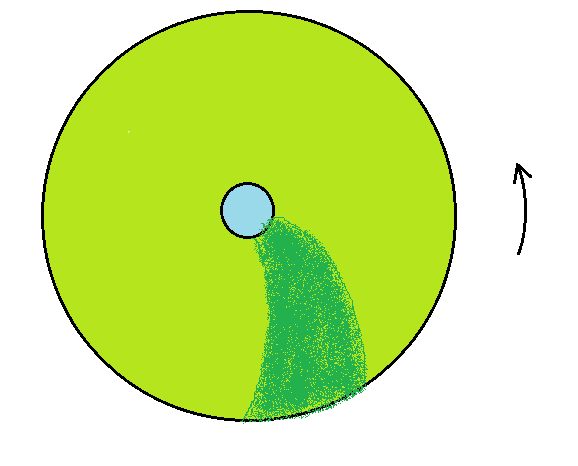}
\caption{Sketch of the model of the circumstellar disk around the Be component with one-armed density enhancement curved in space.}\label{model}
\end{figure}

  The rich observational coverage of spectral, light, and colour variations that we collect here provides a very broad and detailed description that qualitatively supports the model of prograde one-arm oscillation. We recall that both the estimates of the disk inclination by \citet{zorec2016} and our finding that \va shows an inverse correlation in the \ub\ versus \bv\ diagram indicate that we observe the disk more or less equator-on. Therefore, the shadowing effects of the density enhancement inside
the disk are easily possible. However, the absence of detectable binary eclipses means that the inclination is not close to $90^\circ$.
As Figs.~\ref{rvnew}, \ref{time}, \ref{eqwh3}, \ref{eqw}, \ref{fotom},
and \ref{time52} show, the light minima, very pronounced in the ultraviolet passbands,  indeed coincide with the phase, when $V=R$ for the double \ha emission on the descending branch of the $V/R$ curve.
The measured central intensities for both the \ha core and metallic lines show that these shell lines indeed get deeper. However, they become strongest only some
500 days later than the photometric minima. This probably indicates that the revolving region of increased density is curved and lags behind the 'root'
of one-arm oscillation wave during its revolution. This is why it produces the largest absorption against the stellar disk a bit later, as schematically shown
in Fig.~\ref{model}.

A comparison of Figures~\ref{eqwh3} and \ref{time52} also shows that the \ha emission is strongest at the light minima and vice versa.
We suspect that it can be only an~apparent effect, a~consequence of the fact that the observed continuum, to which the line is normalised, gets
smaller during the light minima. Such an~effect
was documented for $\beta$~Lyr by \citet{ak2007}. We note, however, that
the secular increase of the emission strength over about
18000~days has nothing to do with the one-armed oscillation and must be
related to the still unknown mechanism of the formation and dispersal of Be envelopes.

\section{Conclusion}

In this work, we collect, homogenise, and analyse  all the available spectral and photometric observations of the well-known emission-line Be star \ve. We confirm
the binary nature of the object and improve the accuracy of the value of its orbital period, adjusting it to: 214\fd716$\pm$0\fd033. We also provide the most complete description of its cyclic long-term changes. \va is a~rather rare object, which has exhibited cyclic
long-term changes over the whole recorded time interval of about 25000~days, during which
it never lost the Balmer emission. By analysing the mutual relation of the variations
in a number of observed variables, we were able to  qualitatively confirm the tentative conclusion of \citet{mennic97} stating that the long-term variations can be understood as variable
shadowing of the stellar disk by the density enhancement produced by one-armed oscillation with a~prograde revolution within the rotating Be-star disk.

We hope that the future comparison of similar studies for other well-observed Be stars
will lead to a definition of the taxonomy of their variations, which is seemingly quite different
for different objects, and to progress in improving our understanding of this intriguing
class of hot stars. In a future study, we plan to apply a~hydrodynamical model of the disk to a~quantitative modelling of the observed changes.

\begin{acknowledgements}
We acknowledge the use of the program \fotel written by P.~Hadrava, and
the program {\tt reSPEFO} written by A.~Harmanec.  We thank
P.~Hadrava, A.~Kawka, D.~Kor\v{c}\'akov\'a, M.~Kraus, J.~Kub\'at,
B.~Ku\v{c}erov\'a, P.~Nem\'eth, M.~Netolick\'y, J.~Polster, S.~Vennes,
and V.~Votruba, who obtained a number of \ond\ spectra used in this study.
J.R.~Percy kindly provided us with the digital archive of systematic \ubv
\ observations of bright Be stars secured by him and his collaborators.
Our thanks also belong to G. Burki, who provided us with Geneva 7C
observations of a number of Be stars, including \ve, to H.F.~Haupt,
who provided us with their individual \ubv \ observations secured
at the OHP Chiron station, and to J.~Guti\'errez-Soto, who provided us
with the individual OSN \uvby \ observations of \ve. H.~Maehara provided
us with information
about their KWS wide-field photometry and gave us the permission
to use these data. Denis Gillet kindly provided us with copies of
several OHP observing diaries, which allowed us to check the dates and
times of exposures of the OHP spectra used by \citet{deniz94}.
During the corona lockdown in the spring of 2020, the staff of the \ond\ Observatory Library put us into contact with E. Sulistialie from Bosscha Observatory who kindly sent us a copy of the \citet{palmer68} paper. This
allowed us to check the times and RVs of \va\ published there.
This work has made use of the BeSS database, operated at LESIA,
Observatoire de Meudon, France: http://basebe.obspm.fr and we thank
the following amateur observers, who contributed their spectra:
P.~Berardi, Ch.~Buil, A.~de Bruin, V.~Desneoux,  A.~Favaro,
O.~Garde, T.~Garrel, K.~Graham, J.~Guarro Fl\'{o}, F.~Houpert,
P.~Lailly, G.~Martineau, J.~Ribeiro,  O.~Thizy. We also thank
to Jan K\'ara, who helped us to prepare the upper panel of
Figure~\ref{false}. Conctructive criticism and useful suggestions
by an anonymous referee helped to improve the clarity of the paper and
are greatly appreciated.
P.H. and M.W. were supported by the Czech Science Foundation grant GA19-01995S.
H.B. acknowledges financial support from Croatian Science Foundation under the project 6212 ``Solar and Stellar Variability".
D.R. acknowledges financial support from Croatian Science Foundation under the project 7549 ``Millimeter and submillimeter observations of the solar chromosphere with ALMA".
The following internet-based resources were
consulted: the SIMBAD database and the VizieR service operated at
CDS, Strasbourg, France; and the NASA's Astrophysics Data System
Bibliographic Services. This work has made use of data from
the European Space Agency (ESA) mission Gaia
(\url{https://www.cosmos.esa.int/gaia}), processed by the Gaia
Data Processing and Analysis Consortium (DPAC;
\url{https://www.cosmos.esa.int/web/gaia/dpac/consortium}).
Funding for the DPAC has been provided by national institutions,
in particular the institutions participating in
the Gaia Multilateral Agreement.
\end{acknowledgements}

\bibliographystyle{aa.bst}
\bibliography{v923aql.bib}

%%%%%%%%%%%%%%%%%%%%%%%%%%%%%%%%%%%%%%%%%%%%%%%%%%%%%%%%%%%%%%%%%%%%%%%%
%%%%%%%%%%%%%%%%%%%%%%%%%%%%%%%%%   Appendix    %%%%%%%%%%%%%%%%%%%%%%%%
%%%%%%%%%%%%%%%%%%%%%%%%%%%%%%%%%%%%%%%%%%%%%%%%%%%%%%%%%%%%%%%%%%%%%%%%

\begin{appendix}

\section{Some notes on spectra from published papers}\label{apa}
While collecting the RVs and spectrophotometric quantities from published papers, we noted - and if possible - corrected
some misprints and other problems as detailed below.
\begin{itemize}
\item The RVs published by \citet{harp37} generally show  such a~smooth variation that we suspect there is a~misprint
for the spectrum taken on 1930 May 13 (RJD~26109.9755) and that the RV for this spectrum should read as $-$37.4~\ks,
and not the $-$57.4~\kms tabulated by the author.
\item \citet{aab77} published RVs and \ha line profiles for a series of photographic spectra secured in 1974 at
the Special Astrophysical Observatory (SAO). The list of these spectra is in Table~1 of their paper but the problem is
that the dates of exposures quoted there are for one day earlier than the tabulated Julian dates, with the exception of
the last spectrum, where the date and JD agree. Besides, the date of the last but one spectrum is not clear at all.
They also tabulate phases, reportedly calculated for the photometric ephemeris given by \citet{lynds60}, but we were unable
to reproduce these phases, no matter whether we considered the JDs from their table or JDs for one day earlier or later.
They also measured RVs on 1965-1966 Crimean spectra studied earlier by \citet{racko69} but identified them only by
the photometric phases, which we were unable to reproduce. Because of this, we only use mean RVs for mean JD of all those spectra
for the description of the long-term RV changes.
\item \citet{ring81} published an \ha profile from RJD~44450.7 (their Figure~4), which shows a nearly symmetric double \ha emission
with the peak intensity of about 1.5 of the continuum level. The same profile was again reproduced and measured by
\citet{arias2004} but with peak intensities of about 1.9! \citet{arias2004} quote the (obviously) correct JD~2444450.740 for the
first red CTIO spectrum in their Table~2 but give, probably by a mistake, JD~2444450.748 for this spectrum in their Table~3.
This corresponds to the first blue CTIO spectrum. We adopted the value from their Table~2 for the determination
of heliocentric RJD.
\item \citet{deniz94} tabulate JDs of the OHP spectra they used in their Tables II, IV, V, and VI. We noted some obvious problems. Denis Gillet
from OHP kindly provided us with copies of original observing diaries
and we spotted the following misprints in Table II: Correct JD of spectrum GA~7251 is 2446638.443 (not 2446683.443), for GA~7255 it is 2446640.486
(not 2446640.459; the correct value is in Table~IV) and for
GB~9208 the correct JD is 2446640.590 (not 2446640.912). As for other
data sets, we used the mid-exposures from the OPH diaries to calculate
heliocentric Julian dates for all these spectra.
\item \citet{zarf13} used only the first two spectra (out of three) from the paper by \citet{baller87}. There is also
a~misprint in the quotation of the first page of paper by \citet{palmer68}: 71 is given instead of the correct 385.
\item The BeSS database contains one low-resolution (R=6000) red spectrum obtained by French amateur astronomer
Eric Barbotin. The $R$ peak of the \ha profile has an anomalously high intensity and differs dramatically from some
other BeSS spectra taken only a few days apart. Mr. Barbotin informed us upon our inquiry that he is no longer active
in the field of stellar spectroscopy and does not have the software needed to re-reduce the spectrum. As a precaution,
we do not use any quantities measured on his spectrum (file~M of Table~\ref{new}).
\end{itemize}

\section{Details on the photometric data reductions and
some additional data plots}\label{apb}
Since we used photometry from various sources and photometric systems,
both all-sky and differential, relative to several different
comparison stars, we attempted to arrive at some homogenisation and
standardisation. Observations from several observatories were reduced and
transformed to the standard \ubv \ system with the help of the reduction
program \hecdde, which uses non-linear seasonal transformation formul\ae \ and
models changes of the atmospheric extinction in the course of observing nights.
See \citet{hhj94} and \citet{hechor98} for the observational strategy and
data reduction. \footnote{The whole program suite with references to relevant
studies, a detailed manual, examples of data, auxiliary data files, and results is available at
\url{http://astro.troja.mff.cuni.cz/ftp/hec/PHOT}\,.}

Special efforts were made to derive improved all-sky values for the comparison stars used
by observers at different observatories, employing carefully standardised \ubv\ observations secured at Hvar (station 01) over several decades of systematic observations.
The adopted values are collected in Table~\ref{comp},
together with the number of all-sky observations and the rms errors of
one observation. They were added to the respective magnitude differences
for all available differential observations from all observing stations
to obtain directly comparable standard \ubv\ magnitudes. To demonstrate the
internal consistency of all-sky data from other observing stations, whose data were reduced with the program \hecdde, we also show there the
\ubv \ values of observed comparison stars from these stations.

\begin{table*}
\caption[]{ Accurate Hvar all-sky mean \ubv\ values for all comparison stars
used. These were added to the magnitude differences var.$-$comp. and
check$-$comp. for data from all stations. Also shown are the all-sky mean
\ubv \ values for the same comparisons from several other stations, for
which the reductions to the standard system were carried out with the
program \hecdde.}\label{comp}
\begin{center}
\begin{tabular}{crcrcccrr}
\hline\hline\noalign{\smallskip}
Station&Star& HD & No. of& $V$  &  $B$ &  $U$ &$(B-V)$& $(U-B)$ \\
       &    &    &   obs.&(mag)&(mag)&(mag)&(mag) &  (mag) \\
\noalign{\smallskip}\hline\noalign{\smallskip}
01&   HR 7397&183227&1284&5.846&5.861&5.484& 0.015&-0.377\\
04&   HR 7397&183227&   4&5.856&5.868&5.485& 0.012&-0.382\\
30&   HR 7397&183227&  46&5.849&5.855&5.476& 0.006&-0.379\\
66&   HR 7397&183227&   6&5.846&5.852&5.471& 0.006&-0.381\\
\noalign{\smallskip}\hline\noalign{\smallskip}
01& V1431 Aql&183324& 560&5.800&5.888&5.949& 0.088& 0.061\\
04& V1431 Aql&183324&   4&5.806&5.905&5.965& 0.099& 0.060\\
30& V1431 Aql&183324&  72&5.794&5.881&5.952& 0.087& 0.071\\
66& V1431 Aql&183324&  10&5.802&5.887&5.959& 0.085& 0.072\\
\noalign{\smallskip}\hline\noalign{\smallskip}
01&   HR 7404&183387& 775&6.252&7.569&8.931& 1.317& 1.362\\
04&   HR 7404&183387&   2&6.253&7.572&8.937& 1.319& 1.365\\
30&   HR 7404&183387& 775&6.244&7.569&8.924& 1.325& 1.355\\
66&   HR 7404&183387&   6&6.251&7.571&8.922& 1.320& 1.351\\
\noalign{\smallskip}\hline\noalign{\smallskip}
01&   HR 7438&184663& 949&6.374&6.783&6.750& 0.409&-0.033\\
30&   HR 7438&184663& 949&6.368&6.778&6.751& 0.410&-0.027\\
66&   HR 7438&184663& 949&6.366&6.778&6.750& 0.412&-0.028\\
\noalign{\smallskip}\hline\noalign{\smallskip}
\end{tabular}
\end{center}
\end{table*}

Below, we provide some details of the individual data sets and their reductions.
\begin{itemize}
\item {\sl Station 01 -- Hvar:} \ \
 These differential observations have been secured by a number of Croatian and
 Czech observers. Initially, 35~Aql = HD~183324 was used as the principal
 comparison. However, after  \citet{kusch94} reported its microvariability on
 the 0\m01 level and the star obtained a variable-star name V1431~Aql, we used
 HR~7397 = HD~183227 as a new comparison. In several cases, the check star
 HR~7438 = HD~184663 was used as the comparison. The check star was usually
 observed as frequently as the variable. All observations  were transformed
 into the standard \ubv \ system via non-linear transformation formul\ae\ with
 the program \hecdde.
\item {\sl Station 12 -- ESO La Silla:}
\ \  These differential Str\"omgren \uvby\ observations were obtained and
transformed into the standard system in the course of the ESO long-term
monitoring campaign \citep{esofot1,esofot2,esofot3,esofot4}. Comparisons V1431~Aql and HR~7397 were observed along with \ve.
\item {\sl Station 20 -- Toronto:} \ \ These $B$ and $V$ observations were
secured relative to HR~7397 by \citet{percy88} and \citet{percy97} and transformed
into the standard system via linear transformation formul\ae \ by the authors.
We just added the Hvar mean all-sky values of HR~7397 to their magnitude
differences.
\item {\sl Station 30 -- San Pedro M\'artir:} \ \ These \ubv \ observations were
obtained by M.W. and reduced to the standard system with the program \hecdde \ in
a similar way as the Hvar data.
\item {\sl Station 37 -- Jungfraujoch:} \ \ These all-sky seven-colour (7-C)
observations were secured in the Geneva photometric system using the P3
photometer mounted on the 0.76~m reflector. They were transformed into the \ubv \ using the transformation formulae given by \citet{hecboz2001}.
\item{\sl Station 44 -- Mount Palomar:} \ \ These observations were secured by
\citet{lynds59} with a 0.51~m reflector and a photometer with EMI~6094 tube though yellow Corning~3384 filter, relative to HR~7438. Inspecting the magnitude differences
yellow - Johnson $V$ filter for a number of microvariables observed by \citet{lynds59},
we found that the observations are very close to the Johnson $V$ filter. We thus
only added the Hvar $V$ magnitude of HR~7438 to the \citet{lynds60} magnitude
differences \va - HR~7438.
\item {\sl Station 61 -- Hipparcos:} \ \ These all-sky observations were
reduced to the standard $V$ magnitude via the transformation formul\ae\
derived by \citet{hpvb} to verify that no secular light changes in
the system were observed.
%However, for the light-curve solution in \phoebee,
%we consider the Hipparcos transmission curve for the $H_p$ magnitude.
\item {\sl Station 66 -- Tubitak:} \ \ These differential \ubv\ observations were obtained by one of us (H.A.) and transformed to the standard system with
the help of the program \hecdde.
\item {\sl Station 89 -- Canakkale:} \ \ There differential \ubv\ observations
were also obtained by two of us (H.~Bak\i\c{s} and V.~Bak\i\c{s}) and reduced with the program \hecdde.
\item {\sl Station 93 -- ASAS3 $V$ photometry:} \ \ We extracted these
all-sky observations from the ASAS3 public archive \citep{pojm2002},
using the data for diaphragm~1, having, on average, the lowest rms errors.
We omitted all observations of grade D and observations having rms errors
larger than 0\m04. We also omitted a strongly deviating observation at
HJD~2452662.6863.
\item {\sl Station 114 -- ASAS-SN network:} \ \ These $V$ and $g$ band photometries were downloaded from the ASAS-SN database and cleaned
for some deviating data points. We note that the $V$ band photometry is
for about 0\m5 fainter than the usual Johnson $V$.
\item{\sl Station 117 -- KWS $V$ and $I_{\rm c}$ photometry:}
We extracted these CCD photometric observations \citep{maehara2014} from
\centerline{\url{http://kws.cetus-net.org/~maehara/VSdata.py}}
database. It is aperture photometry with fixed aperture radius. The flux
zero-point calibration was carried out by the author, who used stars of
magnitudes 5 to 9 having \bv \ $<$ 1.5 and a scatter smaller than 0\m03 in the
\hp \ magnitude. All the $V$ and $I_{\rm c}$ were published
as instrumental magnitudes. The preliminary transformation formulae to
the standard magnitudes, derived by the author (Maehara, priv.com.), are:
\begin{equation}
V_{\rm inst} - V_{\rm std} = -0.02\cdot(B-V) + const1
\end{equation}
\noindent and
\begin{equation}
I_{\rm c\ inst} - I_{\rm c\ std} = -0.07\cdot(V-I_{\rm c}) + const2,
\end{equation}
\noindent where \bv \ and $I_{\rm c}$ are the standard values
from the Hipparcos catalogue. Since \bv\ $\sim$ 0 for \ve, we
adopted the instrumental $V$ magnitudes for the standard ones.
When using these observations, we omitted a few deviating data
points.
\item {\sl Station 118 -- Sierra Nevada (OSN):} \ \ These differential
\uvby\ observations were obtained by \citet{gutie2007} and kindly
provided by Juan Guti\'errez~Soto as magnitude differences relative
to the comparison star.
\end{itemize}

\begin{figure}
\centering
\resizebox{\hsize}{!}{\includegraphics{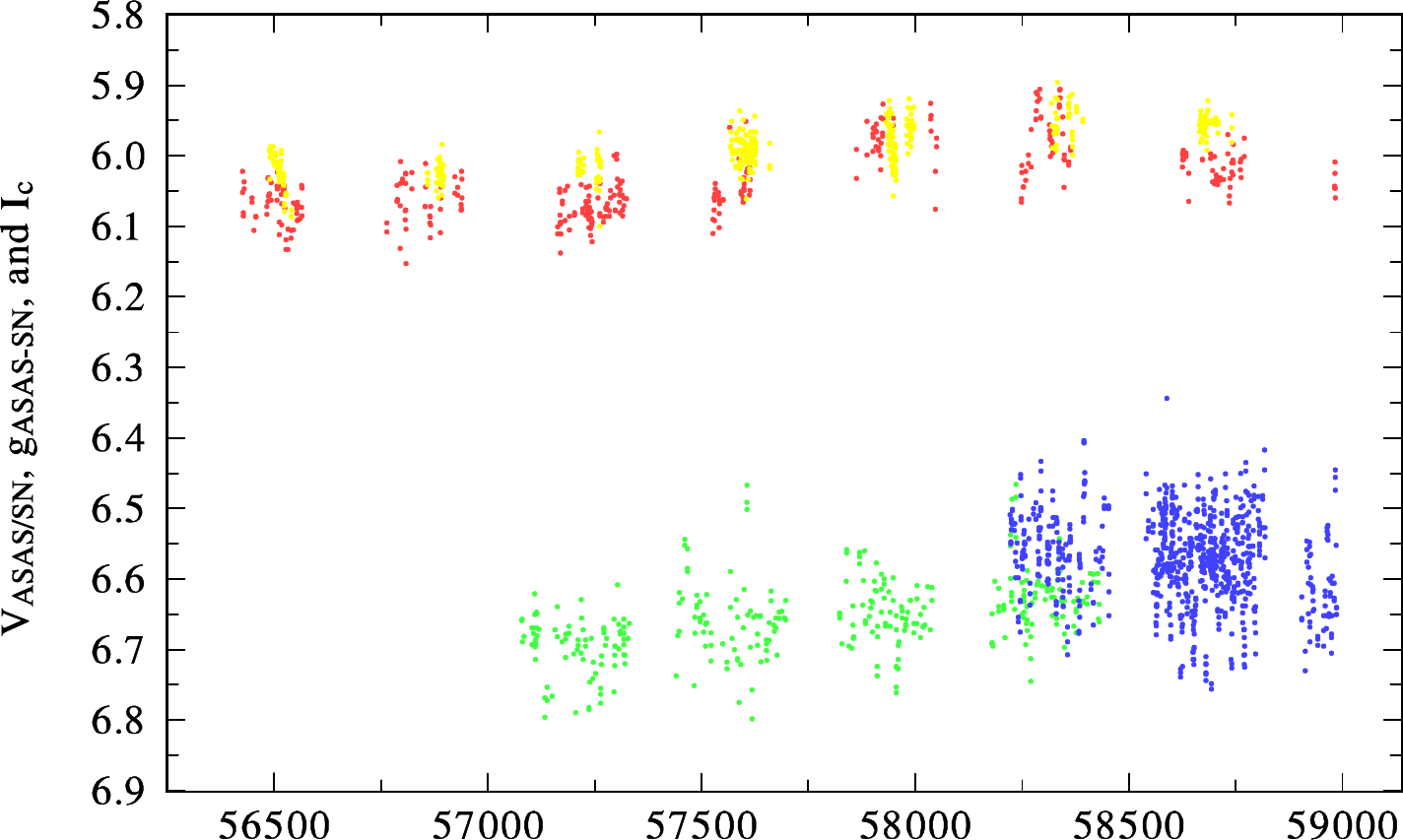}}
\resizebox{\hsize}{!}{\includegraphics{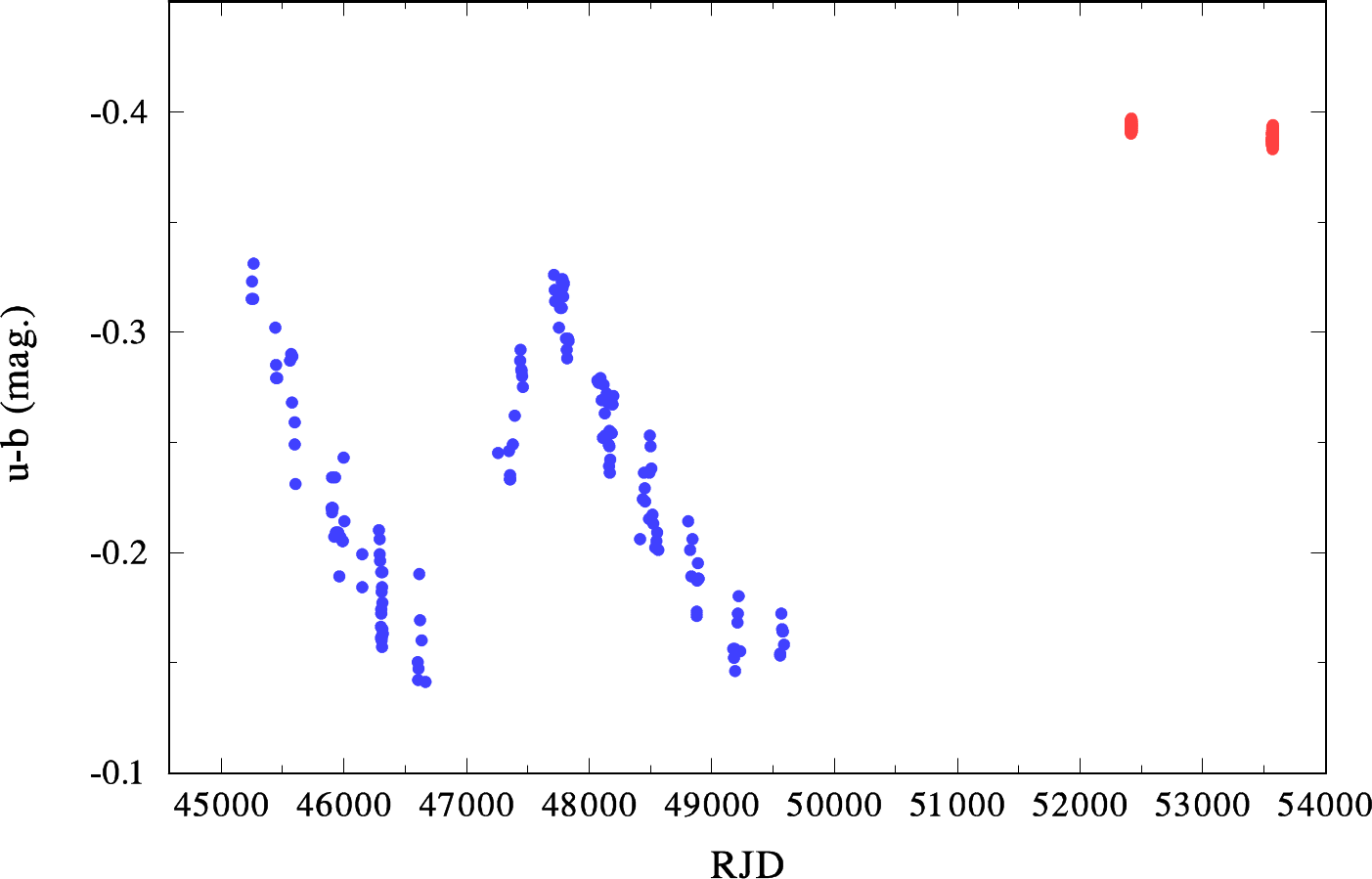}}
\caption{Time evolution of various photometric observations in different
passbands.
Top: Green dots denote the ASAS-SN $V$ magnitude (which is for some 0\m5 fainter
than the usual Johnson $V$ magnitude, blue dots denote the ASAS-SN $g$ filter
observations, red dots denote the KWS Cousins' $I_{\rm c}$ observations, and
yellow dots denote Hvar Johnson $R$ observations.
Bottom:  Stromgren $u-b$ index from ESO observations (blue) and from
Sierra Nevada differential photometry (red). The data generally confirm
the long-term trends seen in the \ubv \ photometry (cf. Figure~\ref{time}),
but with different amplitude and scatter.}\label{fotom}
\end{figure}

\begin{figure}
\centering
\resizebox{\hsize}{!}{\includegraphics{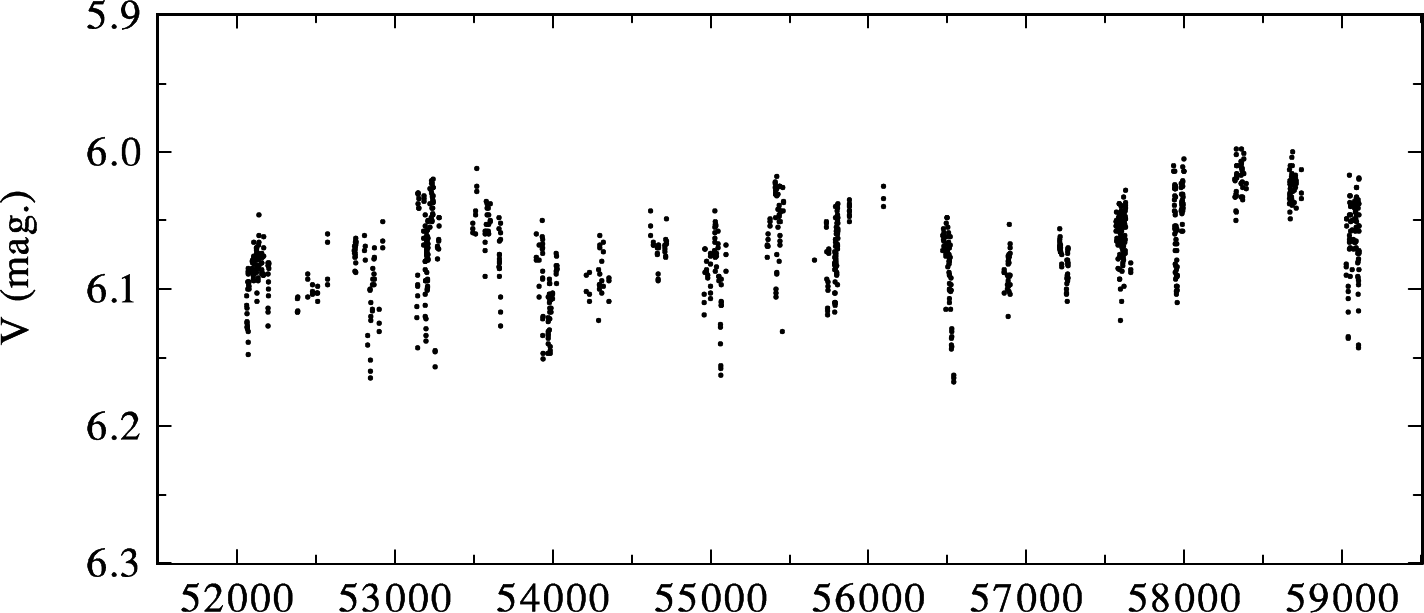}}
\resizebox{\hsize}{!}{\includegraphics{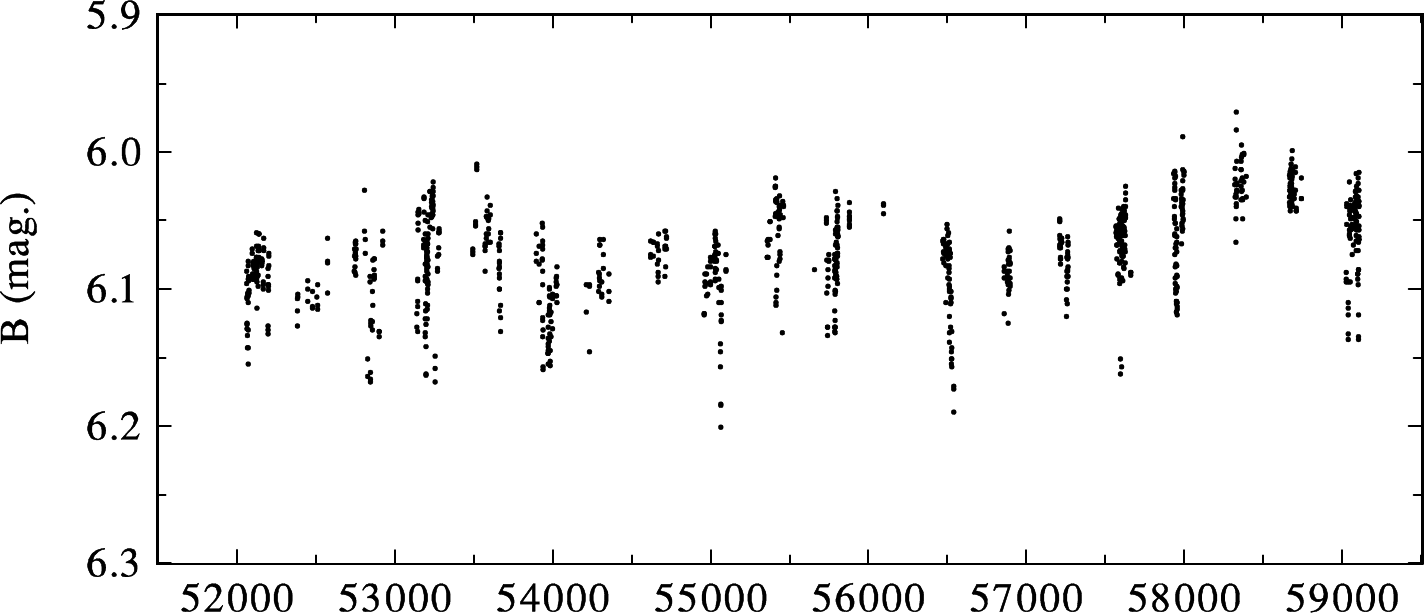}}
\resizebox{\hsize}{!}{\includegraphics{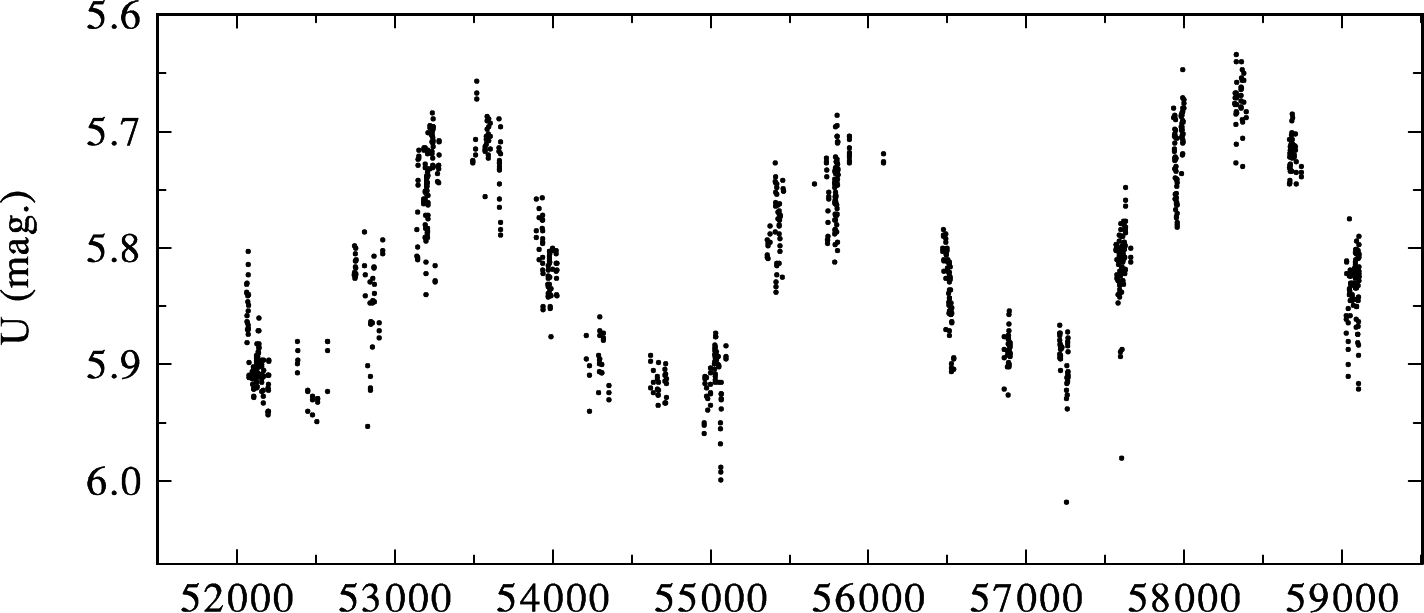}}
\resizebox{\hsize}{!}{\includegraphics{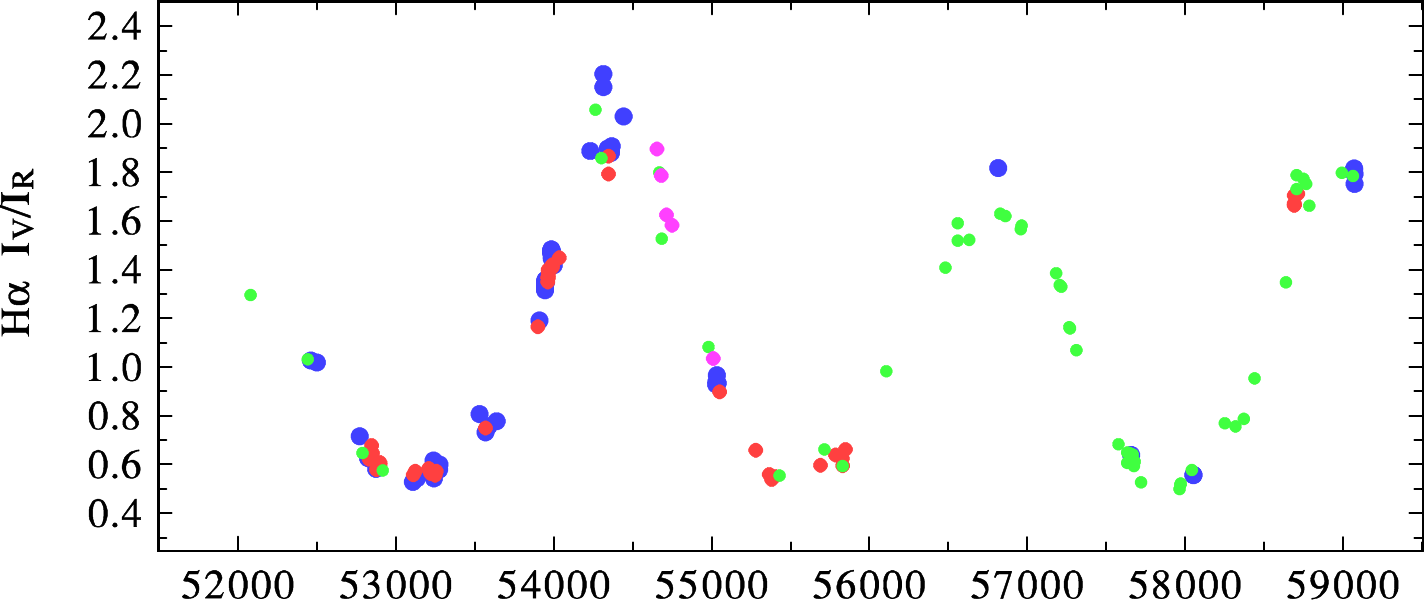}}
\resizebox{\hsize}{!}{\includegraphics{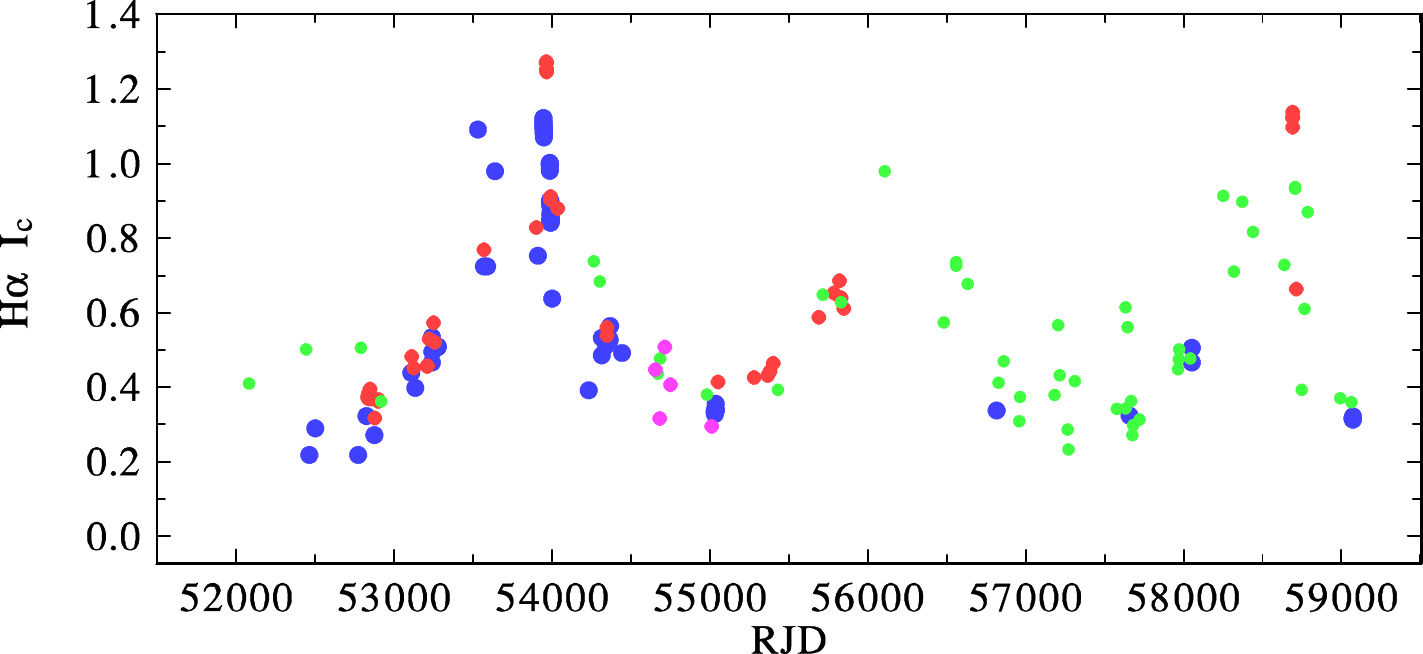}}
\caption{Time evolution of calibrated \ubv\ photometry from stations
1, 20, and 30 (black), and \ha spectrophotometry (V/R and $I_{\rm c}$) over
a limited time interval RJD 51500 to 59500, which is covered by electronic
spectra. Data from the higher-resolution Elodie, Feros, and DAO are shown by blue circles, those from OND by the red circles, those from the BeSS amateur spectra by the green circles, those from the Trieste spectra
re-reduced by us by magenta circles.}\label{time52}
\end{figure}

\begin{figure}[t]
\centering
\resizebox{\hsize}{!}{\includegraphics{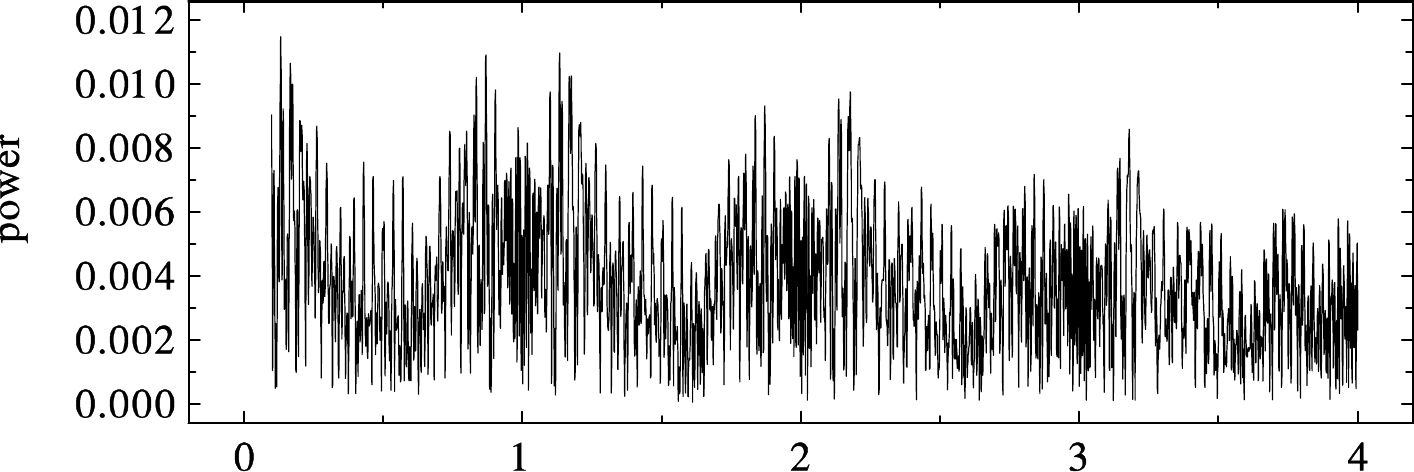}}
\resizebox{\hsize}{!}{\includegraphics{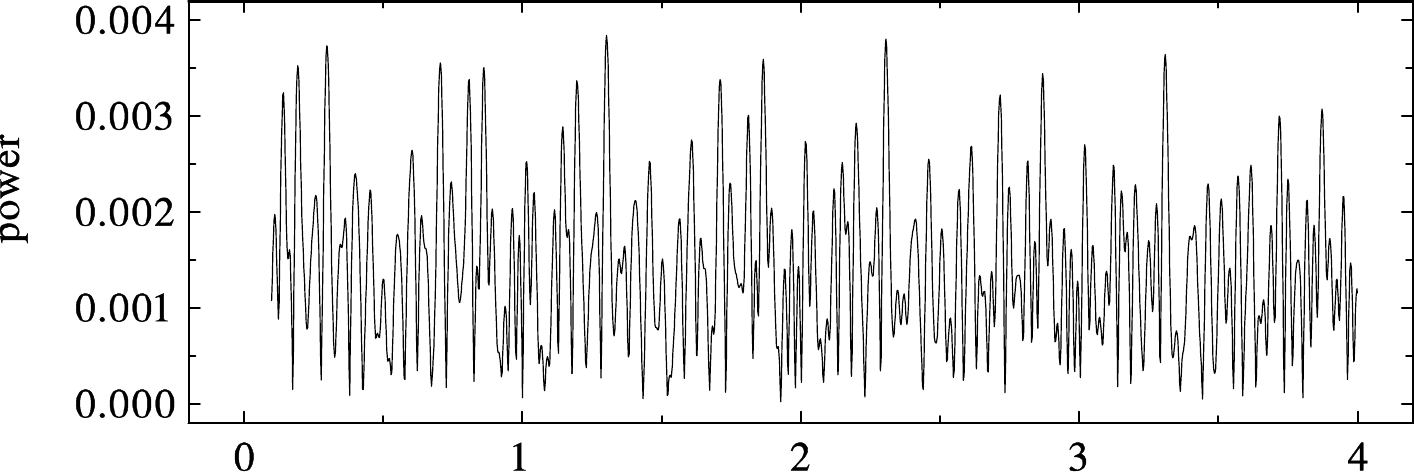}}
\resizebox{\hsize}{!}{\includegraphics{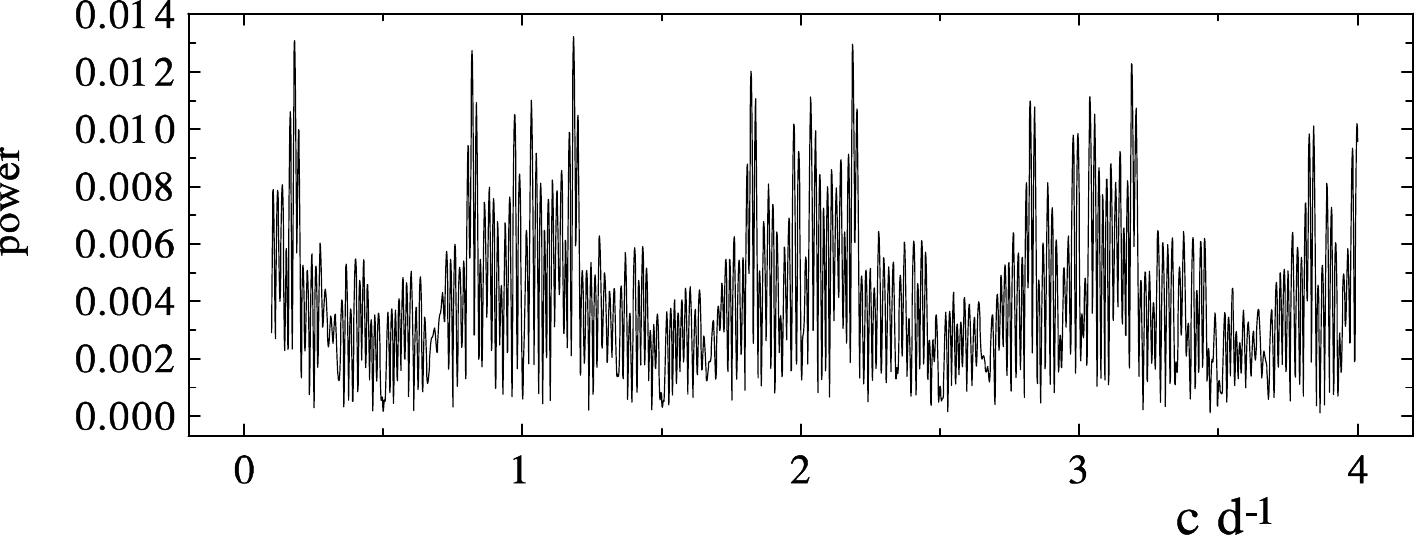}}
\caption{Amplitude periodograms for three $V$ magnitude data subsets
over the range from 0\fd25 to 10\fd0 periods. From top to
bottom: \citet{lynds60} (station 44) photometry; the first season of
Hvar photometry (RJD~44073--44119) and the latest Hvar season
(RJD~59025--59107).}\label{power}
\end{figure}

\clearpage

\section{An explanation for the origin of incorrect semi-amplitudes
of seemingly photopheric lines of Be stars in binary systems}\label{apc}

\begin{figure}[t]
\centering
\includegraphics[angle=0,scale=0.6]{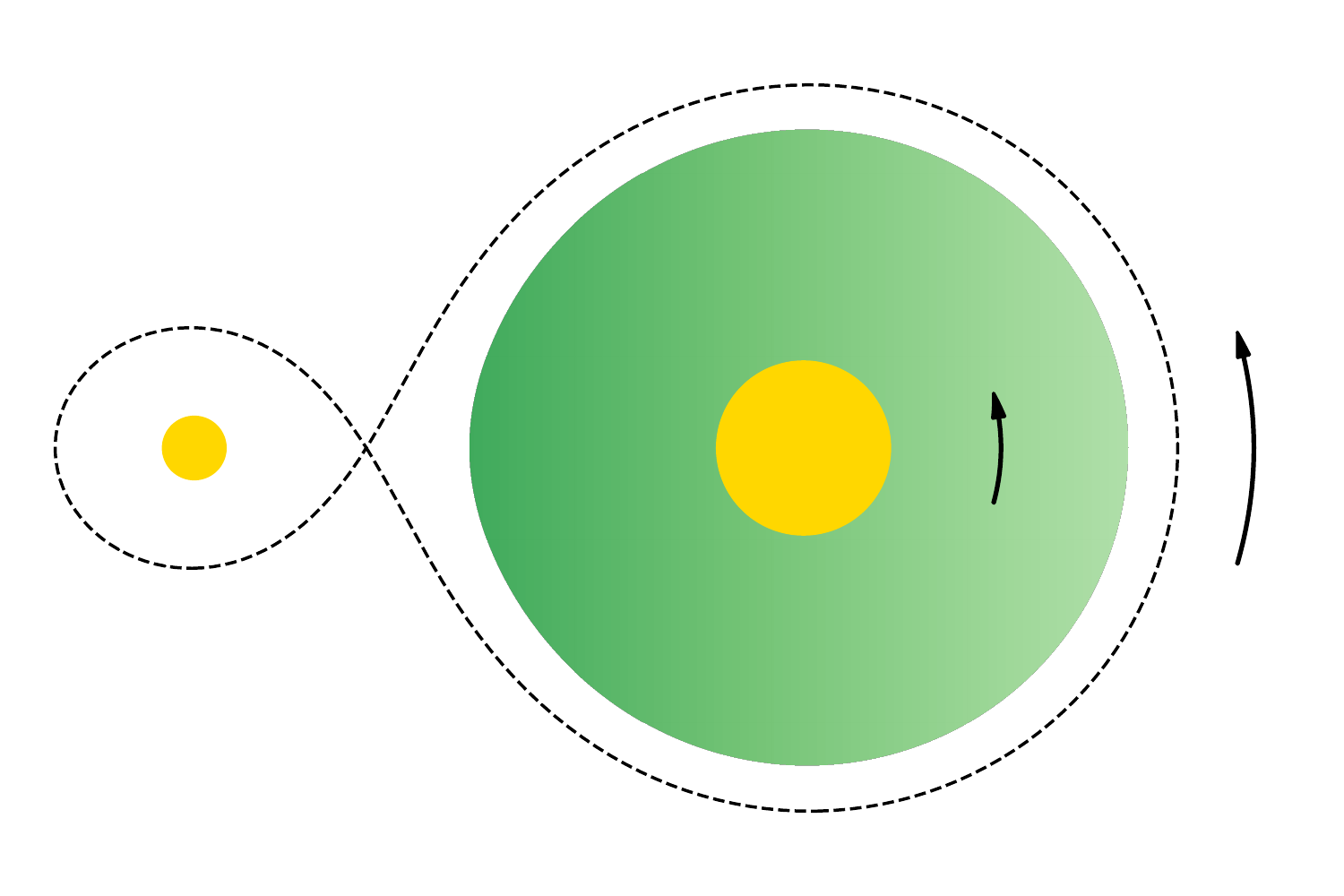}
\resizebox{\hsize}{!}{\includegraphics{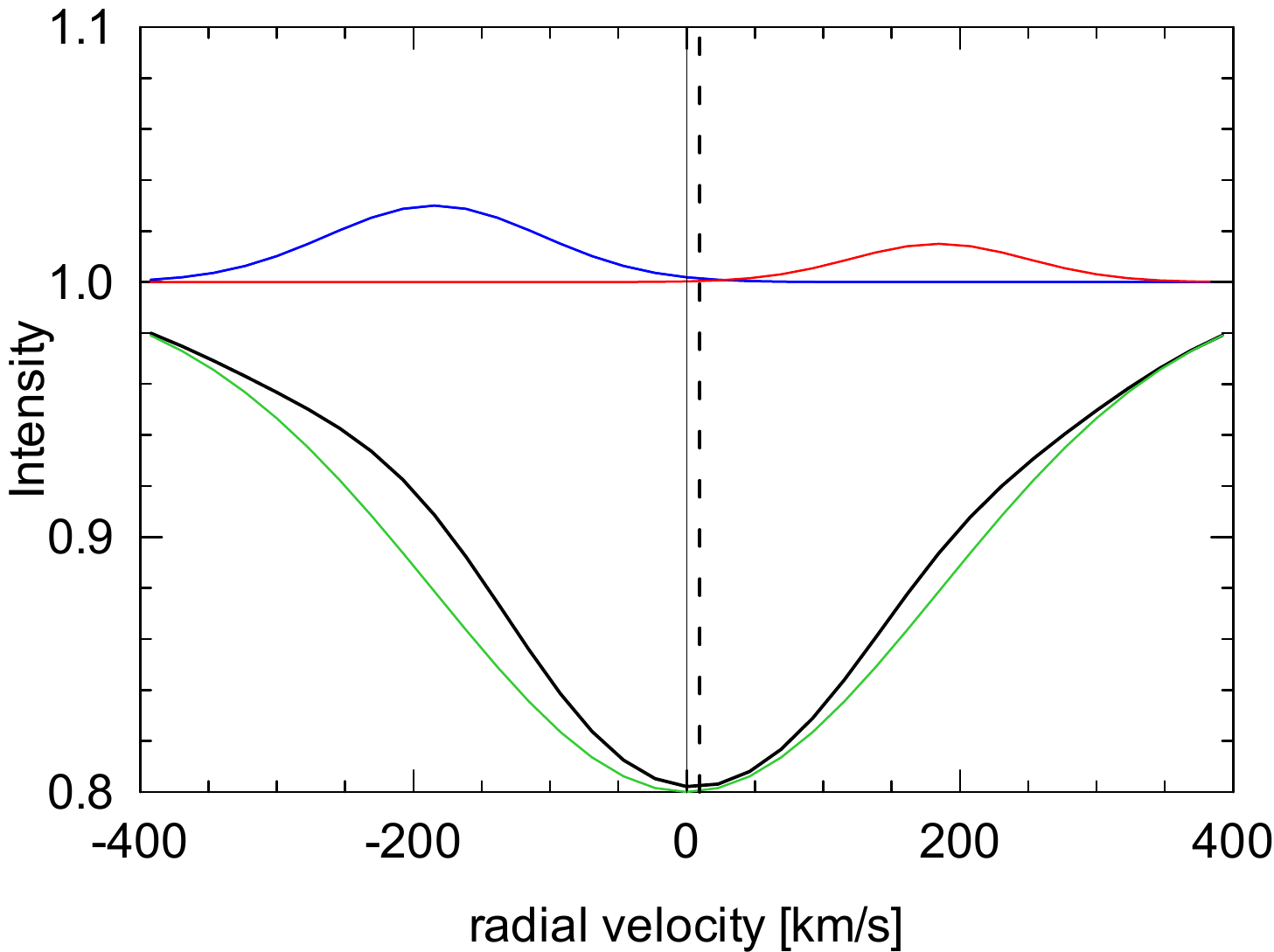}}
\caption{Illustration of the systematic distortion of seemingly
photospheric absorption lines of Be stars in binaries due to unnoticed
weak emission from the envelope with phase-locked $V/R$ changes.
Top: Sketch of the Be-star circumstellar disk located inside the
Roche lobe of a binary system (with a mass ratio of 0.1 in this
example).
Bottom: Plot shows a superposition of the true rotationally
broadened photospheric line profile (green) with a weak double-peak emission having $V/R>1$ (simulated by the blue and red Gaussians). The resulting (black) profile has a more positive RV than the original profile.
This situation corresponds to an observation at the elongation with
the Be component receding from us. Just the opposite happens at the other
elongation.}
\label{false}
\end{figure}

As shown in the upper panel of Figure~\ref{false}, it is probable
that as the Be disk fills a large part of the volume given by
the dimensions of the critical Roche lobe around the Be-star primary and
there will be more emission power in the part of the envelope facing the
secondary. Since the disk rotates, this implies that when the Be star
is receding from the observer and has a maximum RV, also the phase-locked
$V/R$ changes will attain their maximum. If, in addition, the seemingly photospheric
absorption line of the Be star is actually partly blended with a weak
emission contribution from the disk, the phase-locked $V/R$ changes can
affect the measured position of such a line.

  We attempted to model the situation replacing the true spectral lines
by the Gaussian profiles simulating the absorption and double emission
profile. The result of their superposition is shown in the bottom panel of Figure~\ref{false} for the elongation with the Be star receding from us.
It can be seen that this leads to a small but significant shift of the RV measured on the resulting  (and still apparent) absorption line.
In the chosen example, the shift amounts to +10~\kms in the elongation when the $V/R$ is at maximum and the Be star is receding from us. In the opposite
elongation, it is $-10$~\ks, of course. As a consequence, the RV curve measured on such a seemingly photospheric absorption line has a larger
amplitude than what would correspond to the true amplitude of the
orbital RV curve of the Be star itself.

%\pagebreak [4]
%\vfill
%\eject
\newpage

\section{Journals of spectroscopic and photometric observations of V923~Aql}
\begin{table*} [!htbp]
% Toto je tabulka D.1
\caption{Journal of spectroscopic observations of \ve.}\label{new}
\begin{center}
\begin{tabular}{lcrcc}
\hline\hline\noalign{\smallskip}
Spg. & RJD          & No.        & Wavelength &  Spectral  \\
     & range        &            & range      & resolution \\
\hline\noalign{\smallskip}
A FEROS           & 52461.84          & 1  & 6350-6570  & 48000 \\
B OHP             & 52500.44          & 1  & 6530-6600  & 45000  \\
C DAO             & 52771.90--59071.82 & 78 & 6155-6763  & 17200 \\
D OND CCD700      & 52832.39--55849.25 & 46 & 6258-6770  & 12700 \\
E Trieste         & 54653.60--55011.61 & 5  & 6500-6625  & 21000 \\
F OND CCD700      & 58689.47--58714.38 & 5  & 6263-6735  & 12700 \\

\hline\noalign{\smallskip}
\multicolumn{5}{c}{BeSS amateur spectra}\\
\hline\noalign{\smallskip}
G Castanet        & 52080.56--52918.38 & 4 & 6350--6800 &  6000  \\
G Castanet        & 54665.54--55830.33 & 2 & 6500--6700 &  10000 \\
H Atalaia         & 54264.58--55002.64 & 4 & 6520--6610 &  17000 \\
I Paris (Pic du Midi) & 54669.45       & 1 & 6400--6650 &  10000  \\
I Paris           & 55715.58--58747.37 & 2 & 6350--6850 &  4000  \\
I Paris (OHP)     & 57963.47           & 1 & 6400--6640 &  15000 \\
J Piera           & 54685.45--56633.23 & 3 & 6450--6750 &  6000  \\
K Fontcaude       & 55430.53           & 1 & 6500--6620 &  15000 \\
L Dijon           & 56107.57--57577.55 & 4 & 6500--6610 &  13164-18625 \\
M Grande Vallee   & 56484.51           & 1 & 6400--6750 &  6000   \\
N C.A.L.C.        & 56559.42           & 1 & 6480--6600 &  17000   \\
O Bellavista      & 56829.42           & 1 & 6350--6800 &  6139    \\
P St Maurice      & 56860.48           & 1 & 6500--6600 &  11668   \\
Q Canigou         & 56958.27           & 1 & 6520--6680 &  15000   \\
R Verny           & 56963.28--58706.48 & 6 & 6550--6680 &  14686-15414 \\
S Chelles         & 57181.60           & 1 & 6500--6700 &  11000    \\
T Bussum          & 57205.44--58786.26 & 4 & 6400--6750 &  5937-7109 \\
U Alpha           & 57266.69--57719.58 & 6 & 6510--6610 &  17000  \\
V Manhattan       & 57675.58--58765.53 & 3 & 6490--6640 &  10581-12680 \\
W Tourbiere       & 57970.44           & 1 & 6500--6680 &  11000   \\
X Revel           & 57972.35--59062.43 & 5 & 6510--6700 &  11000  \\
Y Desert Wing     & 58991.81      & 1 & 6490--6640 &  17000  \\
\hline\noalign{\smallskip}
\end{tabular}
\tablefoot{The column marked 'Spg' identifies the individual spectrographs and detectors used:
 A: ESO 1.52~m reflector, FEROS spectrograph, EEV 2k$\times$2k detector;
 B: OHP 1.93-m reflector, Elodie spectrograph, CCD Tektronix 1024$\times$1024, observer C.~Catala;
 C: DAO 1.22~m reflector, McKellar spectrograph, 4K$\times$2K SITe CCD detector;
 D: OND 2.00~m reflector, coud\'e grating spg., CCD SITe5 2030 $\times$ 800 pixel detector;
 E: Trieste 0.91-m telescope, REOSC echelle spg., observer G.~Catanzaro;
 F: OND 2.00~m reflector, coud\'e grating spg., CCD Pylon Excelon 2048 $\times$ 512 pixel detector;
 G: Castanet CN212 or C11 LHIRES1 Audine KAF-402ME, observer Ch.~Buil;
 H: Atalaia C14 LHIES3 SBIG ST7 XME, observer J.~Ribeiro;
 I: Paris Saint-Charles C8 or C9 LHIRES3 2400 ATIK460EX, observer V.~Desnoux;
 J: Piera MEADE SC16 LHIRES-B ATIK314L+, observer J.~Guarro Fl\'{o};
 K: Fontcaude CN212 LHIRES3-2400 Atik314L+, observer T.~Garrel;
 L: Dijon C8 LHIRES3 ATIK314L+, observer A.~Favaro;
 M: Grande Vallee TSC225 LHIRES3 ST8XMe, observer E.~Barbotin;
 N: C.A.L.C. C11 LHIRES-2400 QSI683ws, observer J.~Montier;
 O: Bellavista C9 LHIRES3 1200 SXVR-H694, observer P.~Berardi;
 P: St Maurice C11 LHIRES3 2400 QSI583, observer G.~Martineau;
 Q: Canigou SCT LHIRES3-2400 ST8, observer P.~Lailly;
 R: Verny C11 LHIRES3 194-2400t35-QSI516S, observer F.~Houpert;
 S: Chelles C14 eShel ATIK460, observer T.~ Lemoult;
 T: Bussum C11 L200 ATIK 314L+, observer A.~de Bruin;
 U: Alpha 0.51 m Cassegrain, LHIRESIII-2400, observer C.~Sawicki;
 V: Manhattan LX200 12" LHIRES 2400 35u ATIK460EX, obsever K.~Graham;
 W: Tourbiere RC400 Astrosib-Eshel-ATIK460EX, observer O.~Garde;
 X: Revel C11 eShel ATIK460, observer O.~Thizy;
 Y: Desert Wing, C11 LHIRES3-2400 ATIK460EX, observer A.~Stiewing.
 }
\end{center}
\end{table*}

\begin{table*}
\caption{Journal of all RV data sets.}\label{jourv}
\begin{center}
\begin{tabular}{ccrlcccc}
\hline\hline\noalign{\smallskip}
Spg. No. & Time interval & No. of  & Source    \\
         &    (RJD)      &   obs.  &           \\
\hline\noalign{\smallskip}
01 & 25159.66--28076.68 & 24 & \citet{harp37} \\
02 & 33103.32--33248.05 &  2 & \citet{bidel50} \\
03 & 33937.00--34107.00 &  2 & \citet{merrill52} \\
04 & 33110.84--33861.68 &  5 & \citet{gulli76}\\
05 & 33199.70--33861.73 &  9 & \citet{zarf13} file E\\
06 & 35743.52--36053.69 &  3 & \citet{zarf13} file F\\
07 & 41581.56--43357.73 &  9 & \citet{zarf13} file G\\
08 & 40412.84--46216.97 & 30 & \citet{zarf13} file H\\
09 & 37486.51--37830.46 & 12 & \citet{palmer68}\\
10 & 44800.23--44803.21 &  2 & \citet{ring84} \\
11 & 43690.39--47100.26 & 11 & \citet{zarf13} file K\\
12 & 44446.45--47052.50 & 12 & \citet{zarf13} file L\\
13 & 44745.54--47042.36 & 12 & \citet{zarf13} file M\\
14 & 46654.60--47111.46 & 18 & \citet{zarf13} file N\\
15 & 45610.70--45619.68 &  3 & \citet{baller87}\\
\hline\noalign{\smallskip}
16 & 52461.84--58714.38 & 74 & this paper\\
17 & 52080.56--59062.43 & 53 & this paper; amateur spectra\\
\hline\noalign{\smallskip}
18 & 44052.90--44054.90 &  2 & \citet{alduseva80} \\
10 & 44450.75           &  1 & \citet{ring81} \\
19 & 46584.56--48799.61 &  9 & \citet{deniz94}\\
20 & 54653.60--55011.61 &  5 & \citet{catan2013} and this paper\\
\hline\noalign{\smallskip}
\end{tabular}
\end{center}
\tablefoot{Spectrographs 1 - 15 corresponds to files A - O of Table~2 of
\citet{zarf13}. Spectrograph 16 are the Elodie, Feros, DAO, and OND red
spectra and spectrograph 17 are the red BeSS amateur spectra.
Spectrographs 18 to 20 are RVs either not used by \citet{zarf13}, or
published after their publication. One RV obtained by \citet{ring81} was
obtained with the same telescope as the two spectra measured by \citet{ring84}
that were included by \citet{zarf13}. That is why we assign spectrograph No. 10 to it. }
\end{table*}

\begin{table*}
\caption{Journal of available records of the \ha profile.}\label{jouewic}
\begin{center}
\begin{tabular}{ccrlcccc}
\hline\hline\noalign{\smallskip}
 Time interval & No. of  & Source & Note   \\
    (RJD)      &   obs.  &           \\
\hline\noalign{\smallskip}
43007.70 -- 43010.70 & 3 & \citet{font82} & 1 \\
44052.90 -- 44054.90 & 2 & \citet{alduseva80} & 1 \\
44450.75             & 1 & \citet{ring81} & 3 \\
44450.75 -- 51451.00 & 9 & \citet{arias2004} \\
45610.70 -- 45619.68 & 3 & \citet{baller87} \\
46640.49 -- 46990.54 & 3 & \citet{deniz94} \\
47490                & 1 & \citet{doazan91} \\
49096.85 -- 49240.70 & 2 & \citet{hanu96a} & 1 \\
54653.60 -- 55011.61 & 5 & \citet{catan2013} & 2 \\
\hline\noalign{\smallskip}
\end{tabular}
\end{center}
\tablefoot{
1: Fraction of RJD uncertain;
2: Published original spectra re-measured by us;
3: This profile was later re-reduced by \citet{arias2004} \\
}
\end{table*}

\begin{table*}
\caption{Journal of available photometry.}\label{jouphot}
\begin{center}
\begin{tabular}{ccrcccr}
\hline\hline\noalign{\smallskip}
Station & Time interval & No. of & Passbands & Comparison & Check(s) & Source \\
         &    (RJD)      &   obs.  &           &   HD      & HD     \\
\hline\noalign{\smallskip}
44 & 36312.98--36492.70 &  78 & $V$ & 184663 & 183227 & 1\\
26 & 40448.90--40457.90 &   8 &\ubv & 183227 &  --    & 2\\
37 & 43382.63--43755.58 &   2 &7C$\rightarrow$\ubv&all-sky& -- &3 \\
01 & 44073.40--48128.39 & 425 &\ubv & 183324 & 183227, 184663 & 4\\
20 & 44773.75--49951.73 & 118 &$BV$ & 183324 & 183227 & 5 \\
12 & 45249.56--49591.68 & 153 &\uvby& 183324 & 183227 & 6 \\
04 & 45525.50--45573.37 &   5 &\ubv & 183227 & 183324 & 4 \\
61 & 47879.03--48974.17 & 128 &\hp$\rightarrow$$V$ &all-sky&--&  7\\
30 & 52060.81--52752.94 &  43 &\ubv & 183324 & 183227 & 4 \\
01 & 52076.49--53936.53 & 300 &\ubv & 183227 & 184663 & 4 \\
118& 52415.55--52424.56 &  63 &\uvby& 183227 & 183563 & 11\\
93 & 52443.74--55131.52 & 558 & $V$ & all-sky&   --   & 8 \\
66 & 52763.49--52764.54 &   9 &\ubv & 183324 & 184663 & 4 \\
89 & 52850.39--53309.29 &  74 &\ubv & 183324 & 184663 & 4 \\
118& 53565.45--53575.63 & 121 &\uvby& 183227 & 183563 & 11\\
01 & 53937.50--53937.52 &   3 &\ubv & 184663 & 183227 & 4 \\
01 & 53969.36--56475.46 & 293 &\ubv & 183227 & 184663 & 4 \\
117& 55696.30--58983.29 & 454 &$V$  & all-sky& -- & 9 \\
117& 56425.28--58983.29 & 374 &$I_{\rm c}$& all-sky& -- & 9 \\
01 & 56747.35--57116.36 &  57 &\ubvr& 183227 & 184663 & 4  \\
114& 57079.15--58430.69 & 336 &  $V$&all-sky &    --  & 10  \\
114& 58221.90--58986.06 & 821 & $g$ &all-sky &    --  & 10  \\
\hline\noalign{\smallskip}
\end{tabular}
\end{center}
\tablefoot{Column "Station": The running numbers of individual observing
stations they have in the Prague / Zagreb photometric archives: \\
01: Hvar 0.65~m, Cassegrain reflector, EMI6256B (1979-1990), EMI6256S (1990-2003), and EMI9789QB (since 2003) tubes;
04: Ond\v{r}ejov 0.65~m Cassegrain reflector, EMI tube;
12: ESO LaSilla 0.50~m, Bochum 0.61~m, and 0.50~m Danish reflectors;
20: Toronto 0.40~m reflector, DC photometer with EMI6094 tube and
    Optec SSP-3 solid state photometer after 1994;
26: OHP Chiron station 0.60~m reflector, Lallemand tube;
30: San Pedro M\'artir 0.90~m reflector;
37: Jungfraujoch 0.76~m telescope, P3 photometer;
44: Mount Palomar 0.51~m reflector and a photometer with EMI~6094 tube;
61: Hipparcos all-sky \hp \ photometry transformed to Johnson V;
66: Tubitak 0.40~m Cassegrain reflector, Optec SSP5A photometer;
89: Canakkale 0.40~m reflector, Optec SSP5 photometer;
93: ASAS3 photometric survey;
114: ASAS-SN photometric network;
117: Kamogata-Kiso-Kyoto Wide-field Survey (KWS) imager with 0.105~m lens and SBIG ST-8XME CCD Camera;
118: Sierra Nevada 0.90~m reflector, multichannel photometer. \\
Column "Source" gives the source of data: \\
1: \citet{lynds60};
2: \citet{haupt74} and priv.com. to PH;
3: Burki (1980), priv. com. to PH;
4: This paper;
5: \citet{percy88,percy97};
6: \citet{esofot1,esofot2,esofot3,esofot4};
7: Perryman \& ESA (1997);
8: \citet{pojm2002};
9: \citet{maehara2014};
10: \citet{asas2014,asas2017,jay2019};
11: \citet{gutie2007}.
}
\end{table*}
\end{appendix}
\end{document}